\documentclass[]{aa}  

\usepackage{graphicx}
\usepackage{natbib}
\usepackage{derivative}
\usepackage{xcolor}
\usepackage{float}
\usepackage{rotating}
\usepackage{booktabs}
\usepackage{mathtools}
\usepackage{placeins}
\usepackage{bm}
\usepackage{longtable}
\usepackage{adjustbox}
\usepackage{supertabular}
\usepackage{multirow}
\usepackage{tabularx}

\usepackage{txfonts}

\newcommand{\apollinaire}{\texttt{apollinaire} }

\begin{document}

   \title{No swan song for Sun-as-a-star helioseismology: performances of the Solar-SONG prototype for individual mode characterisation}
   \titlerunning{No swan song for Sun-as-a-star helioseismology}

   \author{S.N.~Breton\inst{1}
          \and  
          P.L.~Pallé\inst{2,3}
          \and
          R.A.~Garc\'{i}a\inst{1}
          \and 
          M.~Fredslund Andersen\inst{4}
          \and 
          F.~Grundahl\inst{4}
          \and 
          J.~Christensen-Dalsgaard\inst{4}
          \and
          H.~Kjeldsen\inst{4}
          \and
          S.~Mathur\inst{2,3}
          }
    \institute{AIM, CEA, CNRS, Universit\'e Paris-Saclay, Universit\'e de Paris, Sorbonne Paris Cit\'e, F-91191 Gif-sur-Yvette, France \\
    \email{sylvain.breton@cea.fr} 
    \and Instituto de Astrof\'{i}sica de Canarias, 38205 La Laguna, Tenerife, Spain 
    \and Departamento de Astrof\'{i}sica, Universidad de La Laguna (ULL), 38206 La Laguna, Tenerife, Spain 
    \and Stellar Astrophysics Centre, Aarhus University, 8000 Aarhus C, Denmark 
    }

   \date{}

 \abstract{The GOLF instrument on board SoHO has been in operation for almost 25 years but aging of the instrument has now strongly affected its performance, especially in the low-frequency p-mode region. At the end of the SoHO mission, the ground-based network BiSON will remain the only facility able to perform Sun-integrated helioseismic observations.
 Therefore, we want to assess the helioseismic performances of an {\'e}chelle spectrograph like SONG. Indeed, the high precision of such an instrument and the quality of the data acquired for asteroseismic purpose calls for an evaluation of the instrument ability to perform global radial-velocity measurements of the solar disk. 
  Data acquired during the Solar-SONG 2018 observation campaign at the Teide Observatory are used to study mid- and low-frequency p modes. A Solar-SONG time series of 30-day duration is reduced with a combination of the traditional IDL \texttt{iSONG} pipeline and a new \texttt{Python} pipeline described in this paper. A mode fitting method built around a Bayesian approach is then performed on the Solar-SONG and contemporaneous GOLF, BiSON, and HMI data. 
  For this contemporaneous time series, Solar-SONG is able to characterise p modes at a lower frequency than BiSON and GOLF (1750 $\mu$Hz against 1946 and 2157 $\mu$Hz respectively), while for HMI it is possible to characterise a mode at 1686 $\mu$Hz. 
  The decrease of GOLF sensitivity is then  evaluated through the evolution of its low-frequency p-mode characterisation abilities over the years. A set of 30-day long GOLF time series, considered at the same period of the year, from 1996 to 2017, is therefore analysed. We show that it is more difficult to characterise accurately p modes in the range 1680 to 2160 $\mu$Hz, when considering the most recent time series. By comparing the global power level of different frequency regions, we also observe that the Solar-SONG noise level in the 1000 to 1500 $\mu$Hz region is lower than for any GOLF subseries considered in this work. While the global p-mode power level ratio is larger for GOLF during the first years of the mission, this ratio decreases over the years and is bested by Solar-SONG for every time series after 2000. All these observations strongly suggest that efforts should be made towards deploying more Solar-SONG nodes in order to acquire longer time series with better duty cycles.
 }


   \maketitle

\section{Introduction \label{section:introduction}}

The first detection of oscillations in the Sun \citep{1962ApJ...135..474L,1963ApJ...138..631N} was possibly the event that changed forever the horizon for the study of the dynamics of stellar interiors. A few years later, \citet{1970ApJ...162..993U} and \citet{1971ApL.....7..191L} explained those oscillations in terms of global resonant modes.

The identification of high-degree modal structure in the observed five-minute oscillations \citep{1975A&A....44..371D}, the detections of the 160 minute oscillation by \citet{1976Natur.259...92B} and \citet{1976Natur.259...87S}, identified as a possible solar internal gravity mode (g mode), and claimed oscillations in the solar diameter \citep{1975NYASA.262..472H}, led \citet{1976Natur.259...89C} to point out that such observations would open the way to obtain precise inference about the deep interior of the Sun.
The helioseismic era really began with Sun-as-a-star observations of low-degree p modes by \citet{1979Natur.282..591C} and \citet{1980Natur.288..541G}. Several space missions, namely the
Microvariability and Oscillations of STars mission \citep[MOST,][]{2000ASPC..203...74M}, the Convection, Rotation and planetary Transit satellite \citep[CoRoT,][]{2009A&A...506..411A}, the \textit{Kepler}/K2 mission \citep{2011ApJ...736...19B,2014PASP..126..398H}, and the Transiting Exoplanet Survey Satellite \citep[TESS,][]{2015JATIS...1a4003R} opened the path for asteroseismology. Indeed, over the past two decades, asteroseismology has probed the deep layers of what constitutes now a very large number of solar-like stars \citep[e.g.][]{2019LRSP...16....4G}. Moreover, solar-like oscillations observed in red giants have allowed us to derive their core rotation rate \citep{2011Sci...332..205B,2011Natur.471..608B,2011A&A...532A..86M} and the resulting inferences disrupted the landscape of what was commonly accepted in stellar evolution models concerning angular momentum transport. Combined with previous results obtained with solar data, these observations have been puzzling theoreticians over the last decade \citep[see e.g.][and references therein]{2013LNP...865...23M,2019ARA&A..57...35A}. One of the keys of the enigma resides in the deep-interior dynamics of main-sequence stars: it will be possible to set precise constraints on low-mass stars' core rotation rate only through the detection of individual g modes in those stars. Since the first days of helioseismology, the Sun has always remained the most obvious candidate to observe g modes in a main-sequence star with solar-like pulsations \citep{2013ASPC..478..125A}. Indeed, the fact that we are now able to probe the core dynamics of stars located hundreds of light years away from the Earth while being kept in the dark concerning our own star is somehow incredibly frustrating.   

Large efforts were undertaken in order to characterise solar oscillations with high precision. In 1976, Mark-I, the first node of what would become the Birmingham Solar Oscillations Network \citep[BiSON,][]{1996SoPh..168....1C,doi:10.1093/mnras/stu803,s11207-015-0810-0} was deployed in Tenerife at the Teide Observatory. The IRIS network \citep{2003A&A...408..729S} operated from 1989 to 1999 while the Global Oscillations Network Group \citep[GONG,][]{1996Sci...272.1284H} began operating in 1996. However, the culminating event of the golden era of helioseismology was without doubt the launch of the Solar Heliospheric Observatory \citep[SoHO,][]{1995SoPh..162....1D}.
Bringing to space three instruments dedicated to probe the solar interior, the Global Oscillations at Low Frequency instrument \citep[GOLF,][]{1995SoPh..162...61G}, the Variability of solar IRradiance and Gravity Oscillations instrument \citep[VIRGO,][]{1995SoPh..162..101F} and the Solar Oscillations Investigation's Michelson Doppler Imager instrument \citep[SOI/MDI,][]{1995SoPh..162..129S}, SoHO was thought to encompass all the tools needed to unravel the last mysteries hidden by the core of our star. More recently, the Solar Dynamics Observatory \citep[SDO,][]{2012SoPh..275....3P} was launched, including the SOI/MDI successor, the Helioseismic Magnetic Imager instrument \citep[HMI,][]{2012SoPh..275..207S}.

At the time when SoHO was launched, GOLF was expected to deliver an unambiguous detection of g modes. With its sodium cell and its two photomultipliers, GOLF was designed to perform differential intensity measurements over both wings of the sodium solar doublet. Those intensity measurements allow for an extremely precise radial-velocity (RV) measurement of the upper layers of the Sun. Over the years, several individual g-mode candidates were reported \citep{2002A&A...390.1119G,2004ApJ...604..455T,2011JPhCS.271a2046G} while a global g-mode pattern was identified with a 99.49 \% confidence level \citep{2007Sci...316.1591G}. 
The recent claim of a g-mode detection with GOLF \citep{2017A&A...604A..40F} was reviewed by several groups who could not reproduce it and have raised serious doubts about the validity of the methodology \citep{2018SoPh..293...95S,2019A&A...624A.106A,2019ApJ...877...42S}.

The Stellar Observations Network Group \citep[SONG, ][]{2007CoAst.150..300G} initiative was conceived with the objective to install an asteroseismology dedicated terrestrial network with several operating nodes in order to maximise the observational duty cycle. Stellar observations are performed by a robotic telescope, the light being fed to a high-resolution {\'e}chelle spectrograph. The acquired spectra are then reduced to obtain high-precision radial-velocity measurements.     
The prototype and first node, coupled with the one-meter Hertzsprung telescope, was built at the Teide Observatory and began its operation in 2014 \citep{2016RMxAC..48...54A}. 
In June 2012, as the installation of the telescope was delayed, an optical fibre mounted on a solar tracker, was installed to feed solar light to the spectrograph during the day \citep{2013JPhCS.440a2051P}. This operation represented the first light of the Solar-SONG initiative. 
The approach is aimed at exploiting the fact that the convective noise is expected to be partially decorrelated at different wavelengths while the p-mode signal remains coherent, as highlighted in a short test run with the GOLF-NG prototype \citep{2008JPhCS.118a2044T,2009ASPC..416..341S}.  
Independently from the Solar-SONG initiative, Sun-as-a-star observations were performed with the spectrograph HARPS-N \citep{2015ApJ...814L..21D,2021A&A...648A.103D}. Their purpose is to increase the precision or RV measurements by characterizing and removing the stellar noise injected in the RV signal, using longer observational cadence, not suited for p-mode observation.

Exploiting the outcomes of the 2012 observation campaign, the power spectrum of one week of Solar-SONG observations was compared with GOLF and Mark-I contemporaneous spectra. The power in the 6000 to 8000 $\mu$Hz region, dominated by photon noise, was 2.5 and 4.4 times lower than in Mark-I and GOLF, respectively. A daily low-cadence follow-up has been carried out since 2017. During the 2018 summer, a high-cadence (3.5 seconds) campaign of 57 days was carried out in order to evaluate the helioseismic performance of the instrument. First results of this campaign were presented in \citet{2019A&A...623L...9F}. 

The potential of an {\'e}chelle spectrograph like Solar-SONG to explore the low-frequency regions of the solar power spectrum can be estimated by considering the instrument's ability to detect low-frequency p modes.
The purpose of this work is to complete and extend the previous analyses by assessing Solar-SONG performances in mid- and low-frequency p-mode ranges, using the GOLF observations, as well as BiSON and HMI observations, to evaluate the Solar-SONG capabilities. 
 

The layout of the paper is as follows. Section~\ref{section:reduction} presents the solar time series that were used for this work. We also give some details about the Solar-SONG data reduction method. In Sect.~\ref{sec:peakbagging}, the peakbagging of the power spectral density (PSD) obtained from the time series is described. We use the peakbagging results to compare GOLF and Solar-SONG performances in Sect.~\ref{sec:comparison}. Those results and the potential of Solar-SONG are discussed in Sec.~\ref{sec:discussion}. For a comparison, a detailed analysis of BiSON and HMI data is included in Appendix~\ref{sec:hmi_bison}. 

\section{Data acquisition and reduction \label{section:reduction}}

A Solar-SONG high-cadence observation campaign took place between 27 May and 22 July, 2018. Observations were carried out in a fully automatic way and the scheduling was handled by an automation software \citep["the Conductor", described in][]{2019PASP..131d5003F}. In the work presented here, we consider only thirty days of observations, spanning 3 June to 2 July, the interval of time with the best set of consecutive days leading to a 47\% duty cycle. This time range yields a good balance between spectral resolution and windowing effects due to the low duty cycle.

\subsection{GOLF data reduction \label{sec:golf_data_selection}}

Due to a loss in the counting rates measured by the photomultipliers resulting from normal aging, the GOLF mean noise level has been increasing over the years \citep{2005A&A...442..385G,2018A&A...617A.108A} in the high and medium frequency regions. Above 4 mHz and around 1 mHz, the photon noise power contribution dominates the PSD. Not only do we want to compare Solar-SONG performances to what GOLF performances are now, but also to what it used to be. We therefore select 22 time series of same length, all at similar epoch of the year in order to ensure that SoHO position on its orbit is each time comparable to what it was during the summer 2018.

We use GOLF time series calibrated using the method described in \citet{2005A&A...442..385G}. GOLF measurements are obtained using the instrument's own time reference and not the SoHO main on-board time reference. However, the GOLF clock experienced on several occasions unexpected events that resulted in time lags \citep[e.g.][]{2018A&A...617A.108A}. {VIRGO is synchronised on SoHO's main on-board clock. It has been used as a reference to cross-correlate GOLF measurements and correct the timing issue of the GOLF data.} 

\subsection{Solar-SONG data reduction}

In high-cadence mode, the SONG spectrograph acquires a spectrum approximately every 3.5 seconds, with an exposure time below one second. The acquired spectrum covers 4400 to 6900 \AA, with a pixel scale of 0.022 \AA, over 51 orders. However, for solar observations, we have used an iodine cell to provide precise wavelength calibration, and for those observations, hence only 24 orders were used, covering the 4994 to 6208 \AA\, range where the iodine cell imprint is present. The IDL \texttt{iSONG} \citep{2012A&A...537A...9C,2013MNRAS.435.1563A} has been used to compute the radial velocities for the solar data. The method that has been developed over the last decades consists in dividing each order of the spectrum into so-called \textit{chunks} and to compute an RV for each of those chunks \citep[see e.g.][]{1996PASP..108..500B}. 22 chunks per order are used for Solar-SONG spectra. For each spectrum, \texttt{iSONG} produces data outputs denoted as \textit{cubes}. These cubes are built as 24x22x27 arrays.
Indeed, 27 parameters are related to each chunk, these include the identifiers of the chunk (given by the order number and the rank of the chunk among the order), the computed RV, the photon flux level, and the observation time. 

The \texttt{iSONG} pipeline is able to carry out the data processing and produces an integrated RV over the chunks, but we introduce in this paper a complementary code as an open source Python module called \texttt{songlib}, which is part of the \texttt{apollinaire}\footnote{The source code is available at \\ \texttt{https://gitlab.com/sybreton/apollinaire}} helio- and asteroseismic library (see Breton et al. in prep and appendix~\ref{appendic:apollinaire_songlib}). The new code is dedicated to obtain the integrated RV starting from the \texttt{iSONG} cubes. It has the advantage of extending the original \texttt{iSONG} abilities by being able to reduce unequally sampled SONG data (with one spectrum acquired every $\sim$ 3.5 s) into regularly sampled velocity measurements.
For this work, we produced data sampled at 20 seconds. 

Starting from the cubes output provided by the \texttt{iSONG} pipeline, each day of observation is then reduced individually.
The first step is to integrate the chunk-relative RV to get a one-dimensional RV vector. Weights are attributed to each chunk by considering
\begin{equation}
    w_{ij} = \frac{1}{\sigma_{ij}^2} \;,
\end{equation}
where $\sigma_{ij}$ is the robust standard deviation of the RV measurements of the $j^\mathrm{th}$ chunk of the  $i^\mathrm{th}$ order. The robust standard deviation $\sigma$ is computed from the median absolute deviation, MAD, as follows:
\begin{equation}
    \sigma \approx \frac{\mathrm{MAD}}{\Phi^{-1}(3/4)} \approx 1.4826 \; \mathrm{MAD} \;,
\end{equation}
where $\Phi^{-1}$ is the normal inverse cumulative distribution function evaluated at probability 3/4\footnote{See the \texttt{astropy} \citep{astropy:2013, astropy:2018} documentation at: \\
\texttt{https://docs.astropy.org/en/stable/ \\
api/astropy.stats.mad\_std.html}}. If $\sigma_{ij}>1$ km/s or $\sigma_{ij}<3$ m/s, the corresponding weight is set to 0. We check that we obtain the same results with \texttt{songlib} and \texttt{iSONG}.
Using the $\sigma_{ij}$, the one-dimensional RV vector is computed as the weighted average of the 528 RV vectors. 
Considering the rms of the point-to-point difference of these daily time series (that is the difference between two consecutive measurements), we compute a proxy $u$ for the spectrograph single-point precision {(that is the typical RV uncertainty related to the acquisition of a single spectrum)}. The proxy is taken as the mean of the obtained rms values 
\begin{equation}
   u = \frac{1}{\sqrt{2}} \underset{i}{\overline {\mathrm{rms}}} \, (v_{i+1} - v_i) \;,
\end{equation}
where $v_{i}$ and $v_{i+1}$ are consecutive RV measurements; we get $0.88 \pm 0.13$ m.s$^{-1}$

The second step is to correct and re-sample this vector.
In order to have RV measurements regularly sampled, so-called \textit{boxes} of 20 seconds are computed.
According to its timestamp, each cube is attributed to a box. The mean RV inside each box is computed. Values beyond three standard deviations are considered as outliers and removed. The same process is repeated with the remaining values, this time considering a two-standard-deviation threshold. To get rid of measurements that would be inconsistent with a longer trend, some outliers are again removed by considering means over a neighborhood of 50 boxes (1000 seconds). Values are again removed in two steps, the first time if they are outside of eight standard deviations, the second time if they are outside six standard deviations.
The RV inside each box is computed as the mean of the remaining cubes values.

\begin{figure}
    \centering
    \includegraphics[width=.48\textwidth]{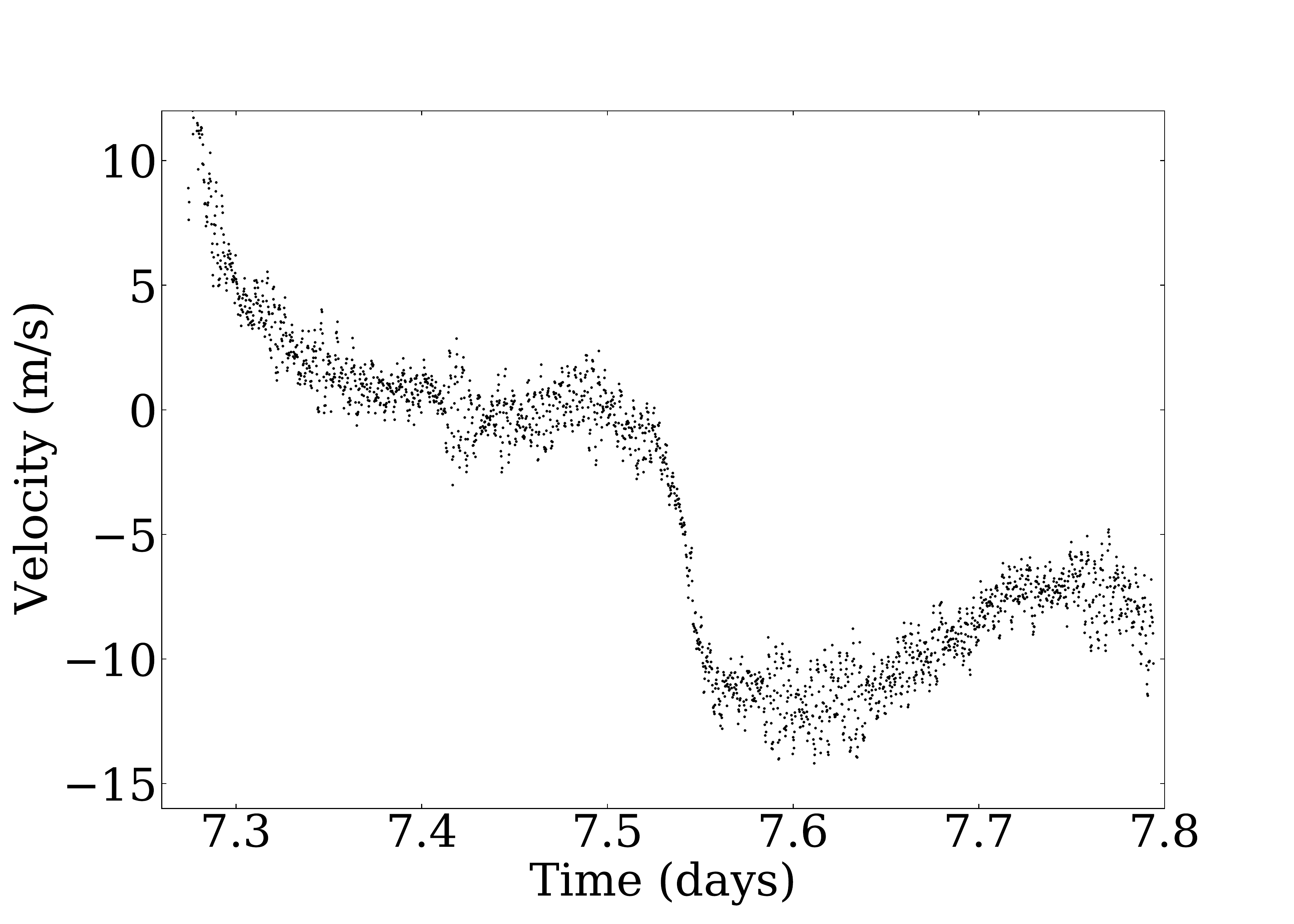}
    \caption{Velocity residual showing the remaining trend before application of the FIR filter in the seventh day of the considered Solar-SONG time series (10 July 2018).}
    \label{fig:daily_trend}
\end{figure}

The ephemeris velocity (including a barycentric correction), obtained from the IMCCE Solar system portal\footnote{Available at: \\ \texttt{https://ssp.imcce.fr/forms/visibility}}, is finally substracted from each box measurement. The Julian time noon velocity value is also substracted from every measurement so that the residual velocity after the ephemeris correction is zero at noon (see Fig.~\ref{fig:daily_trend}).

\begin{figure}
    \centering
    \includegraphics[width=.48\textwidth]{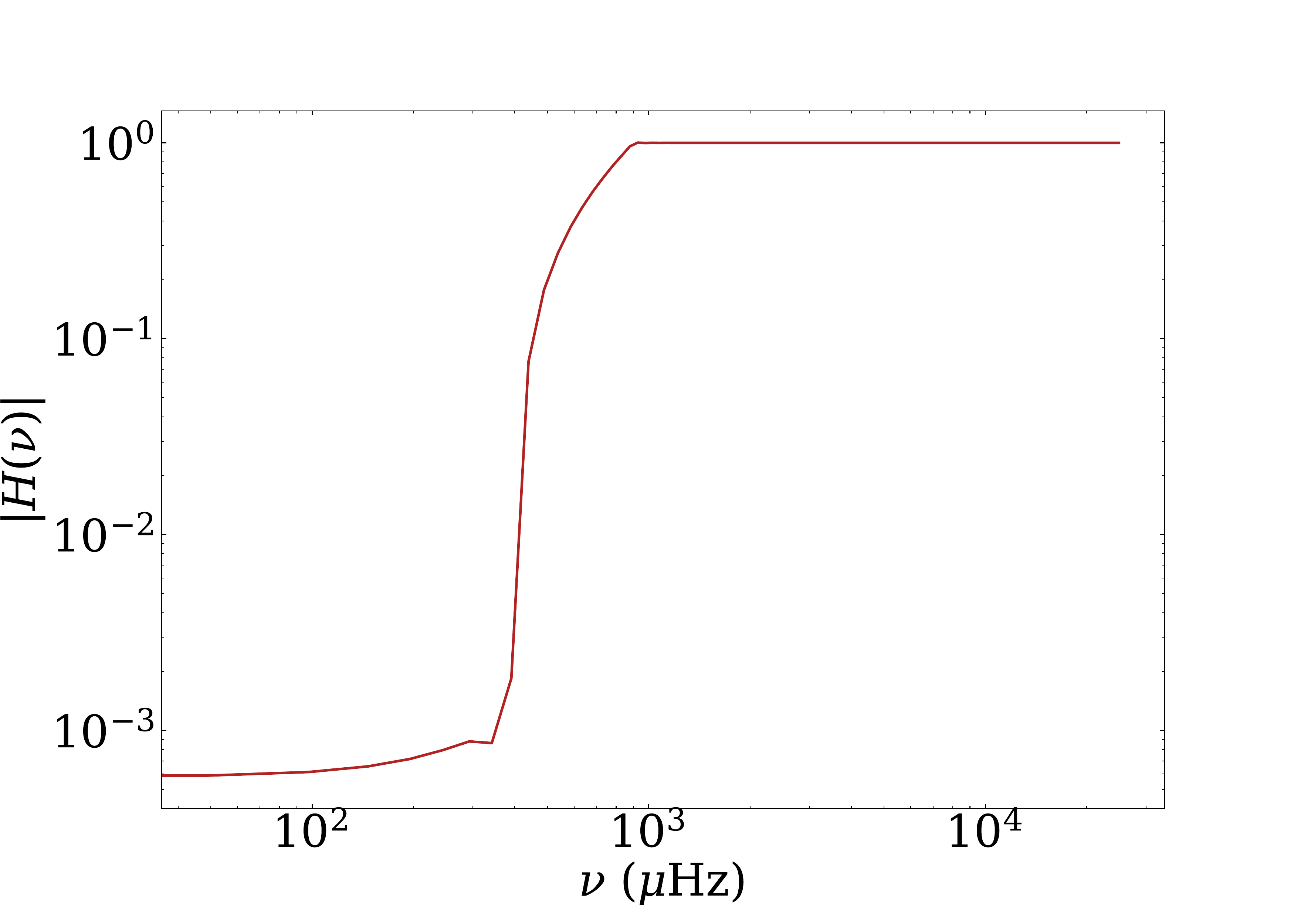}
    \caption{Modulus of the transfer function $H$ of the FIR filter applied on the Solar-SONG time series.}
    \label{fig:transfer_function}
\end{figure}

The last step consists in high-pass filtering and some final corrections. During the observation campaign, the alto-azimuthal solar tracker was set to follow a pre-computed solar ephemeris without any servo correction. This introduced low-frequency daily fluctuations in the RV measurements, especially around the time of the solar meridian crossing. To filter out the harmonics that these fluctuations introduce in the spectrum below 800 $\mu$Hz we use a finite impulse response (FIR) filter.
The transfer function of the FIR filter is represented in Fig.~\ref{fig:transfer_function}. The time series have been extensively visually inspected and time intervals with brutal drops of absolute values of the RV measurements, clearly related to clouds obscuring the instrument line-of-sight, are set to zero at this stage\footnote{The exhaustive list of corrections can be found at: \\ \texttt{https://gitlab.com/sybreton/apollinaire/-/blob/master/\\apollinaire/songlib/interval\_nan.py}}.
Considering the mean photon flux level, measurements below a threshold of 12,000 ADU are also set to zero.
We finally compute the RV mean values over the entire day. RV values beyond 3.5 standard deviation are removed. The daily RV mean is then computed again and values beyond three standard deviations are removed.

\begin{figure*}[ht]
    \centering
    \includegraphics[width=1.\textwidth]{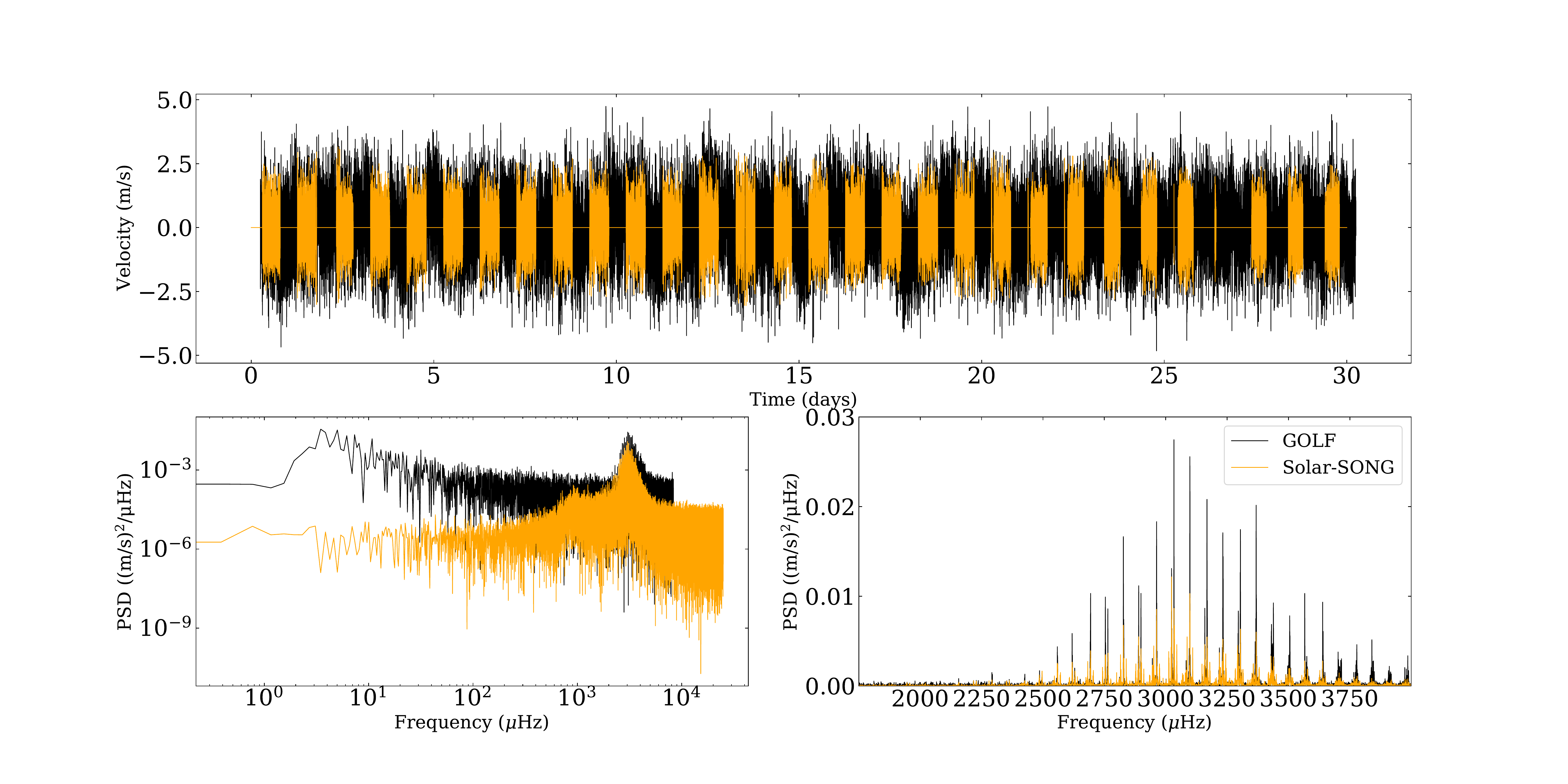}
    \caption{\textit{Top panel:} Solar-SONG (orange) and GOLF (black) complete time series from 3 June to 2 July 2018. \textit{Bottom left:} corresponding PSD. \textit{Bottom right:} Zoom into the p-mode region.}
    \label{fig:series}
\end{figure*}

\begin{figure}[ht]
    \centering
    \includegraphics[width=.48\textwidth]{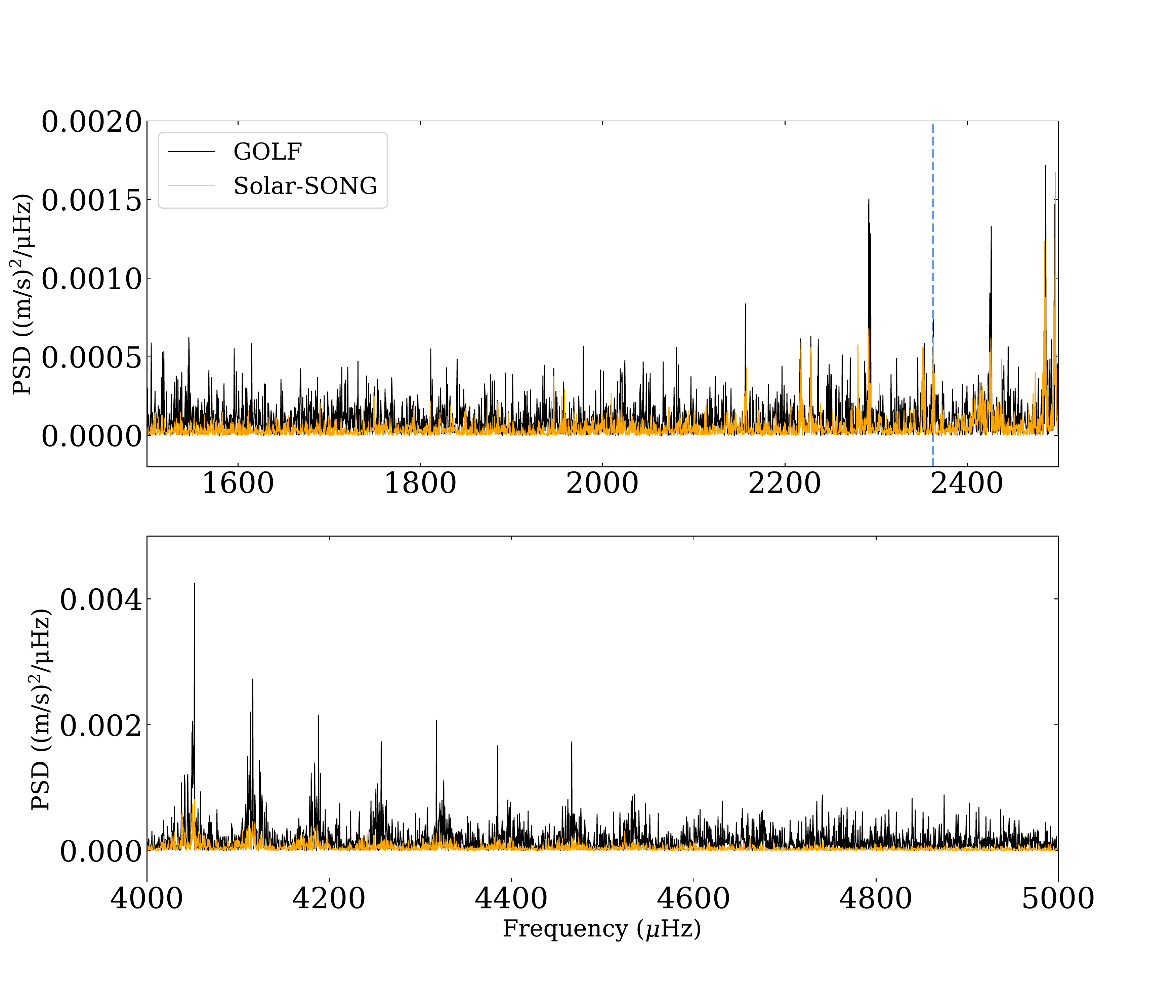}
    \caption{GOLF (black) and Solar-SONG PSD in the 1500-2500 $\mu$Hz (\textit{top}) and 4000-5000 $\mu$Hz (\textit{bottom}) regions. The blue line marks the $n=16$, $\ell=0$ mode which has been fitted in Solar-SONG PSD and not in GOLF PSD.}
    \label{fig:zoom_on_psd}
\end{figure}

\begin{figure}[ht]
    \centering
    \includegraphics[width=0.48 \textwidth]{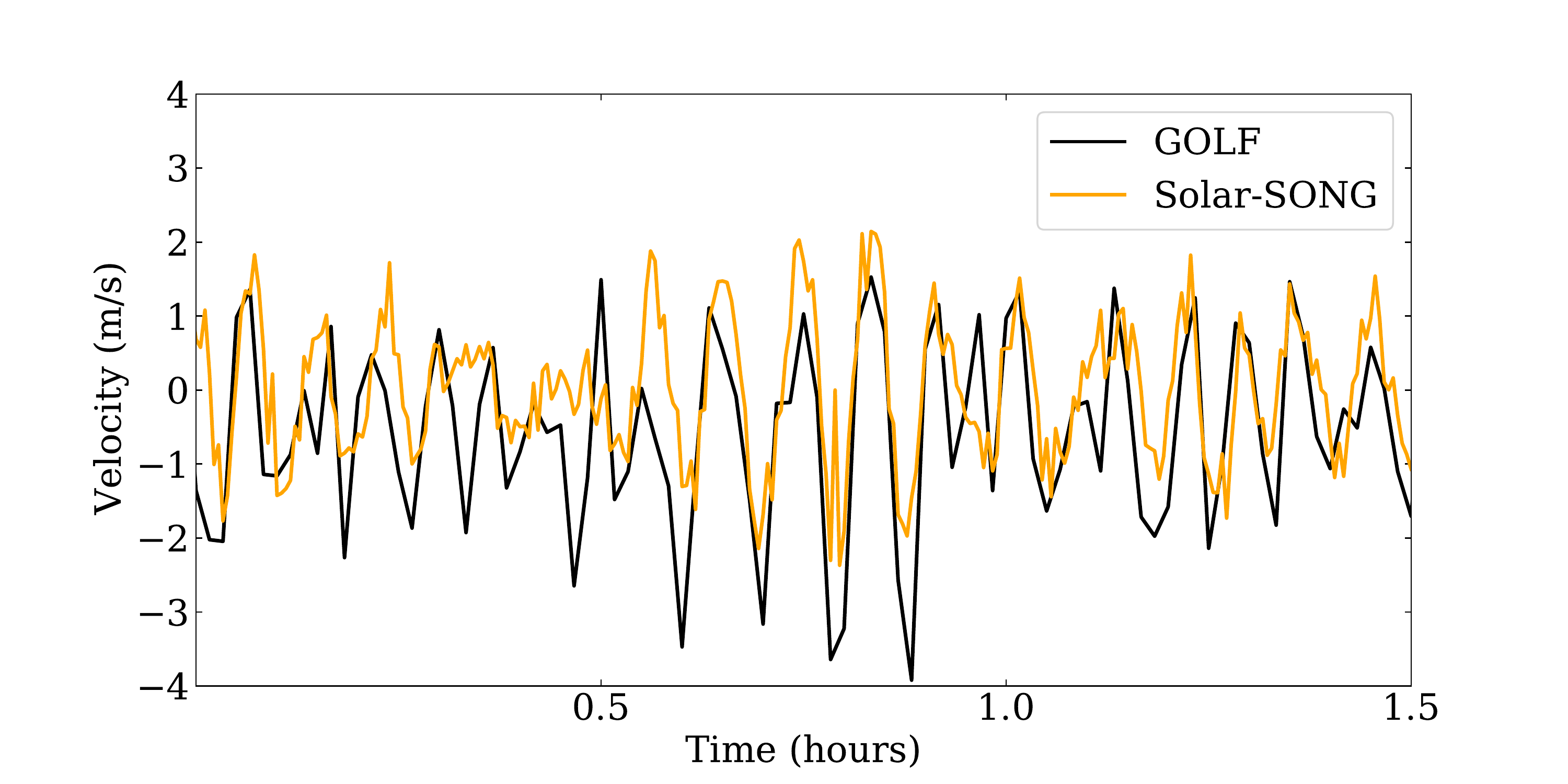}
    \caption{One hour and a half of the Solar-SONG (orange) and GOLF (black) RV time series. The Solar-SONG time series is sampled at 20s and the GOLF time series is sampled at 60s.}
    \label{fig:series_hours}
\end{figure}

\begin{figure*}[ht]
    \centering
    \includegraphics[width=1.\textwidth]{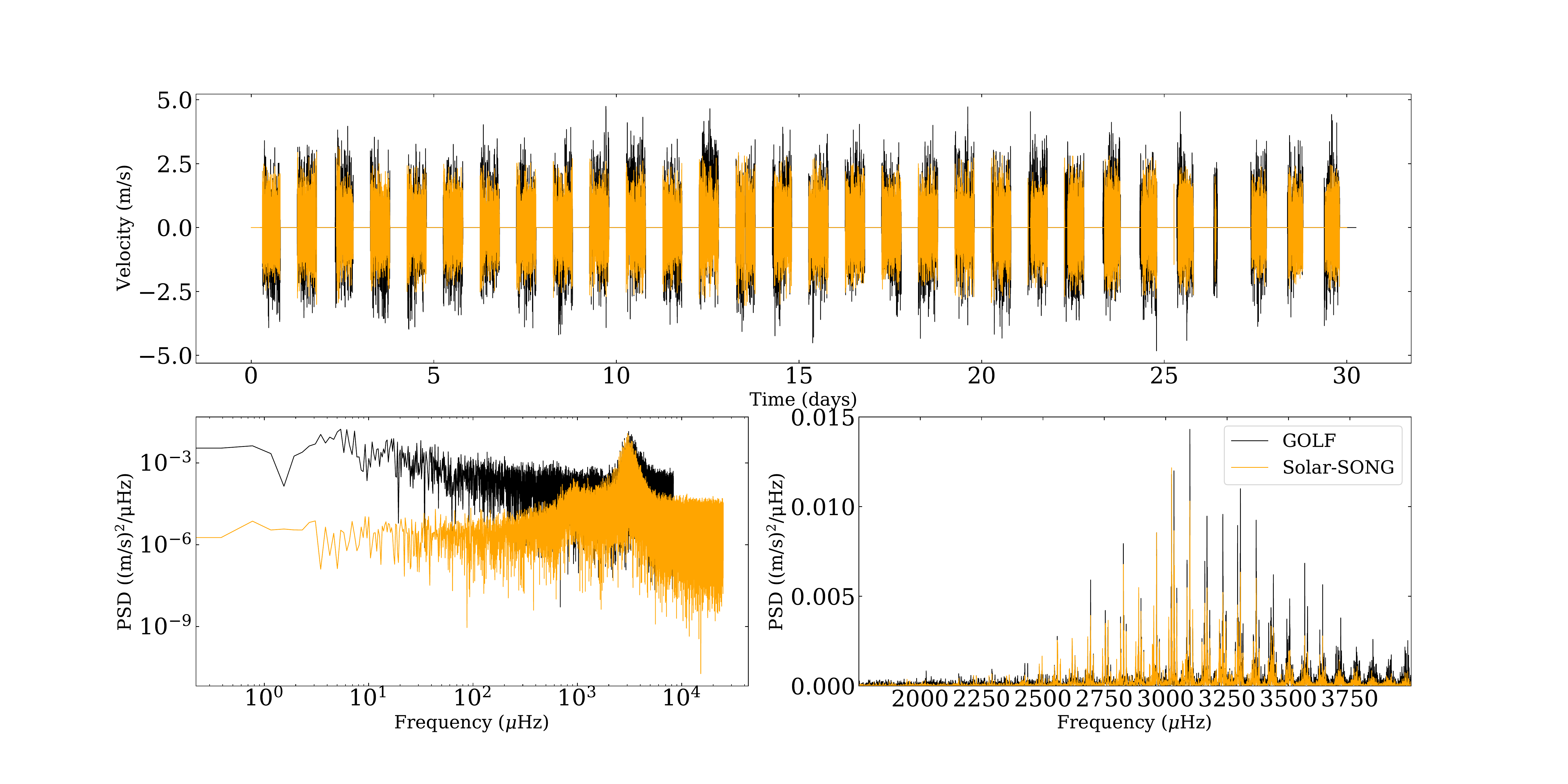}
    \caption{\textit{Top panel:} Solar-SONG (orange) and GOLF (black) time series from 3 June to 2 July 2018, like on Fig~\ref{fig:series} but with an observational window identical to the one of Solar-SONG applied on GOLF time series. \textit{Bottom left:} corresponding PSD. \textit{Bottom right:} Zoom into the p-mode region.}
    \label{fig:series_wdw}
\end{figure*}

Figure~\ref{fig:series} shows the GOLF and Solar-SONG time series from 3 June to 2 July and the corresponding PSD.
For the sake of clarity of the rest of the manuscript, we will only compare Solar-SONG with GOLF. The analysis and comparisons done with BiSON and HMI are given in Appendix~\ref{sec:hmi_bison}. The results obtained with these last two instruments are qualitatively the same as with GOLF. 
The main difference found is between HMI and the disk-integrated instruments. The p-mode power level observed with HMI is lower than the others, which is a natural consequence of integrating the power to mimic full-disk Sun-as-a-star observations.
Figure~\ref{fig:zoom_on_psd} shows the 1500-2500 $\mu$Hz and 4000-5000 $\mu$Hz regions of the PSD.
The time series have been restricted to one hour and a half in Fig.~\ref{fig:series_hours} in order to highlight the presence of the five-minute oscillations in the signal. Figure~\ref{fig:series_wdw} presents the same panels as in Fig.~\ref{fig:series} but with the observational window of Solar-SONG data applied on GOLF time series. Due to the convolution by the observational window, the power of the p-mode peaks is redistributed between the main peak and the side-lobes. The 800 $\mu$Hz cut of the Solar-SONG time series FIR filter is visible. The comparison of the PSD in Fig~\ref{fig:series_wdw} also clearly shows that the Solar-SONG mean noise level below 2000 $\mu$Hz and above 6000 $\mu$Hz is lower than the one in GOLF. The GOLF excess of power at the high-end frequency range of the p modes is explained by the chromospheric contribution in the sodium lines used by GOLF. \citep{2007ApJ...654.1135J}. 

\section{Peakbagging method \label{sec:peakbagging}}

\citet{1984PhDT........34W} showed that the PSD follows a $\chi^2$ law with two degrees of freedom. Assuming that the frequency bins are independent from each other, the corresponding likelihood function is given by:

\begin{equation}
    \label{eq:likelihood}
    \mathcal{L} (\mathbf{S_x}, \theta) = \prod\limits_{i=1}^{k} \frac{1}{S(\nu_i, \theta)} \exp \left[ - \frac{S_{x_i}}{S(\nu_i, \theta)} \right]  \;,
\end{equation}
where  $S$ denotes the limit spectrum parametrised by a set $\theta$ of parameters. $\mathbf{S_x}$ is the observed spectrum evaluated at a given set of $k$ frequency bins $\nu_i$.

Fits are processed using a Bayesian approach through the use of Monte Carlo Markov Chains \citep[MCMC,][]{Sokal1997,2009_book_liu,2010CAMCS...5...65G}. MCMC properties are exploited to evaluate the shape of the posterior probability distribution of our model, using:
\begin{equation}
    p (\theta | \mathbf{S_x}) = \frac{p (\mathbf{S_x} | \theta) p (\theta)}{p (\mathbf{S_x})} \;,
\end{equation}
where $p (\mathbf{S_x} | \theta)$ is the likelihood $\mathcal{L}(\mathbf{S_x}, \theta)$, $p(\theta)$ the prior probability and $p (\mathbf{S_x})$ a normalisation factor. In practice, only the numerator $p (\mathbf{S_x} | \theta) p (\theta)$ of the posterior probability distribution is sampled by the MCMC. In this work, all priors $p (\theta)$ have been taken as uniform distributions.  

MCMC are sampled with the \texttt{emcee}\footnote{The module documentation is available at: \texttt{https://emcee.readthedocs.io/en/stable}} \citep{Foreman-Mackey2016} implementation through the \texttt{apollinaire}\footnote{The module documentation is available at: \texttt{https://apollinaire.readthedocs.io/en/latest}} (Breton et al. in prep) peakbagging library, which has been designed to perform analysis of both astero- and helioseismic PSD, from stellar background profile characterisation to individual p-mode characterisation. In this work, in order to ensure their convergence, MCMC have been sampled considering 500 walkers and 2000 iterations, with the 200 first iterations being discarded as burned-in to avoid biasing the sampled distribution. Consequently, each sampled MCMC is constituted of 900,000 points after the discarding step.  The uncertainties $\sigma_+$ and $\sigma_-$ over each parameter are computed considering the 16th and 84th centiles of the sampled distribution (which approximates the standard deviation in case of a Gaussian distribution).

Our fitting strategy is the following: first, a global background model is adjusted to the PSD. In the second step, the PSD is divided by this background model to obtain a spectrum with a SNR scale (the so-called \textit{signal-to-noise spectrum}) that we use to fit the p modes. Solar-oscillation modes can be modelled as randomly excited and damped harmonic oscillators \citep{1977ApJ...212..243G,1988ApJ...326..462G}, therefore modes are fitted using a Lorentzian profile, by odd ($\ell=\{1, 3\}$) and even ($\ell=\{0, 2\}$) pairs, that is, for each pair of modes, we perform the fit considering a segment of the spectrum containing only those modes. The Lorentzian profile equation is given by:
\begin{equation}
    L_{n,\ell} (\nu, \nu_{n,\ell}, H_{n,\ell}, \Gamma_{n,\ell}) = \frac{H_{n,\ell}}{1 + \frac{4 (\nu - \nu_{n,\ell})^2}{\Gamma_{n,\ell}^2}} \;,
\end{equation}
where $\nu_{n,\ell}$ is the central mode frequency, $H_{n,\ell}$ the mode height, and $\Gamma_{n,\ell}$ the mode width.
Due to the low resolution of the spectrum, we do not include splittings or asymmetries in our model.
We also include a flat background parameter $b$ to take into account any residual local background contribution in the fitted window. For a given pair, our p-mode model $M_n (\nu)$ is therefore described by the following equations for even and odd pairs, respectively:
\begin{equation}
\begin{split}
\label{eq:pair_model}
    M_n (\nu) = L_{n,0} (\nu, \nu_{n,0}, H_{n,0}, \Gamma_{n,0}) + L_{n-1,2} (\nu, \nu_{n-1,2}, H_{n-1,2}, \Gamma_{n-1,2}) + b \;, \\
    M_n (\nu) = L_{n,1} (\nu, \nu_{n,1}, H_{n,1}, \Gamma_{n,1}) + L_{n-1,3} (\nu, \nu_{n-1,3}, H_{n-1,3}, \Gamma_{n-1,3}) + b \;.
\end{split}
\end{equation}
Following \citet{1994A&A...289..649T}, we fit the natural logarithm of the height and width parameters. Hence, for each pair of modes, we fit seven parameters: $\{ \nu_{n,0}, \log H_{n,0}, \log \Gamma_{n,0}, \nu_{n-1,2}, \log H_{n-1,2}, \log \Gamma_{n-1,2}, b \}$ for even pairs, $\{ \nu_{n,1}, \log H_{n,1}, \log \Gamma_{n,1}, \nu_{n-1,3}, \log H_{n-1,3}, \log \Gamma_{n-1,3}, b \}$ for odd pairs.

The background for the GOLF spectrum is fitted considering the sum of one Harvey profile \citep{1985ESASP.235..199H} and a high-frequency noise parameter $P$, according to the following equation:
\begin{equation}
    B (\nu, A, \nu_c, \gamma, P) = \frac{A}{1 + \left(\frac{\nu}{\nu_c}\right)^\gamma} + P \;,
\end{equation}
with $A$ the amplitude, $\nu_c$ the characteristic frequency and $\gamma$ a power exponent. The four parameters that we fit are therefore $A$, $\nu_c$, $\gamma$ and $P$. 
Since the Solar-SONG time series have been filtered with a high-pass filter, set to a 800-$\mu$Hz cut-off frequency, we do not fit any background on the spectrum and, in order to estimate the signal-to-noise spectrum, we only divide the PSD by the mean value of the high frequency noise (above 8 mHz). 

In this work, each fitted MCMC was double checked using the corresponding corner plots. Fits for which we do not learn anything from the priors have been rejected, that is, fits where the marginalisation over each parameter of the sampled posterior probability distribution still has a uniform distribution shape. After this first step, for each fitted mode, we computed a proxy of the natural logarithm of the Bayes factor $\ln K$, related to the rejection of the null hypothesis H0 \citep{1995_kass_raftery,2016MNRAS.456.2183D}. 
{
In our case, the H0 null hypothesis is the absence of mode. For each fitted mode, we select a given number $N$ of sets of parameters among the values explored by the MCMC sampling. Those sets of parameters are selected by regularly thinning the MCMC in order to conserve the same parameter distribution in the thinned chain. For each set of parameters, the corresponding model likelihood, that is computed with a spectrum model including modes, is compared to the H0 likelihood (that is computed with a spectrum model without modes). Defining $N_\mathrm{H1}$ as the number of times the model likelihood is greater than the H0 likelihood, we have: 
\begin{equation}
    \ln K \approx \ln \frac{N_\mathrm{H1}}{N} \;. 
\end{equation}
The main interest of the thinning step is to save computing time. In the work presented here, we thinned the MCMC from 900,000 to 9,000 sets of parameters.} The interpretation of the $\ln K$ is recalled in Table~\ref{table:ln_k__}.

\begin{table} 
\centering
{
\caption{Interpretation of the $\ln K$ values}
\begin{tabular}{cr}
\hline\hline
    $\ln K$ & Interpretation \\
    \hline
    < 0 \quad & favours H0 \\
     0 to 1 \quad & not worth more than a bare mention \\
     1 to 3 \quad & positive \\
     3 to 5 \quad & strong \\
     > 5 \quad & very strong \\
\end{tabular}
\label{table:ln_k__}
}
\end{table}

\subsection{Accounting for the observational windows \label{sec:observational_window}}

Since the Solar-SONG project is still a single-site ground-based instrument, its observational duty cycle is constrained by the day-night cycle. The consequence of the gaps in the time series is the convolution of the PSD by the Fourier transform of the window function \citep[e.g.][]{2002A&A...390..717S,2004A&A...413.1135S,2015EAS....73..193G}.
Therefore the PSD does not follow a $\chi^2$ with two degrees of freedom statistics, as the bins in the PSD are no longer independent from each other.

However, as mentioned by \citet{1994A&A...287..685G}, the formulation of the likelihood that takes into account time series with gaps \citet{1998A&AS..132..107A} is impracticable to use. As a consequence, the $\chi^2$ likelihood has to be used as a good approximation. 

\begin{figure}[ht]
    \centering
    \includegraphics[width=.48\textwidth]{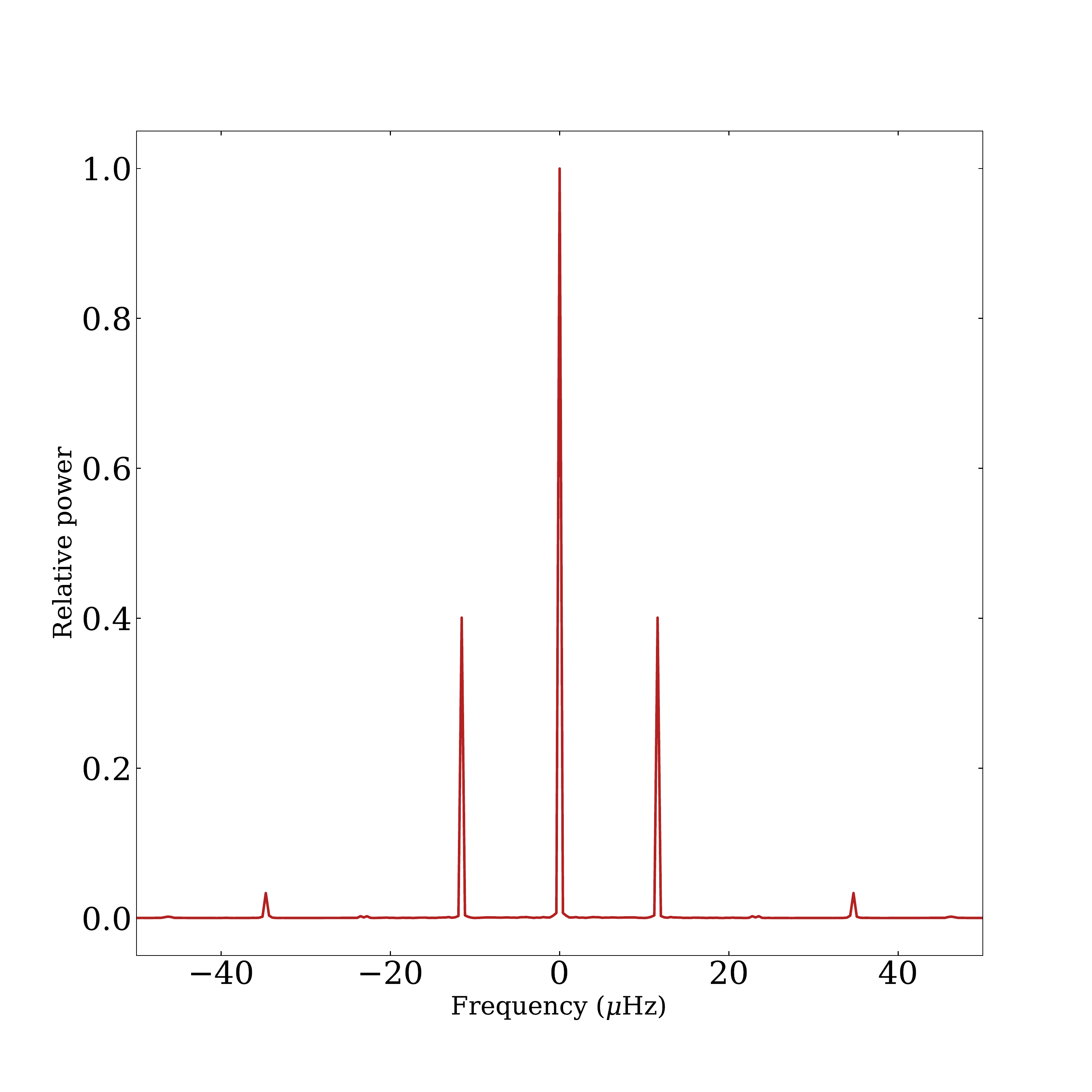}
    \caption{Power spectrum $|\tilde{W}|^2$ of the window function $W$, normalised to one at zero frequency. 
    }
    \label{fig:tf_wdw}
\end{figure}

In order to take into account the effect of the window function in the PSD, we use an \textit{ad hoc} correction to our model. First, we define the observational window vector $W$ as a boolean vector of the same length as the actual time series. Considering a given time stamp, the value of $W$ is 1 if the RV value at this time stamp is non-zero and 0 otherwise. The Fourier transform of this window function, $\tilde{W}$, is then computed (see Fig.~\ref{fig:tf_wdw}). The peaks above 1\% of the height of the zero-frequency peak in $|\tilde{W}|^2$ are selected in order to modify Eq.~\ref{eq:pair_model} as follows:
\begin{equation}
    \begin{split}
        \label{eq:pair_model_wdw}
    M_n (\nu) = \sum_k &\bigg[ L_{n,0} (\nu, \nu_{n,0}+\delta\nu_k, a_k H_{n,0}, \Gamma_{n,0}) + \\ &L_{n-1,2} (\nu, \nu_{n-1,2}+\delta\nu_k, a_k H_{n-1,2}, \Gamma_{n-1,2}) \bigg] + b \;, \\
    M_n (\nu) = \sum_k &\bigg[ L_{n,1} (\nu, \nu_{n,1}+\delta\nu_k, a_k H_{n,1}, \Gamma_{n,1}) + \\ &L_{n-1,3} (\nu, \nu_{n-1,3}+\delta\nu_k, a_k H_{n-1,3}, \Gamma_{n-1,3}) \bigg] + b \;,
    \end{split}
\end{equation}
where $\delta\nu_k$ is the frequency of the $k^\mathrm{th}$ selected peak in $|\tilde{W}|^2$ and $a_k$ is the ratio between the height of the $k^\mathrm{th}$ selected peak in $|\tilde{W}|^2$ and the sum of the heights of all selected peaks.

\begin{figure}[ht]
    \centering
    \includegraphics[width=.48\textwidth]{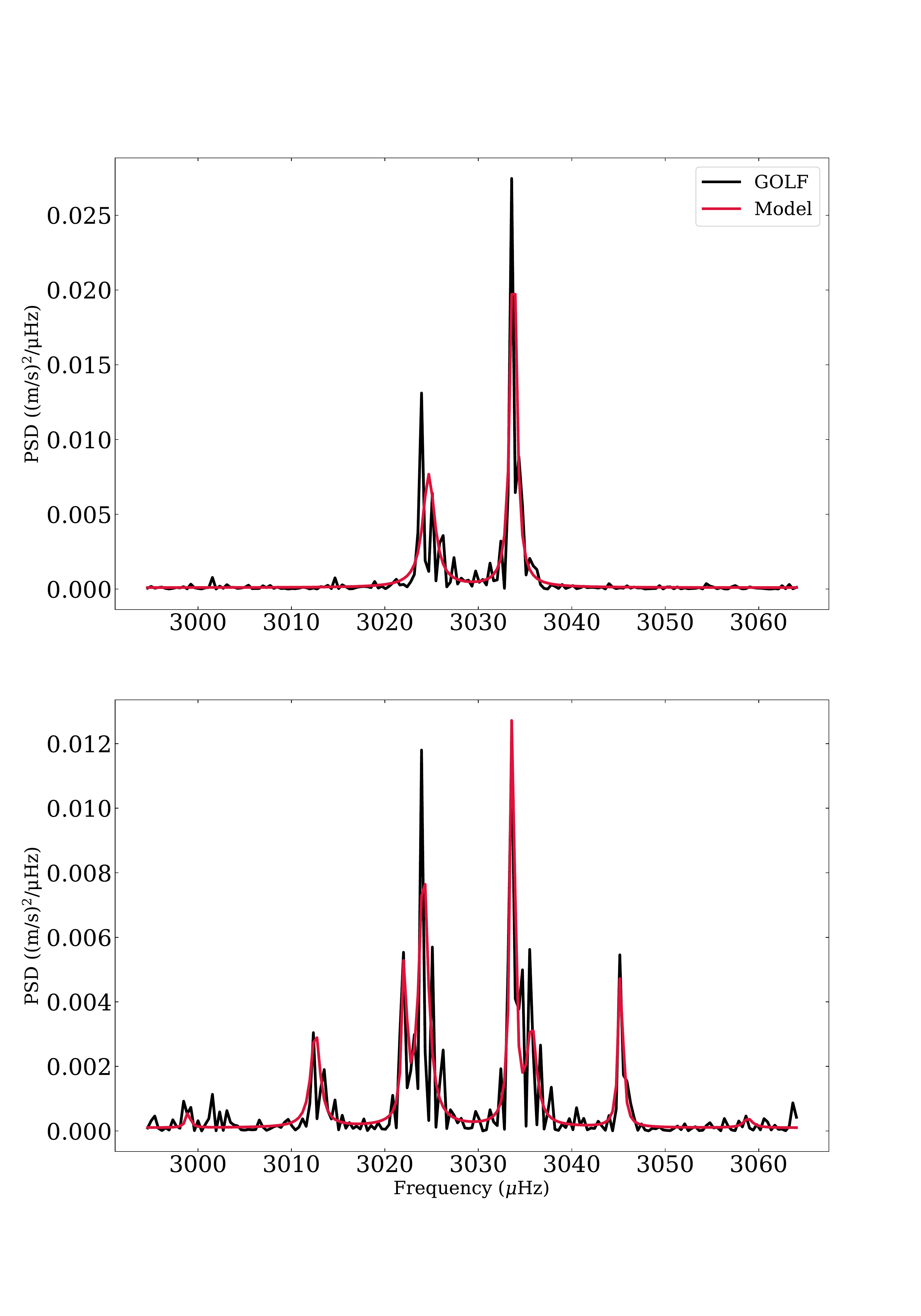}
    \caption{Sections of power spectra for the GOLF time series (\textit{top}) and the same GOLF time series truncated by Solar-SONG observational window (\textit{bottom}), centered around the $n = 21$ modes for the even pair (in black). The profiles corresponding to the fits are shown in red.}
    \label{fig:modes_wdw}
\end{figure}

The comparison of the structure of the $n=21$ even pair in GOLF data, with and without the Solar-SONG-like observational window, is shown in Fig.~\ref{fig:modes_wdw}. The method presented above enables to accurately model the mode profile when the observational window has daily gaps. It is also interesting to note that one of the $\ell=0$ side-lobe power excesses lies very close to the $\ell=2$ central frequency, and reciprocally for one of the $\ell=2$ side-lobes.

\section{Solar-SONG compared to GOLF \label{sec:comparison}}

\begin{figure}[ht]
    \centering
    \includegraphics[width=.48\textwidth]{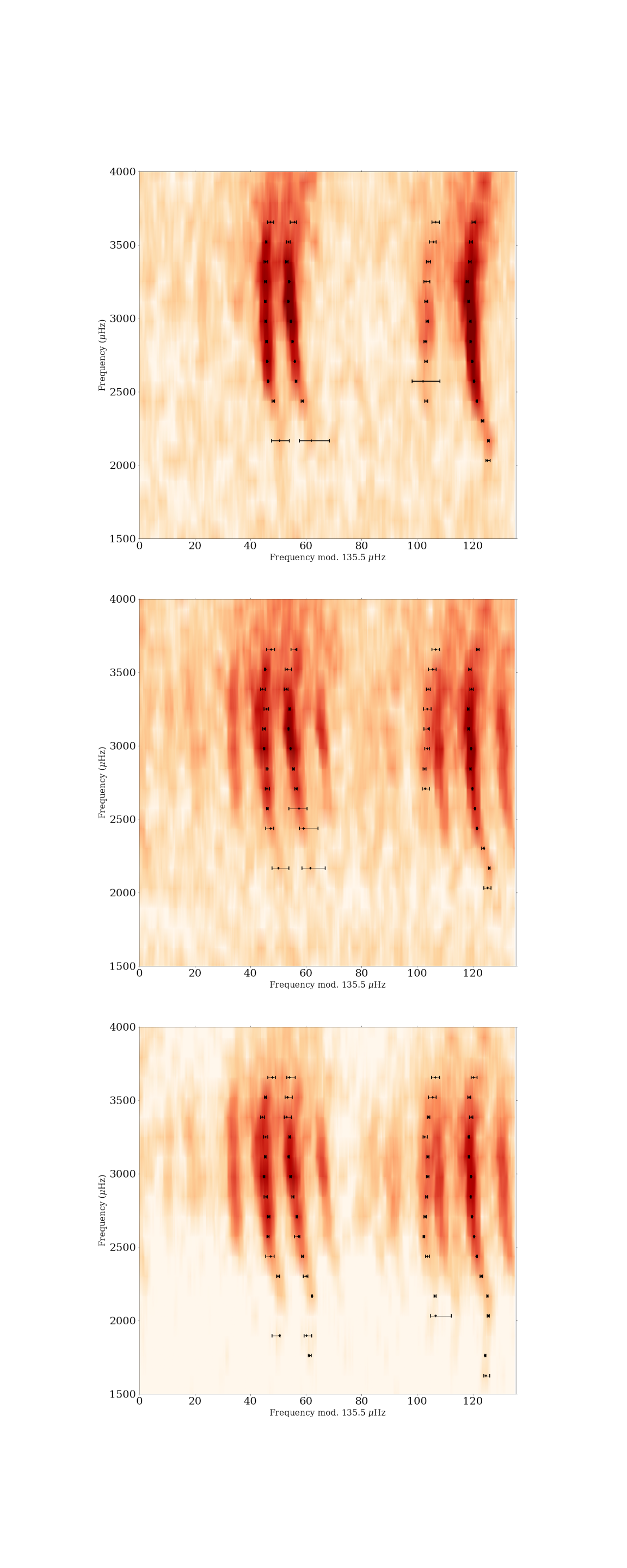}
    \caption{{\'E}chelle diagram for GOLF (\textit{top}), GOLF with Solar-SONG-like observational window (\textit{middle}) and Solar-SONG (\textit{bottom}). Fitted mode frequencies are represented in black.}
    \label{fig:echelle_diagrams}
\end{figure}

\begin{figure*}[ht]
    \centering
    \includegraphics[width=1.\textwidth]{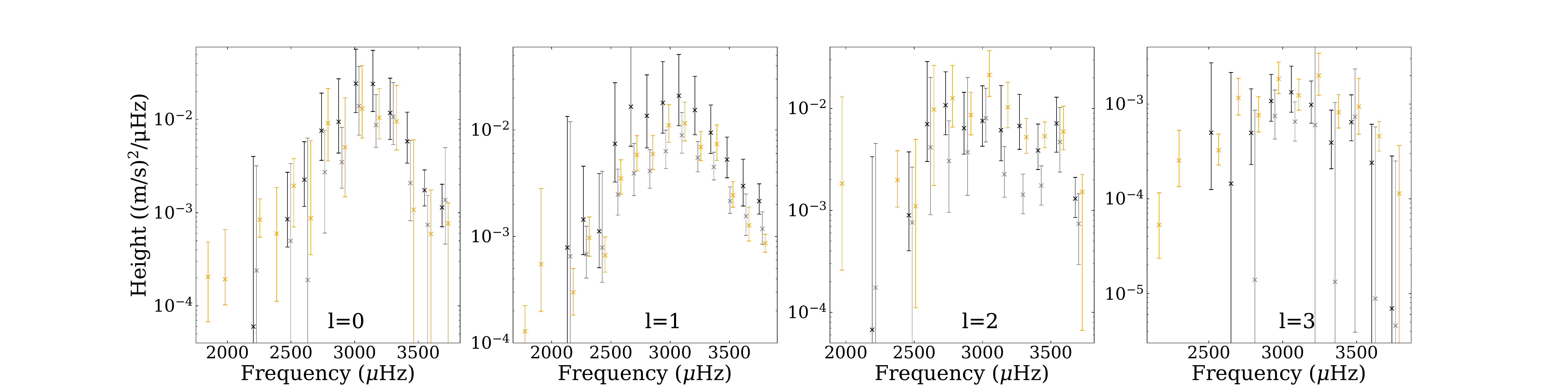}
    \caption{Heights, $H$, of the fitted mode for GOLF (black), GOLF with the Solar-SONG window (grey) and Solar-SONG (orange) spectra. The horizontal position of the markers has been slightly shifted for visualisation convenience.}
    \label{fig:heights}
\end{figure*}

\begin{figure*}[ht]
    \centering
    \includegraphics[width=1.\textwidth]{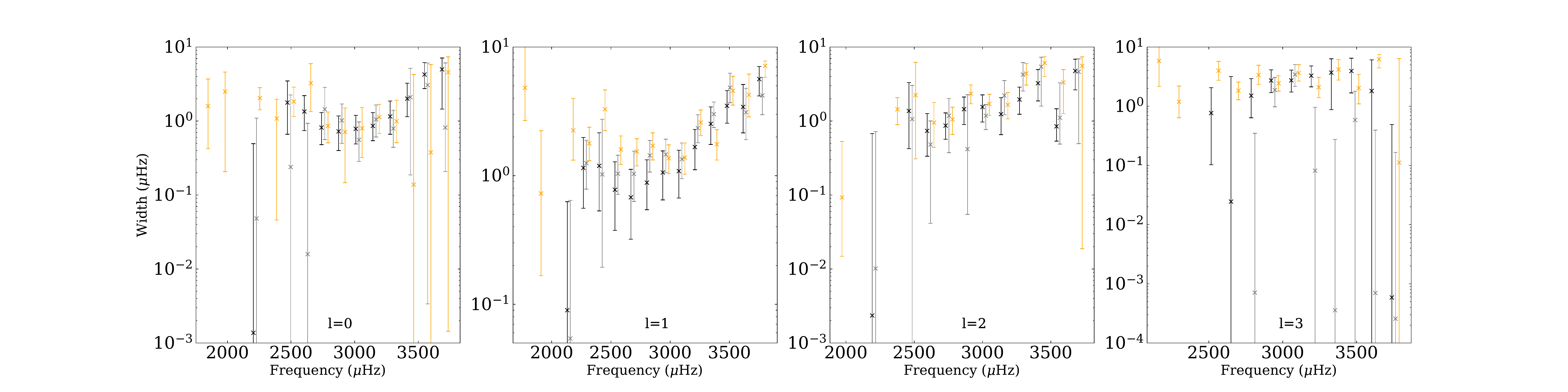}
    \caption{Same as in Fig.~\ref{fig:heights} but for mode widths, $\Gamma$.}
    \label{fig:widths}
\end{figure*}

Considering the 30-day contemporaneous series, we are able to fit modes in Solar-SONG spectrum at lower frequencies than in GOLF (even when considering the GOLF time series with full duty cycle). Indeed, the lowest-frequency fitted Solar-SONG mode is $n=11$, $\ell=1$ at $1749.67\pm1.36$~$\mu$Hz, while for GOLF it is $n=14$, $\ell=1$ at $2156.57\pm0.86$~$\mu$Hz. 
All the fitted frequencies are superimposed to the {\'e}chelle diagrams shown in Fig.~\ref{fig:echelle_diagrams}. The side-lobes of the $\ell=0$ and $\ell=1$ modes appear clearly in the middle and bottom panel. It should be stressed that several $\ell=3$ frequencies could not be fitted when applying the Solar-SONG-like window to GOLF, although those modes have been successfully fitted in the real GOLF spectrum and in the Solar-SONG spectrum. Figure~\ref{fig:heights} and \ref{fig:widths} show our estimates of the fitted modes height $H$ and width $\Gamma$ as a function of frequency.
At high frequency, as it was already visible in Fig.~\ref{fig:series_wdw}, the height of the modes observed by GOLF are larger due to the chromospheric contribution to the solar sodium doublet. Most of the width values are in agreement within the error bars, except for $\ell=1$ modes, where Solar-SONG observed widths seem overestimated below 3 mHz, although the fitted values remain compatible within the error bars with what we have measured with GOLF.

Another interesting aspect of the comparison between the two instruments is the inability of the code to fit the $n=16$, $\ell=0$ mode in the GOLF spectrum while this same mode is well characterized using Solar-SONG. 
{
From Fig.~\ref{fig:zoom_on_psd} top panel, it appears that during the time of observation, this mode was less excited than its $\ell=0$ and $\ell=1$ neighbours, making it more difficult to detect with both instruments.
The mode structure is also difficult to distinguish from the surrounding noise in the GOLF PSD, while the SNR appears to be higher in the Solar-SONG PSD. This can be explained by both the higher level of noise in the GOLF PSD and the different spatial sensitivity of GOLF in its single-wing configuration.
As shown by \citet{1998ApJ...504L..51G} and \citet{1999PhDT.........8H}, the sensitivity of GOLF depends on the observation wing (blue or red) and on the time of the year (due to the non-zero orbital velocity). Thus, excited modes can have different amplitudes in GOLF than in other instruments with an homogeneous response window.
}

{
We include in the appendix (see Tables~\ref{table:pkb_golf}, \ref{table:pkb_golf_wdw} and \ref{table:pkb_song}) all the fitted mode frequencies, heights, and widths, as well as their uncertainties and the corresponding value of $\ln K$. We note that the smallest frequency uncertainties estimates are comparable to the spectral resolution of 0.39 $\mu$Hz
}

\subsection{Evaluating GOLF aging over the years \label{sec:golf_aging}}

Based on the total power in the 1000-1500 $\mu$Hz region of the GOLF data, \citet{2018A&A...617A.108A} showed evidence of an increase of the noise at low frequency over the past two decades, noise that is due to the instrument photon noise and the contribution of solar convection to the RV signal.
This increase is most likely due to the aging of the two photomultipliers, of the entrance window, and of the interference filter, as already pointed out by \citet{2005A&A...442..385G}, based on the increase of instrumental photon noise between 1996 and 2004.
It is straightforward to check that, considering the 2018 observing campaign, the mean power density in the 1000-1500 $\mu$Hz region is in favour of Solar-SONG: this value is $29.1$~${\rm m^2 s^{-2} Hz^{-1}}$ {versus} $104$~${\rm m^2 s^{-2} Hz^{-1}}$ for GOLF. We note that the same comparison in the 5000-6000 $\mu$Hz region yields $14$~${\rm m^2 s^{-2} Hz^{-1}}$ for Solar-SONG and $103$~${\rm m^2 s^{-2} Hz^{-1}}$ for GOLF.

\begin{figure}[ht]
    \centering
    \includegraphics[width=.48\textwidth]{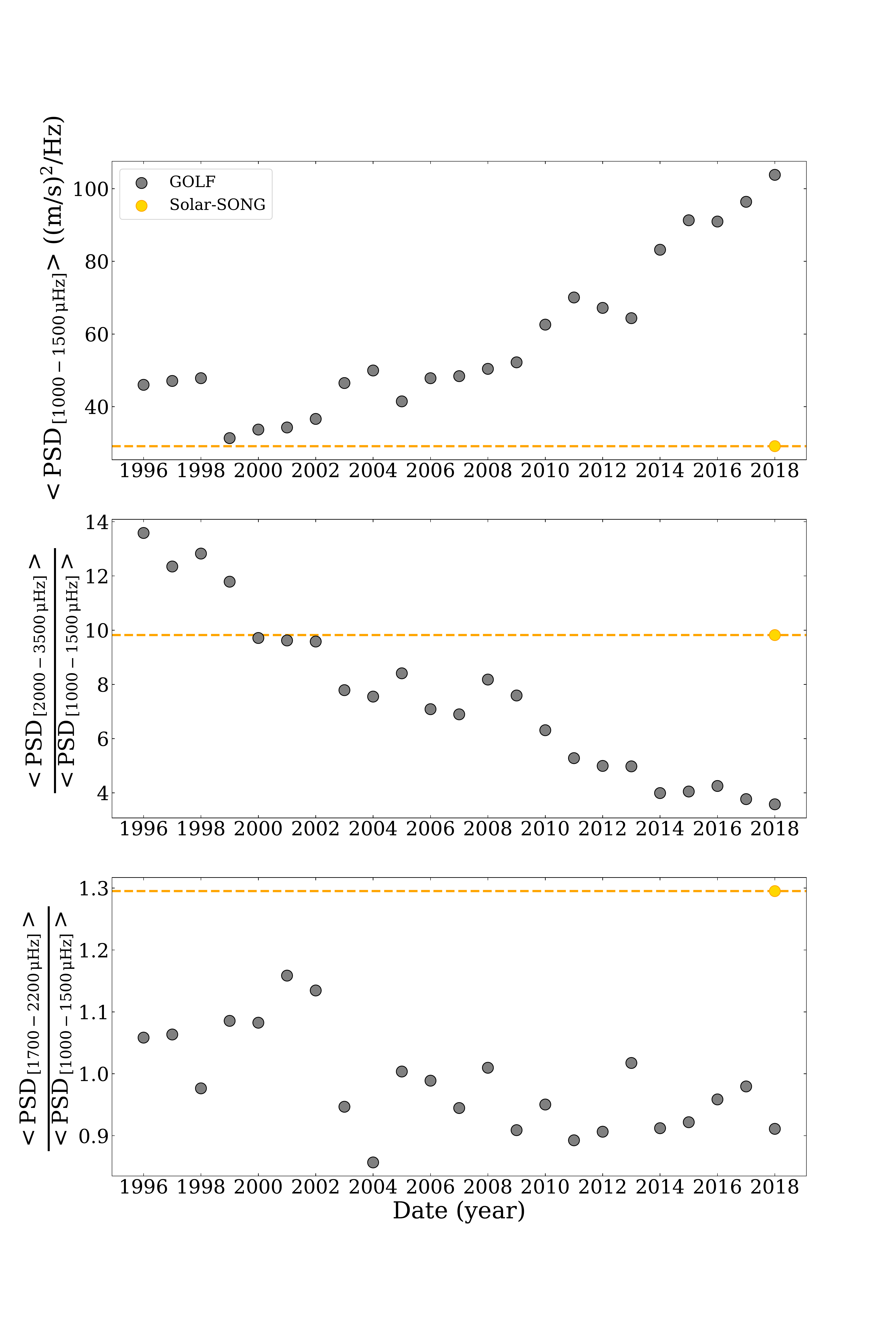}
    \caption{\textit{Top}: mean power density in the 1000-1500 $\mu$Hz region of the considered 30-day GOLF and Solar-SONG PSD. \textit{Middle}: Mean power density ratio computed as the ratio between the mean power in the 2000-3500 $\mu$Hz p-mode region and the 1000-1500 $\mu$Hz region. \textit{Bottom}: Mean power density ratio computed as the ratio between the mean power in the 1700-2200 $\mu$Hz region and the 1000-1500 $\mu$Hz region. The yellow dotted line represents the value obtained with Solar-SONG during the 2018 campaign (represented by the yellow dot).}
    \label{fig:lf_power}
\end{figure}

\begin{figure}[ht]
    \centering
    \includegraphics[width=.48\textwidth]{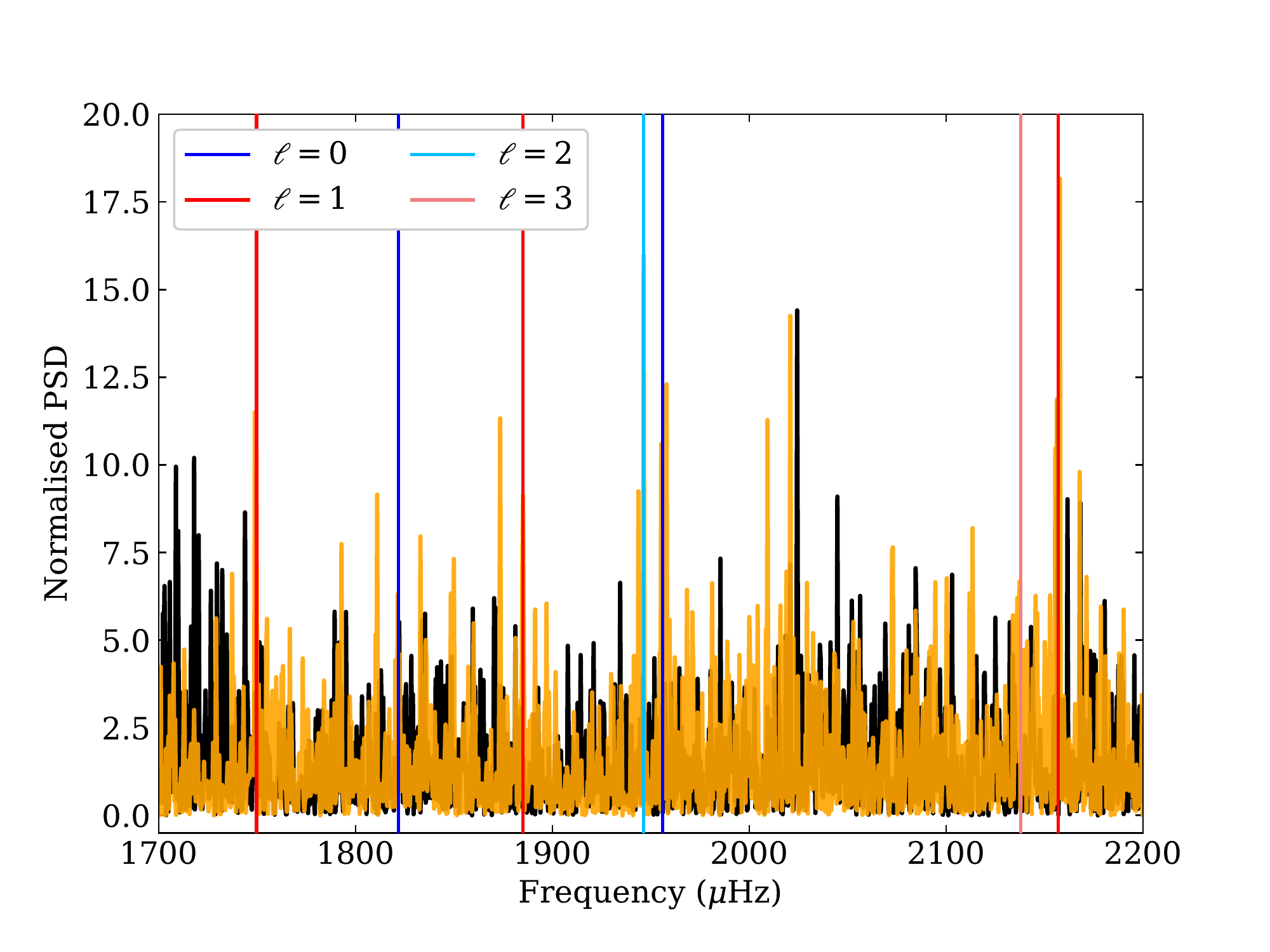}
    \caption{GOLF (with Solar-SONG observational window, black) and Solar-SONG (orange) normalised PSD between 1700 and 2200 $\mu$Hz. The normalisation has been performed by dividing each PSD by its median value in the 1700-2200 $\mu$Hz region. 
    The frequencies fitted in the Solar-SONG spectrum are marked by vertical lines with a color coding related to the mode degree: blue, red, light blue, and pink for $\ell = 0$, 1, 2, and 3, respectively.
    }
    \label{fig:normalised_psd_lf}
\end{figure}

{
The top panel of Fig.~\ref{fig:lf_power} shows the mean power density in the 1000-1500 $\mu$Hz region for each 30-day GOLF time series considered in this work (see Sect.~\ref{sec:golf_data_selection}). The middle panel shows the ratio between the mean power density in the 2000-3500 $\mu$Hz region and the mean power density in the 1000-1500 $\mu$Hz region. The bottom panel shows the same ratio for the 1700-2200 $\mu$Hz region (that is the lowest frequency region where we were able to fit modes for Solar-SONG). In each panel, the value we obtain with the Solar-SONG time series is also represented. 

The temporal evolution of the mean power density in the 1000-1500 $\mu$Hz region unveils evidence that the Solar-SONG noise level in this region, is comparable, if not smaller, to what it was for GOLF in the best years of the instrument. The power decrease observed from 1996 to 1999 can probably be linked to the minimum of magnetic activity reached at this time. After this date, ignoring some yearly modulations, the mean power density in this frequency region increases continually. 

Concerning the mean power density ratio between 2000 and 3500 $\mu$Hz, for the 2018 time series, we obtain a 9.8 ratio for Solar-SONG versus 3.6 for GOLF. However, we note that in the first years of GOLF operations, this value was much higher (13.6 in 1996). During the year 2000, it reached the Solar-SONG level and then kept on decreasing.

In the 1700-2200 $\mu$Hz region, we find a maximal ratio of 1.16 for GOLF (in 2001) and 0.9 in 2018 while we have 1.3 for Solar-SONG.} To help visually assess the signification in this difference in ratio, we represent in Fig.~\ref{fig:normalised_psd_lf}, the normalised PSD of the 1700-2200 $\mu$Hz region for GOLF (considering the time series with Solar-SONG-like window) and Solar-SONG. The normalisation is performed by dividing each PSD by their median value in the 1700-2200 $\mu$Hz region. With this normalisation, it appears that, in this specific region, most of the p modes have a relative height that is higher in Solar-SONG than in GOLF. These elements combined with our ability to fit several modes for Solar-SONG in this frequency region therefore strongly suggests that the SNR is also in favour of Solar-SONG in this frequency range.

\begin{sidewaysfigure*}[ht]
    \centering
    \includegraphics[width=1.\textwidth]{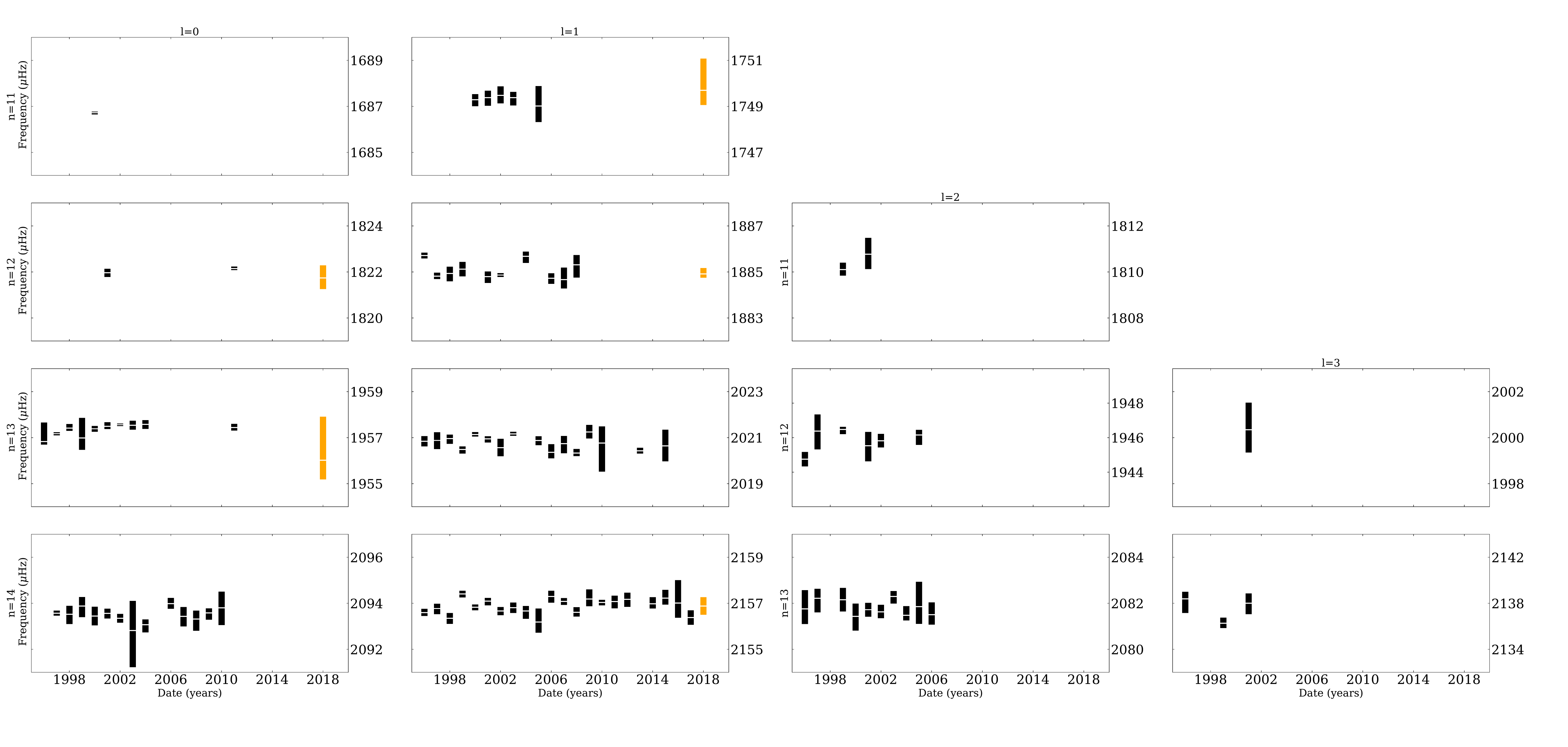}
    \caption{Fitted mode frequencies and uncertainties for yearly GOLF 30-day series (black). The uncertainties on mode frequencies fitted within Solar-SONG spectrum are represented in orange for comparison purpose. The white tick signals the median value of the fitted distribution. For both instruments, only modes with an uncertainty below 2 $\mu$Hz have been represented.}
    \label{fig:golf_months}
\end{sidewaysfigure*}

The second step of this GOLF yearly evolution analysis is to consider the mode orders for which, considering the 2018 GOLF series, we were not able to provide mode parameters although some of those modes were fitted considering Solar-SONG data. With Solar-SONG, we were able to fit the $n=11$, $\ell=1$ mode while for GOLF we had to stop at the $n=14$, $\ell=1$ mode. We therefore decide to perform our peakbagging process for odd and even pairs of order 11 to 14 on each GOLF 30-day series. The results are summarised order by order and degree by degree in Fig.~\ref{fig:golf_months}. The mode frequency variations are related to the magnetic solar activity \citep{1985Natur.318..449W,1989A&A...224..253P}. Modes that are not represented in this figure could not be fitted or the uncertainty on fitted frequency was above 2 $\mu$Hz. The $n=11$, $\ell=1$ could not be fitted in the considered GOLF series after 2005. For this order, we were not able to fit any $\ell=2$ or $\ell=3$ modes. The only mode we were able to fit almost every time until 2018 is the $n=14$, $\ell=1$. It should be reminded that for such short time series, our ability to fit a given mode is not only dependent to the instrumental SNR ratio but also to the excitation state of the mode. This explains why for certain 30-day series, some modes could not be fitted although GOLF instrumental noise did not increase drastically or was even smaller. 
It should be noted that this GOLF performance analysis over time is only valid for 30-day long time series. By considering longer GOLF time series, it is of course possible to obtain much better constraints for the mode parameters in the frequency region considered in Fig.~\ref{fig:golf_months} \citep[see e.g.][which used 365-day long GOLF time series to probe the p-mode temporal frequency variation]{2015A&A...578A.137S}.

\section{The future of Solar-SONG: discussion and conclusion \label{sec:discussion}}

In this work, we presented a new reduction pipeline for Solar-SONG data. We compared the contemporaneous GOLF (as well as BiSON and HMI) and Solar-SONG observations by performing a peakbagging analysis with a Bayesian approach. On the one hand, by studying the PSD of the Solar-SONG data, we were able to identify modes at lower frequency than in the GOLF PSD. On the other hand, we evaluated the effect of the aging of GOLF on its performance by considering the yearly mean power evolution in the 1000 to 1500 $\mu$Hz region with 30-day long series. For each considered series, the mean power density was above the mean power density obtained in 2018 with Solar-SONG. 
However, the GOLF global p-mode power density ratio in the 2000-3500 $\mu$Hz was above the Solar-SONG level from 1996 to 1999. This power density ratio decreased over the years: in 2018, Solar-SONG power density ratio was almost three times higher than GOLF power density ratio. Considering the 1700-2200 $\mu$Hz region only, Solar-SONG power density ratio appears higher than GOLF power density ratio at any time. 

We then performed another peakbagging analysis on these series focusing on the low-frequency p modes (below 2200 $\mu$Hz) for which we were able to provide more precise mode frequencies for Solar-SONG and not for GOLF in the 2018 comparison. We were able to provide frequencies for many of these modes in the first years of SoHO's operations. However, after 2005, the decrease in SNR reduced the number of modes we were able to fit inside each subseries. 

Despite its aging, GOLF remains an invaluable asset for helioseismology: it has been almost continuously collecting data over the last 25 years and will carry on in its mission in the years to come.
However, the promising helioseismic measurements obtained
during the Solar-SONG 2018 summer high-cadence run show the potential of longer observations with a better duty cycle. This second condition can only be achieved if other SONG nodes are available. 
Presently, a new node of the SONG network is being commissioned at the Mt. Kent Observatory in Queensland, Australia. 
Performing observations with both Australian and Canarian SONG nodes would allow a significant improvement to the time series duty cycle, although it would still be necessary to keep on extending the SONG network in order to reach a constant duty cycle above 80\%. 

In order to improve the SNR of the low-frequency regions of the Solar-SONG PSD, the solar tracker used for the 2018 observation campaign should also be modified. The current Solar-SONG set-up uses an azimuthal commercial mount which is not optimal for stability in {frequency} regions below 800 $\mu$Hz. Indeed, the daily Earth-motion RV residual that appears in Fig.~\ref{fig:daily_trend} creates a high-amplitude harmonic pattern in the PSD if the low-frequency trend of the Solar-SONG time series has not been properly filtered out with the FIR filter. It should be noted that the servo guidance was not active during the 2018 campaign. The solar tracker followed the Sun's motion using only pre-computed ephemeris. However, we noticed during a short test run performed in 2019 that turning on the servo introduced additional low-frequency trends to the RV signal. In order to overcome this limitation and to extend the scientific objectives of the Solar-SONG initiative, funding was obtained for a new project baptised Magnetrometry Unit for SOLar-SONG (MUSOL) which plans to upgrade the Solar-SONG Teide node with both an equatorial mount allowing improved guidance 
and a new polarimetric unit.
Indeed, the dipolar and quadrupolar components of the solar global magnetic field can only be measured by detecting the weak polarization signal induced in some spectral lines by the Hanle effect. Long-term and continuous solar observations with this new polarimetric unit should in principle be sensitive enough to measure the dipolar component of the global solar magnetic field and its variation along the solar activity cycle \citep[see][]{2017MNRAS.465.4414V}.


\begin{acknowledgements}
The authors want to  thank  the  anonymous  referee  for  valuable  comments  that helped  in  improving  the  manuscript.
This work is based on observations made at the Hertzsprung SONG telescope operated at the Spanish Observatorio del Teide on the island of Tenerife by  the  Aarhus  and  Copenhagen  Universities  and  by  the  Instituto  de Astrofísica de Canarias. SoHO is a mission of international collaboration between ESA and NASA.
S.N.B. and R.A.G acknowledge the support from PLATO and GOLF CNES grants.
Funding for Solar-SONG was provided by the Excellence 'Severo Ochoa' programme at the IAC and the Ministry MINECO under the program AYA-2016-76378-P. Funding for the Stellar Astrophysics Centre is provided by The Danish National Research Foundation (Grant DNRF106).
We would like to acknowledge the contribution of the engineers at the IAC, Felix Gracia (optical), Ezequiel Ballesteros (electronics) and the support from the day-time operator at the SolarLab, Teide Observatory, on this Solar-SONG initiative and during the campaign. We thank Dr. Kun Wang (China West Normal University) for his dedicated effort during the campaign with  data handling and analysis of the observations. We also thank Thierry Appourchaux for his relevant comments during the TASOC review. 
The solar visibility calculations were carried out by the ephemeris calculation service of the IMCCE through its Solar System portal (\texttt{https://ssp.imcce.fr}).    
\\
\textit{Software:} \texttt{Python} \citep{10.5555/1593511}, \texttt{numpy} \citep{numpy,Harris_2020}, \texttt{pandas} \citep{reback2020pandas, mckinney-proc-scipy-2010}, \texttt{matplotlib} \citep{4160265}, \texttt{emcee} \citep{2013PASP..125..306F}, \texttt{scipy} \citep{2020SciPy-NMeth}, \texttt{corner} \citep{Foreman-Mackey2016}, \texttt{astropy }\citep{astropy:2013, astropy:2018}.
\\
The source code used to obtain the present results can be found at \texttt{gitlab.com/sybreton/apollinaire}.
\end{acknowledgements}

\bibliographystyle{aa} 
\bibliography{biblio.bib} 

\begin {appendix}

\section{Fitting results for GOLF and Solar-SONG}

The full summary of the GOLF and Solar-SONG fits performed for this work is presented in this section. Table~\ref{table:pkb_golf} presents the mode parameters obtained with GOLF, Table~\ref{table:pkb_golf_wdw} the mode parameters obtained with GOLF when applying a Solar-SONG-like window and Table~\ref{table:pkb_song} the mode parameters obtained with Solar-SONG.

\begin{table*}[!ht]
\centering
\caption{Parameters of the modes fitted in the GOLF spectrum.} 
\begingroup
\renewcommand{\arraystretch}{1.3}
\begin{adjustbox}{totalheight = .8\textheight}
\begin{tabular}{rrrrrrrrr}
\hline\hline
  $n$ &  $\ell$ &  $\nu$ & $H$ & $\Gamma$ & $\ln K$  \\
  &   &  ($\mu$Hz) &  (${\rm m^2 s^{-2} \mu Hz^{-1}}$) &  ($\mu$Hz) &   \\
\hline

 14 &  1 &  $2156.56_{-0.61}^{+1.01}$ &  $7.87 \times 10^{-4} ~_{-7.80 \times 10^{-4}}^{+1.26 \times 10^{-2}}$ &  $0.09_{-0.09}^{+0.54}$ &   > 6 \\
 14 &  2 &  $2217.25_{-2.94}^{+3.47}$ &  $6.75 \times 10^{-5} ~_{-6.73 \times 10^{-5}}^{+3.29 \times 10^{-3}}$ &  $0.00_{-0.00}^{+0.68}$ &  0.16 \\
 15 &  0 &  $2228.57_{-4.21}^{+6.54}$ &  $6.00 \times 10^{-5} ~_{-5.98 \times 10^{-5}}^{+3.96 \times 10^{-3}}$ &  $0.00_{-0.00}^{+0.49}$ &   > 6 \\
 15 &  1 &  $2292.30_{-0.32}^{+0.34}$ &  $1.44 \times 10^{-3} ~_{-7.66 \times 10^{-4}}^{+3.13 \times 10^{-3}}$ &  $1.15_{-0.59}^{+0.83}$ &   > 6 \\
 16 &  1 &  $2425.61_{-0.42}^{+0.40}$ &  $1.12 \times 10^{-3} ~_{-6.10 \times 10^{-4}}^{+2.78 \times 10^{-3}}$ &  $1.19_{-0.66}^{+0.96}$ &   > 6 \\
 16 &  2 &  $2485.79_{-0.44}^{+0.36}$ &  $8.96 \times 10^{-4} ~_{-4.94 \times 10^{-4}}^{+2.85 \times 10^{-3}}$ &  $1.37_{-0.95}^{+1.94}$ &  5.92 \\
 17 &  0 &  $2496.23_{-0.49}^{+0.39}$ &  $8.52 \times 10^{-4} ~_{-4.24 \times 10^{-4}}^{+1.86 \times 10^{-3}}$ &  $1.78_{-1.11}^{+1.71}$ &   > 6 \\
 16 &  3 &  $2540.76_{-0.55}^{+0.54}$ &  $4.98 \times 10^{-4} ~_{-3.73 \times 10^{-4}}^{+2.23 \times 10^{-3}}$ &  $0.77_{-0.66}^{+1.29}$ &  2.33 \\
 17 &  1 &  $2558.96_{-0.20}^{+0.22}$ &  $7.41 \times 10^{-3} ~_{-4.16 \times 10^{-3}}^{+2.03 \times 10^{-2}}$ &  $0.78_{-0.40}^{+0.50}$ &   > 6 \\
 17 &  2 &  $2619.30_{-0.22}^{+0.21}$ &  $7.01 \times 10^{-3} ~_{-4.00 \times 10^{-3}}^{+2.18 \times 10^{-2}}$ &  $0.74_{-0.40}^{+0.53}$ &   > 6 \\
 18 &  0 &  $2629.42_{-0.33}^{+0.33}$ &  $2.26 \times 10^{-3} ~_{-1.10 \times 10^{-3}}^{+3.52 \times 10^{-3}}$ &  $1.34_{-0.60}^{+0.86}$ &   > 6 \\
 17 &  3 &  $2675.06_{-3.94}^{+6.00}$ &  $1.44 \times 10^{-4} ~_{-1.44 \times 10^{-4}}^{+2.01 \times 10^{-3}}$ &  $0.02_{-0.02}^{+3.14}$ &  0.97 \\
 18 &  1 &  $2693.39_{-0.20}^{+0.18}$ &  $1.65 \times 10^{-2} ~_{-9.51 \times 10^{-3}}^{+5.11 \times 10^{-2}}$ &  $0.68_{-0.36}^{+0.44}$ &   > 6 \\
 18 &  2 &  $2754.44_{-0.20}^{+0.20}$ &  $1.07 \times 10^{-2} ~_{-5.23 \times 10^{-3}}^{+1.21 \times 10^{-2}}$ &  $0.87_{-0.30}^{+0.42}$ &   > 6 \\
 19 &  0 &  $2764.35_{-0.21}^{+0.20}$ &  $7.60 \times 10^{-3} ~_{-3.94 \times 10^{-3}}^{+1.16 \times 10^{-2}}$ &  $0.82_{-0.34}^{+0.49}$ &   > 6 \\
 18 &  3 &  $2811.52_{-0.47}^{+0.47}$ &  $4.97 \times 10^{-4} ~_{-2.67 \times 10^{-4}}^{+9.50 \times 10^{-4}}$ &  $1.51_{-0.87}^{+1.41}$ &   > 6 \\
 19 &  1 &  $2828.19_{-0.19}^{+0.20}$ &  $1.36 \times 10^{-2} ~_{-6.79 \times 10^{-3}}^{+1.93 \times 10^{-2}}$ &  $0.88_{-0.34}^{+0.45}$ &   > 6 \\
 19 &  2 &  $2889.57_{-0.26}^{+0.27}$ &  $6.41 \times 10^{-3} ~_{-2.85 \times 10^{-3}}^{+7.97 \times 10^{-3}}$ &  $1.45_{-0.55}^{+0.66}$ &   > 6 \\
 20 &  0 &  $2898.94_{-0.20}^{+0.19}$ &  $9.46 \times 10^{-3} ~_{-5.09 \times 10^{-3}}^{+1.79 \times 10^{-2}}$ &  $0.72_{-0.32}^{+0.45}$ &   > 6 \\
 19 &  3 &  $2946.68_{-0.48}^{+0.52}$ &  $1.08 \times 10^{-3} ~_{-4.17 \times 10^{-4}}^{+9.82 \times 10^{-4}}$ &  $2.72_{-1.02}^{+1.40}$ &   > 6 \\
 20 &  1 &  $2963.01_{-0.22}^{+0.21}$ &  $1.80 \times 10^{-2} ~_{-8.65 \times 10^{-3}}^{+2.58 \times 10^{-2}}$ &  $1.06_{-0.41}^{+0.50}$ &   > 6 \\
 20 &  2 &  $3024.69_{-0.28}^{+0.30}$ &  $7.55 \times 10^{-3} ~_{-3.28 \times 10^{-3}}^{+9.10 \times 10^{-3}}$ &  $1.55_{-0.58}^{+0.71}$ &   > 6 \\
 21 &  0 &  $3033.76_{-0.19}^{+0.18}$ &  $2.43 \times 10^{-2} ~_{-1.24 \times 10^{-2}}^{+3.26 \times 10^{-2}}$ &  $0.78_{-0.29}^{+0.41}$ &   > 6 \\
 20 &  3 &  $3082.84_{-0.46}^{+0.45}$ &  $1.33 \times 10^{-3} ~_{-5.11 \times 10^{-4}}^{+1.18 \times 10^{-3}}$ &  $2.74_{-1.00}^{+1.34}$ &   > 6 \\
 21 &  1 &  $3098.37_{-0.22}^{+0.21}$ &  $2.09 \times 10^{-2} ~_{-9.96 \times 10^{-3}}^{+3.02 \times 10^{-2}}$ &  $1.09_{-0.42}^{+0.49}$ &   > 6 \\
 21 &  2 &  $3160.00_{-0.29}^{+0.31}$ &  $6.12 \times 10^{-3} ~_{-3.05 \times 10^{-3}}^{+1.06 \times 10^{-2}}$ &  $1.24_{-0.58}^{+0.83}$ &   > 6 \\
 22 &  0 &  $3168.26_{-0.20}^{+0.19}$ &  $2.41 \times 10^{-2} ~_{-1.19 \times 10^{-2}}^{+3.11 \times 10^{-2}}$ &  $0.86_{-0.32}^{+0.44}$ &   > 6 \\
 21 &  3 &  $3217.81_{-0.55}^{+0.57}$ &  $9.81 \times 10^{-4} ~_{-3.53 \times 10^{-4}}^{+7.69 \times 10^{-4}}$ &  $3.28_{-1.16}^{+1.50}$ &   > 6 \\
 22 &  1 &  $3233.12_{-0.26}^{+0.26}$ &  $1.54 \times 10^{-2} ~_{-6.34 \times 10^{-3}}^{+1.65 \times 10^{-2}}$ &  $1.67_{-0.56}^{+0.62}$ &   > 6 \\
 22 &  2 &  $3295.47_{-0.34}^{+0.37}$ &  $6.73 \times 10^{-3} ~_{-2.76 \times 10^{-3}}^{+6.99 \times 10^{-3}}$ &  $1.95_{-0.72}^{+0.92}$ &   > 6 \\
 23 &  0 &  $3304.00_{-0.25}^{+0.25}$ &  $1.18 \times 10^{-2} ~_{-5.67 \times 10^{-3}}^{+1.60 \times 10^{-2}}$ &  $1.15_{-0.49}^{+0.71}$ &   > 6 \\
 22 &  3 &  $3353.27_{-0.80}^{+1.24}$ &  $3.91 \times 10^{-4} ~_{-1.84 \times 10^{-4}}^{+4.70 \times 10^{-4}}$ &  $3.70_{-2.82}^{+2.63}$ &  3.75 \\
 23 &  1 &  $3367.99_{-0.36}^{+0.34}$ &  $9.42 \times 10^{-3} ~_{-3.43 \times 10^{-3}}^{+7.74 \times 10^{-3}}$ &  $2.53_{-0.78}^{+0.90}$ &   > 6 \\
 23 &  2 &  $3430.96_{-0.60}^{+0.75}$ &  $3.87 \times 10^{-3} ~_{-1.35 \times 10^{-3}}^{+3.15 \times 10^{-3}}$ &  $3.24_{-1.38}^{+1.72}$ &   > 6 \\
 24 &  0 &  $3438.61_{-0.56}^{+0.41}$ &  $5.84 \times 10^{-3} ~_{-2.43 \times 10^{-3}}^{+6.14 \times 10^{-3}}$ &  $2.01_{-0.86}^{+1.24}$ &   > 6 \\
 23 &  3 &  $3489.45_{-0.68}^{+0.89}$ &  $6.45 \times 10^{-4} ~_{-2.35 \times 10^{-4}}^{+6.05 \times 10^{-4}}$ &  $3.94_{-2.26}^{+2.55}$ &   > 6 \\
 24 &  1 &  $3504.45_{-0.45}^{+0.44}$ &  $5.27 \times 10^{-3} ~_{-1.71 \times 10^{-3}}^{+3.30 \times 10^{-3}}$ &  $3.49_{-0.93}^{+1.09}$ &   > 6 \\
 24 &  2 &  $3566.53_{-0.21}^{+0.24}$ &  $7.12 \times 10^{-3} ~_{-3.40 \times 10^{-3}}^{+5.77 \times 10^{-3}}$ &  $0.85_{-0.31}^{+0.62}$ &   > 6 \\
 25 &  0 &  $3574.54_{-0.73}^{+0.69}$ &  $1.74 \times 10^{-3} ~_{-5.58 \times 10^{-4}}^{+1.14 \times 10^{-3}}$ &  $4.25_{-1.51}^{+1.91}$ &   > 6 \\
 24 &  3 &  $3626.70_{-1.51}^{+1.06}$ &  $2.40 \times 10^{-4} ~_{-2.39 \times 10^{-4}}^{+3.72 \times 10^{-4}}$ &  $1.81_{-1.81}^{+4.25}$ &  1.87 \\
 25 &  1 &  $3640.23_{-0.50}^{+0.49}$ &  $2.98 \times 10^{-3} ~_{-1.04 \times 10^{-3}}^{+2.33 \times 10^{-3}}$ &  $3.42_{-1.27}^{+1.70}$ &   > 6 \\
 25 &  2 &  $3703.51_{-1.08}^{+1.18}$ &  $1.30 \times 10^{-3} ~_{-4.52 \times 10^{-4}}^{+8.03 \times 10^{-4}}$ &  $4.76_{-2.12}^{+2.09}$ &   > 6 \\
 26 &  0 &  $3712.14_{-1.57}^{+0.72}$ &  $1.14 \times 10^{-3} ~_{-4.32 \times 10^{-4}}^{+8.84 \times 10^{-4}}$ &  $4.99_{-3.54}^{+2.12}$ &   > 6 \\
 25 &  3 &  $3762.94_{-1.34}^{+1.39}$ &  $6.95 \times 10^{-6} ~_{-6.87 \times 10^{-6}}^{+2.76 \times 10^{-4}}$ &  $0.00_{-0.00}^{+0.49}$ &  0.44 \\
 26 &  1 &  $3776.70_{-0.67}^{+0.64}$ &  $2.15 \times 10^{-3} ~_{-5.27 \times 10^{-4}}^{+9.73 \times 10^{-4}}$ &  $5.63_{-1.47}^{+1.41}$ &   > 6 \\
 
\hline
\end{tabular}
\end{adjustbox}
\endgroup
\label{table:pkb_golf}
\end{table*}
\begin{table*}[!ht]
\centering
\caption{Same as Table~\ref{table:pkb_golf} but for the GOLF spectrum obtained with the series multiplied by the Solar-SONG-like window.}
\begingroup
\renewcommand{\arraystretch}{1.3}
\begin{adjustbox}{totalheight = .8\textheight}
\begin{tabular}{rrrrrrrrr}
\hline\hline
  $n$ &  $\ell$ &  $\nu$ & $H$ & $\Gamma$ & $\ln K$  \\
  &   &  ($\mu$Hz) &  (${\rm m^2 s^{-2} \mu Hz^{-1}}$) &  ($\mu$Hz) &   \\
\hline

 14 &  1 &  $2156.62_{-1.30}^{+1.21}$ &  $1.22 \times 10^{-3} ~_{-1.21 \times 10^{-3}}^{+2.11 \times 10^{-2}}$ &  $0.05_{-0.05}^{+0.59}$ &    > 6 \\
 14 &  2 &  $2216.78_{-2.24}^{+3.75}$ &  $3.26 \times 10^{-4} ~_{-3.25 \times 10^{-4}}^{+8.14 \times 10^{-3}}$ &  $0.01_{-0.01}^{+0.71}$ &   0.50 \\
 15 &  0 &  $2228.28_{-3.07}^{+5.29}$ &  $4.48 \times 10^{-4} ~_{-4.47 \times 10^{-4}}^{+5.51 \times 10^{-3}}$ &  $0.05_{-0.05}^{+1.05}$ &    > 6 \\
 15 &  1 &  $2292.62_{-0.33}^{+0.30}$ &  $1.27 \times 10^{-3} ~_{-5.14 \times 10^{-4}}^{+1.06 \times 10^{-3}}$ &  $1.25_{-0.47}^{+0.64}$ &    > 6 \\
 16 &  1 &  $2426.08_{-0.69}^{+0.19}$ &  $1.47 \times 10^{-3} ~_{-7.77 \times 10^{-4}}^{+6.18 \times 10^{-3}}$ &  $1.02_{-0.83}^{+1.72}$ &    > 6 \\
 16 &  2 &  $2484.91_{-1.84}^{+1.01}$ &  $1.42 \times 10^{-3} ~_{-1.38 \times 10^{-3}}^{+3.53 \times 10^{-3}}$ &  $1.06_{-1.06}^{+1.96}$ &   1.72 \\
 17 &  0 &  $2496.71_{-1.54}^{+5.07}$ &  $9.31 \times 10^{-4} ~_{-9.29 \times 10^{-4}}^{+5.37 \times 10^{-3}}$ &  $0.24_{-0.24}^{+2.01}$ &    > 6 \\
 17 &  1 &  $2558.99_{-0.18}^{+0.18}$ &  $4.63 \times 10^{-3} ~_{-1.66 \times 10^{-3}}^{+3.39 \times 10^{-3}}$ &  $1.04_{-0.32}^{+0.41}$ &    > 6 \\
 17 &  2 &  $2619.03_{-0.30}^{+0.29}$ &  $7.76 \times 10^{-3} ~_{-6.06 \times 10^{-3}}^{+2.98 \times 10^{-2}}$ &  $0.48_{-0.44}^{+0.52}$ &   1.97 \\
 18 &  0 &  $2630.51_{-3.78}^{+2.80}$ &  $3.54 \times 10^{-4} ~_{-3.54 \times 10^{-4}}^{+1.15 \times 10^{-2}}$ &  $0.02_{-0.02}^{+0.92}$ &    > 6 \\
 18 &  1 &  $2693.69_{-0.18}^{+0.16}$ &  $7.32 \times 10^{-3} ~_{-2.80 \times 10^{-3}}^{+6.62 \times 10^{-3}}$ &  $1.03_{-0.40}^{+0.52}$ &    > 6 \\
 18 &  2 &  $2754.27_{-0.55}^{+0.91}$ &  $5.70 \times 10^{-3} ~_{-3.91 \times 10^{-3}}^{+8.40 \times 10^{-3}}$ &  $1.18_{-0.81}^{+0.83}$ &   3.90 \\
 19 &  0 &  $2764.99_{-0.71}^{+0.48}$ &  $5.09 \times 10^{-3} ~_{-3.96 \times 10^{-3}}^{+9.18 \times 10^{-3}}$ &  $1.44_{-0.88}^{+1.41}$ &    > 6 \\
 18 &  3 &  $2811.30_{-1.16}^{+1.43}$ &  $2.60 \times 10^{-5} ~_{-2.58 \times 10^{-5}}^{+1.58 \times 10^{-3}}$ &  $0.00_{-0.00}^{+0.35}$ &   1.06 \\
 19 &  1 &  $2828.25_{-0.19}^{+0.20}$ &  $7.71 \times 10^{-3} ~_{-2.41 \times 10^{-3}}^{+4.48 \times 10^{-3}}$ &  $1.44_{-0.37}^{+0.44}$ &    > 6 \\
 19 &  2 &  $2889.99_{-0.56}^{+0.11}$ &  $6.94 \times 10^{-3} ~_{-4.32 \times 10^{-3}}^{+3.07 \times 10^{-2}}$ &  $0.42_{-0.36}^{+1.86}$ &    > 6 \\
 20 &  0 &  $2899.28_{-0.27}^{+0.31}$ &  $6.52 \times 10^{-3} ~_{-3.10 \times 10^{-3}}^{+8.88 \times 10^{-3}}$ &  $1.02_{-0.52}^{+0.68}$ &    > 6 \\
 19 &  3 &  $2946.46_{-0.52}^{+0.54}$ &  $1.40 \times 10^{-3} ~_{-6.01 \times 10^{-4}}^{+1.23 \times 10^{-3}}$ &  $1.88_{-0.90}^{+1.19}$ &   5.84 \\
 20 &  1 &  $2962.95_{-0.18}^{+0.18}$ &  $1.18 \times 10^{-2} ~_{-3.65 \times 10^{-3}}^{+6.82 \times 10^{-3}}$ &  $1.46_{-0.41}^{+0.46}$ &    > 6 \\
 20 &  2 &  $3024.14_{-0.26}^{+0.25}$ &  $1.50 \times 10^{-2} ~_{-6.22 \times 10^{-3}}^{+1.45 \times 10^{-2}}$ &  $1.18_{-0.41}^{+0.49}$ &    > 6 \\
 21 &  0 &  $3033.66_{-0.18}^{+0.20}$ &  $2.61 \times 10^{-2} ~_{-1.34 \times 10^{-2}}^{+4.34 \times 10^{-2}}$ &  $0.56_{-0.27}^{+0.42}$ &    > 6 \\
 20 &  3 &  $3082.76_{-0.80}^{+0.81}$ &  $1.22 \times 10^{-3} ~_{-4.59 \times 10^{-4}}^{+7.56 \times 10^{-4}}$ &  $3.43_{-1.26}^{+1.62}$ &   5.70 \\
 21 &  1 &  $3098.64_{-0.18}^{+0.17}$ &  $1.67 \times 10^{-2} ~_{-5.34 \times 10^{-3}}^{+1.06 \times 10^{-2}}$ &  $1.34_{-0.40}^{+0.45}$ &    > 6 \\
 21 &  2 &  $3159.64_{-0.53}^{+0.49}$ &  $4.21 \times 10^{-3} ~_{-1.70 \times 10^{-3}}^{+3.66 \times 10^{-3}}$ &  $2.22_{-1.00}^{+1.31}$ &    > 6 \\
 22 &  0 &  $3168.30_{-0.22}^{+0.23}$ &  $1.63 \times 10^{-2} ~_{-6.91 \times 10^{-3}}^{+1.83 \times 10^{-2}}$ &  $1.05_{-0.44}^{+0.60}$ &    > 6 \\
 21 &  3 &  $3218.70_{-1.66}^{+0.30}$ &  $1.12 \times 10^{-3} ~_{-1.12 \times 10^{-3}}^{+1.02 \times 10^{-2}}$ &  $0.08_{-0.08}^{+0.88}$ &   2.04 \\
 22 &  1 &  $3233.12_{-0.28}^{+0.27}$ &  $1.03 \times 10^{-2} ~_{-2.70 \times 10^{-3}}^{+4.29 \times 10^{-3}}$ &  $2.34_{-0.54}^{+0.63}$ &    > 6 \\
 22 &  2 &  $3295.79_{-0.92}^{+0.81}$ &  $2.66 \times 10^{-3} ~_{-9.31 \times 10^{-4}}^{+1.57 \times 10^{-3}}$ &  $4.24_{-1.70}^{+1.94}$ &   4.81 \\
 23 &  0 &  $3304.16_{-0.22}^{+0.19}$ &  $2.00 \times 10^{-2} ~_{-9.90 \times 10^{-3}}^{+2.67 \times 10^{-2}}$ &  $0.79_{-0.35}^{+0.61}$ &    > 6 \\
 22 &  3 &  $3353.68_{-1.37}^{+1.34}$ &  $2.48 \times 10^{-5} ~_{-2.46 \times 10^{-5}}^{+1.92 \times 10^{-3}}$ &  $0.00_{-0.00}^{+0.27}$ &   1.10 \\
 23 &  1 &  $3368.36_{-0.29}^{+0.28}$ &  $8.39 \times 10^{-3} ~_{-2.05 \times 10^{-3}}^{+3.14 \times 10^{-3}}$ &  $3.01_{-0.65}^{+0.73}$ &    > 6 \\
 23 &  2 &  $3429.81_{-0.63}^{+0.95}$ &  $3.28 \times 10^{-3} ~_{-1.17 \times 10^{-3}}^{+1.83 \times 10^{-3}}$ &  $5.42_{-3.83}^{+1.85}$ &   4.11 \\
 24 &  0 &  $3438.40_{-0.79}^{+0.52}$ &  $3.89 \times 10^{-3} ~_{-2.35 \times 10^{-3}}^{+7.26 \times 10^{-3}}$ &  $2.10_{-1.91}^{+3.06}$ &    > 6 \\
 23 &  3 &  $3489.34_{-0.57}^{+0.79}$ &  $1.37 \times 10^{-3} ~_{-1.36 \times 10^{-3}}^{+3.03 \times 10^{-3}}$ &  $0.59_{-0.59}^{+1.21}$ &   2.10 \\
 24 &  1 &  $3504.99_{-0.64}^{+0.58}$ &  $4.02 \times 10^{-3} ~_{-9.49 \times 10^{-4}}^{+1.43 \times 10^{-3}}$ &  $4.84_{-1.15}^{+1.43}$ &    > 6 \\
 24 &  2 &  $3566.21_{-0.31}^{+0.21}$ &  $8.76 \times 10^{-3} ~_{-4.33 \times 10^{-3}}^{+1.03 \times 10^{-2}}$ &  $1.11_{-0.62}^{+2.18}$ &    > 6 \\
 25 &  0 &  $3574.11_{-0.83}^{+1.44}$ &  $1.38 \times 10^{-3} ~_{-1.31 \times 10^{-3}}^{+1.47 \times 10^{-3}}$ &  $3.06_{-3.05}^{+3.03}$ &    > 6 \\
 24 &  3 &  $3626.40_{-1.43}^{+1.30}$ &  $1.65 \times 10^{-5} ~_{-1.63 \times 10^{-5}}^{+1.06 \times 10^{-3}}$ &  $0.00_{-0.00}^{+0.39}$ &   0.28 \\
 25 &  1 &  $3639.76_{-0.42}^{+0.52}$ &  $2.90 \times 10^{-3} ~_{-9.85 \times 10^{-4}}^{+1.77 \times 10^{-3}}$ &  $3.11_{-1.21}^{+1.67}$ &    > 6 \\
 25 &  2 &  $3703.81_{-1.75}^{+1.19}$ &  $1.38 \times 10^{-3} ~_{-8.27 \times 10^{-4}}^{+1.34 \times 10^{-3}}$ &  $4.63_{-4.13}^{+2.32}$ &   2.75 \\
 26 &  0 &  $3712.76_{-1.89}^{+0.23}$ &  $2.56 \times 10^{-3} ~_{-1.70 \times 10^{-3}}^{+6.79 \times 10^{-3}}$ &  $0.82_{-0.61}^{+5.29}$ &    > 6 \\
 25 &  3 &  $3762.95_{-1.34}^{+1.38}$ &  $8.53 \times 10^{-6} ~_{-8.39 \times 10^{-6}}^{+4.60 \times 10^{-4}}$ &  $0.00_{-0.00}^{+0.17}$ &  -0.09 \\
 26 &  1 &  $3778.16_{-0.54}^{+0.43}$ &  $2.21 \times 10^{-3} ~_{-6.25 \times 10^{-4}}^{+9.78 \times 10^{-4}}$ &  $4.20_{-1.22}^{+1.62}$ &    > 6 \\

\hline
\end{tabular}
\end{adjustbox}
\endgroup
\label{table:pkb_golf_wdw}
\end{table*}
\begin{table*}[h!]
\centering
\caption{Same as Table~\ref{table:pkb_golf}, but for Solar-SONG spectrum.}
\begingroup
\renewcommand{\arraystretch}{1.3}
\begin{adjustbox}{totalheight = 1.\textheight}
\begin{tabular}{rrrrrr}
\hline\hline
  $n$ &  $\ell$ &  $\nu$ & $H$ & $\Gamma$ & $\ln K$  \\
  &   &  ($\mu$Hz) &  (${\rm m^2 s^{-2} \mu Hz^{-1}}$) &  ($\mu$Hz) &   \\
\hline

 11 &  1 &  $1749.70_{-0.64}^{+1.38}$ &  $1.29 \times 10^{-4} ~_{-6.47 \times 10^{-5}}^{+9.50 \times 10^{-5}}$ &   $4.82_{-2.14}^{+6.68}$ &    > 6 \\
 12 &  0 &  $1821.74_{-0.48}^{+0.55}$ &  $2.05 \times 10^{-4} ~_{-1.38 \times 10^{-4}}^{+2.81 \times 10^{-4}}$ &   $1.59_{-1.17}^{+2.11}$ &   2.84 \\
 12 &  1 &  $1884.92_{-0.16}^{+0.26}$ &  $5.50 \times 10^{-4} ~_{-3.51 \times 10^{-4}}^{+2.27 \times 10^{-3}}$ &   $0.73_{-0.56}^{+1.51}$ &   5.09 \\
 12 &  2 &  $1946.35_{-2.68}^{+0.12}$ &  $1.83 \times 10^{-3} ~_{-1.57 \times 10^{-3}}^{+1.11 \times 10^{-2}}$ &   $0.09_{-0.08}^{+0.44}$ &   2.46 \\
 13 &  0 &  $1956.02_{-0.83}^{+1.89}$ &  $1.93 \times 10^{-4} ~_{-9.03 \times 10^{-5}}^{+4.69 \times 10^{-4}}$ &   $2.50_{-2.29}^{+2.09}$ &   5.97 \\
 13 &  3 &  $2137.99_{-1.80}^{+5.50}$ &  $5.29 \times 10^{-5} ~_{-2.93 \times 10^{-5}}^{+6.25 \times 10^{-5}}$ &  $5.83_{-3.66}^{+14.65}$ &   3.85 \\
 14 &  1 &  $2156.88_{-0.37}^{+0.39}$ &  $2.99 \times 10^{-4} ~_{-1.16 \times 10^{-4}}^{+2.01 \times 10^{-4}}$ &   $2.25_{-0.93}^{+1.73}$ &   5.46 \\
 15 &  0 &  $2228.84_{-0.28}^{+0.28}$ &  $8.42 \times 10^{-4} ~_{-2.95 \times 10^{-4}}^{+5.69 \times 10^{-4}}$ &   $2.04_{-0.62}^{+0.80}$ &    > 6 \\
 14 &  3 &  $2273.12_{-0.35}^{+0.33}$ &  $2.54 \times 10^{-4} ~_{-1.19 \times 10^{-4}}^{+2.74 \times 10^{-4}}$ &   $1.19_{-0.55}^{+1.00}$ &    > 6 \\
 15 &  1 &  $2291.96_{-0.22}^{+0.22}$ &  $9.70 \times 10^{-4} ~_{-3.18 \times 10^{-4}}^{+5.54 \times 10^{-4}}$ &   $1.78_{-0.48}^{+0.60}$ &    > 6 \\
 15 &  2 &  $2352.09_{-0.56}^{+0.51}$ &  $1.98 \times 10^{-3} ~_{-9.06 \times 10^{-4}}^{+1.87 \times 10^{-3}}$ &   $1.44_{-0.55}^{+0.63}$ &    > 6 \\
 16 &  0 &  $2362.11_{-0.98}^{+0.65}$ &  $5.98 \times 10^{-4} ~_{-4.86 \times 10^{-4}}^{+1.27 \times 10^{-3}}$ &   $1.08_{-1.04}^{+0.89}$ &    > 6 \\
 16 &  1 &  $2425.13_{-0.45}^{+0.42}$ &  $6.65 \times 10^{-4} ~_{-2.03 \times 10^{-4}}^{+3.23 \times 10^{-4}}$ &   $3.28_{-1.05}^{+1.35}$ &    > 6 \\
 16 &  2 &  $2484.81_{-1.83}^{+1.25}$ &  $1.10 \times 10^{-3} ~_{-9.87 \times 10^{-4}}^{+3.87 \times 10^{-3}}$ &   $2.24_{-1.93}^{+3.97}$ &   2.69 \\
 17 &  0 &  $2496.32_{-0.44}^{+0.37}$ &  $1.94 \times 10^{-3} ~_{-1.24 \times 10^{-3}}^{+1.86 \times 10^{-3}}$ &   $1.83_{-0.68}^{+1.04}$ &    > 6 \\
 16 &  3 &  $2541.18_{-0.61}^{+0.69}$ &  $3.26 \times 10^{-4} ~_{-9.86 \times 10^{-5}}^{+1.57 \times 10^{-4}}$ &   $3.99_{-1.24}^{+1.74}$ &    > 6 \\
 17 &  1 &  $2558.99_{-0.18}^{+0.18}$ &  $3.51 \times 10^{-3} ~_{-1.02 \times 10^{-3}}^{+1.74 \times 10^{-3}}$ &   $1.60_{-0.37}^{+0.44}$ &    > 6 \\
 17 &  2 &  $2619.22_{-0.29}^{+0.42}$ &  $9.77 \times 10^{-3} ~_{-8.02 \times 10^{-3}}^{+1.66 \times 10^{-2}}$ &   $0.95_{-0.51}^{+0.82}$ &   3.45 \\
 18 &  0 &  $2630.22_{-1.48}^{+0.57}$ &  $8.72 \times 10^{-4} ~_{-5.18 \times 10^{-4}}^{+5.03 \times 10^{-3}}$ &   $3.25_{-1.91}^{+2.71}$ &    > 6 \\
 17 &  3 &  $2675.37_{-0.26}^{+0.29}$ &  $1.16 \times 10^{-3} ~_{-3.97 \times 10^{-4}}^{+7.03 \times 10^{-4}}$ &   $1.84_{-0.55}^{+0.72}$ &    > 6 \\
 18 &  1 &  $2693.40_{-0.17}^{+0.17}$ &  $5.83 \times 10^{-3} ~_{-1.71 \times 10^{-3}}^{+3.00 \times 10^{-3}}$ &   $1.54_{-0.35}^{+0.40}$ &    > 6 \\
 18 &  2 &  $2754.96_{-0.56}^{+0.38}$ &  $1.26 \times 10^{-2} ~_{-6.04 \times 10^{-3}}^{+1.38 \times 10^{-2}}$ &   $1.05_{-0.39}^{+0.49}$ &    > 6 \\
 19 &  0 &  $2765.04_{-0.30}^{+0.24}$ &  $9.13 \times 10^{-3} ~_{-5.53 \times 10^{-3}}^{+1.25 \times 10^{-2}}$ &   $0.86_{-0.35}^{+0.47}$ &    > 6 \\
 18 &  3 &  $2811.24_{-0.48}^{+0.46}$ &  $7.63 \times 10^{-4} ~_{-2.55 \times 10^{-4}}^{+4.29 \times 10^{-4}}$ &   $3.36_{-1.04}^{+1.56}$ &    > 6 \\
 19 &  1 &  $2828.02_{-0.19}^{+0.19}$ &  $5.95 \times 10^{-3} ~_{-1.71 \times 10^{-3}}^{+2.91 \times 10^{-3}}$ &   $1.71_{-0.38}^{+0.44}$ &    > 6 \\
 19 &  2 &  $2889.39_{-0.65}^{+0.48}$ &  $8.61 \times 10^{-3} ~_{-3.10 \times 10^{-3}}^{+5.81 \times 10^{-3}}$ &   $2.34_{-0.63}^{+0.73}$ &    > 6 \\
 20 &  0 &  $2899.11_{-0.35}^{+0.34}$ &  $5.04 \times 10^{-3} ~_{-3.56 \times 10^{-3}}^{+1.21 \times 10^{-2}}$ &   $0.71_{-0.57}^{+0.79}$ &    > 6 \\
 19 &  3 &  $2947.19_{-0.34}^{+0.36}$ &  $1.84 \times 10^{-3} ~_{-5.56 \times 10^{-4}}^{+9.24 \times 10^{-4}}$ &   $2.44_{-0.64}^{+0.83}$ &    > 6 \\
 20 &  1 &  $2963.07_{-0.16}^{+0.16}$ &  $1.11 \times 10^{-2} ~_{-3.42 \times 10^{-3}}^{+6.23 \times 10^{-3}}$ &   $1.37_{-0.33}^{+0.37}$ &    > 6 \\
 20 &  2 &  $3024.12_{-0.31}^{+0.30}$ &  $2.13 \times 10^{-2} ~_{-8.18 \times 10^{-3}}^{+1.57 \times 10^{-2}}$ &   $1.70_{-0.51}^{+0.60}$ &    > 6 \\
 21 &  0 &  $3033.63_{-0.22}^{+0.34}$ &  $1.31 \times 10^{-2} ~_{-6.78 \times 10^{-3}}^{+2.49 \times 10^{-2}}$ &   $0.80_{-0.47}^{+0.73}$ &    > 6 \\
 20 &  3 &  $3082.99_{-0.52}^{+0.51}$ &  $1.24 \times 10^{-3} ~_{-3.75 \times 10^{-4}}^{+6.00 \times 10^{-4}}$ &   $3.64_{-0.98}^{+1.40}$ &    > 6 \\
 21 &  1 &  $3098.54_{-0.17}^{+0.16}$ &  $1.15 \times 10^{-2} ~_{-3.66 \times 10^{-3}}^{+6.75 \times 10^{-3}}$ &   $1.38_{-0.35}^{+0.41}$ &    > 6 \\
 21 &  2 &  $3159.98_{-0.34}^{+0.30}$ &  $1.03 \times 10^{-2} ~_{-3.79 \times 10^{-3}}^{+7.78 \times 10^{-3}}$ &   $1.66_{-0.59}^{+0.77}$ &    > 6 \\
 22 &  0 &  $3168.33_{-0.22}^{+0.25}$ &  $1.05 \times 10^{-2} ~_{-4.25 \times 10^{-3}}^{+1.10 \times 10^{-2}}$ &   $1.13_{-0.45}^{+0.55}$ &    > 6 \\
 21 &  3 &  $3218.43_{-0.41}^{+0.36}$ &  $2.00 \times 10^{-3} ~_{-7.60 \times 10^{-4}}^{+1.43 \times 10^{-3}}$ &   $2.09_{-0.70}^{+0.97}$ &    > 6 \\
 22 &  1 &  $3233.19_{-0.28}^{+0.27}$ &  $6.92 \times 10^{-3} ~_{-1.77 \times 10^{-3}}^{+2.76 \times 10^{-3}}$ &   $2.59_{-0.54}^{+0.65}$ &    > 6 \\
 22 &  2 &  $3295.56_{-0.85}^{+0.72}$ &  $5.23 \times 10^{-3} ~_{-1.61 \times 10^{-3}}^{+2.74 \times 10^{-3}}$ &   $4.39_{-1.33}^{+1.60}$ &   5.41 \\
 23 &  0 &  $3304.19_{-0.28}^{+0.27}$ &  $9.55 \times 10^{-3} ~_{-4.84 \times 10^{-3}}^{+1.38 \times 10^{-2}}$ &   $0.99_{-0.48}^{+0.92}$ &    > 6 \\
 22 &  3 &  $3352.78_{-0.60}^{+0.81}$ &  $8.19 \times 10^{-4} ~_{-2.60 \times 10^{-4}}^{+4.39 \times 10^{-4}}$ &   $4.18_{-1.41}^{+1.99}$ &    > 6 \\
 23 &  1 &  $3368.58_{-0.21}^{+0.20}$ &  $7.37 \times 10^{-3} ~_{-2.20 \times 10^{-3}}^{+3.74 \times 10^{-3}}$ &   $1.76_{-0.43}^{+0.52}$ &    > 6 \\
 23 &  2 &  $3429.70_{-0.56}^{+0.75}$ &  $5.34 \times 10^{-3} ~_{-1.21 \times 10^{-3}}^{+2.03 \times 10^{-3}}$ &   $6.09_{-2.10}^{+1.29}$ &   0.26 \\
 24 &  0 &  $3438.58_{-0.99}^{+1.57}$ &  $1.08 \times 10^{-3} ~_{-1.07 \times 10^{-3}}^{+4.99 \times 10^{-3}}$ &   $0.14_{-0.14}^{+4.11}$ &   0.26 \\
 23 &  3 &  $3489.57_{-0.49}^{+0.45}$ &  $9.40 \times 10^{-4} ~_{-4.59 \times 10^{-4}}^{+9.32 \times 10^{-4}}$ &   $2.03_{-0.93}^{+1.42}$ &    > 6 \\
 24 &  1 &  $3504.93_{-0.61}^{+0.55}$ &  $2.44 \times 10^{-3} ~_{-5.58 \times 10^{-4}}^{+8.29 \times 10^{-4}}$ &   $4.57_{-1.03}^{+1.36}$ &    > 6 \\
 24 &  2 &  $3566.33_{-0.41}^{+0.37}$ &  $5.92 \times 10^{-3} ~_{-2.00 \times 10^{-3}}^{+4.57 \times 10^{-3}}$ &   $3.32_{-2.06}^{+1.63}$ &   2.69 \\
 25 &  0 &  $3574.31_{-0.98}^{+1.64}$ &  $5.93 \times 10^{-4} ~_{-5.91 \times 10^{-4}}^{+1.16 \times 10^{-3}}$ &   $0.38_{-0.38}^{+5.44}$ &   2.69 \\
 24 &  3 &  $3626.45_{-1.45}^{+1.20}$ &  $4.59 \times 10^{-4} ~_{-1.39 \times 10^{-4}}^{+1.94 \times 10^{-4}}$ &   $6.21_{-1.77}^{+1.25}$ &    > 6 \\
 25 &  1 &  $3639.53_{-0.51}^{+0.55}$ &  $1.27 \times 10^{-3} ~_{-3.68 \times 10^{-4}}^{+6.14 \times 10^{-4}}$ &   $4.25_{-1.38}^{+1.91}$ &    > 6 \\
 25 &  2 &  $3704.25_{-1.66}^{+0.94}$ &  $1.52 \times 10^{-3} ~_{-1.45 \times 10^{-3}}^{+7.30 \times 10^{-4}}$ &   $5.56_{-5.54}^{+1.82}$ &  -0.10 \\
 26 &  0 &  $3710.29_{-0.94}^{+2.18}$ &  $7.66 \times 10^{-4} ~_{-7.50 \times 10^{-4}}^{+5.11 \times 10^{-4}}$ &   $4.57_{-4.57}^{+2.81}$ &   0.41 \\
 25 &  3 &  $3762.77_{-1.30}^{+1.53}$ &  $1.14 \times 10^{-4} ~_{-1.13 \times 10^{-4}}^{+2.52 \times 10^{-4}}$ &   $0.11_{-0.11}^{+6.28}$ &   1.54 \\
 26 &  1 &  $3776.68_{-1.04}^{+1.12}$ &  $8.64 \times 10^{-4} ~_{-1.53 \times 10^{-4}}^{+1.86 \times 10^{-4}}$ &   $7.14_{-1.33}^{+0.64}$ &   4.01 \\
 
\hline
\end{tabular}
\end{adjustbox}
\endgroup
\label{table:pkb_song}
\end{table*}

\section{Comparison with HMI and BiSON \label{sec:hmi_bison}}

\begin{figure*}[ht]
    \centering
    \includegraphics[width=1.\textwidth]{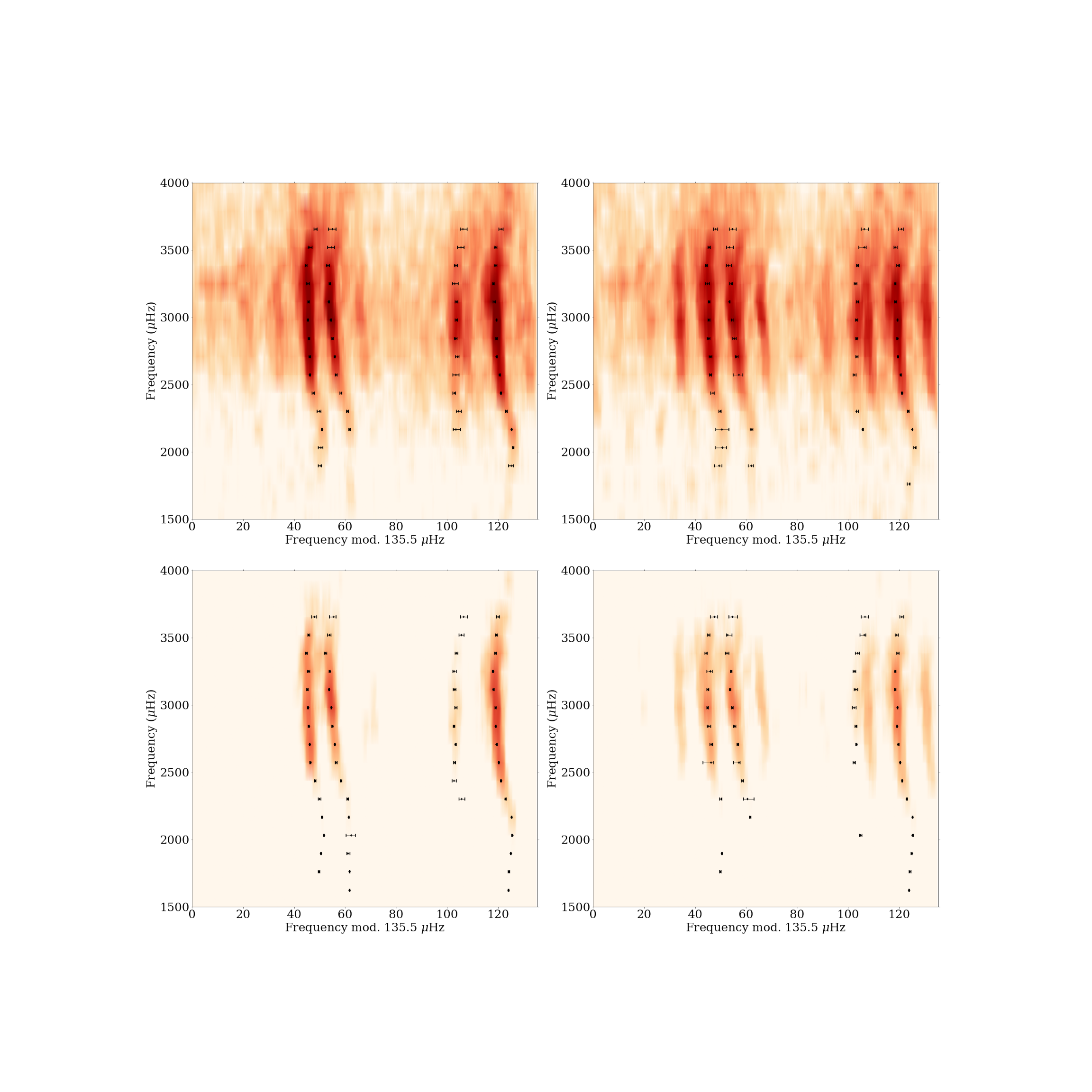}
    \caption{{\'E}chelle diagram for BiSON (\textit{top left}), BiSON with Solar-SONG-like observational window (\textit{top right}), HMI (\textit{bottom left}) and HMI with Solar-SONG-like observational window (\textit{bottom right}). Fitted modes frequencies are represented in black.}
    \label{fig:echelle_diagrams_hmi_bison}
\end{figure*}

\begin{figure*}[ht]
    \centering
    \includegraphics[width=1.\textwidth]{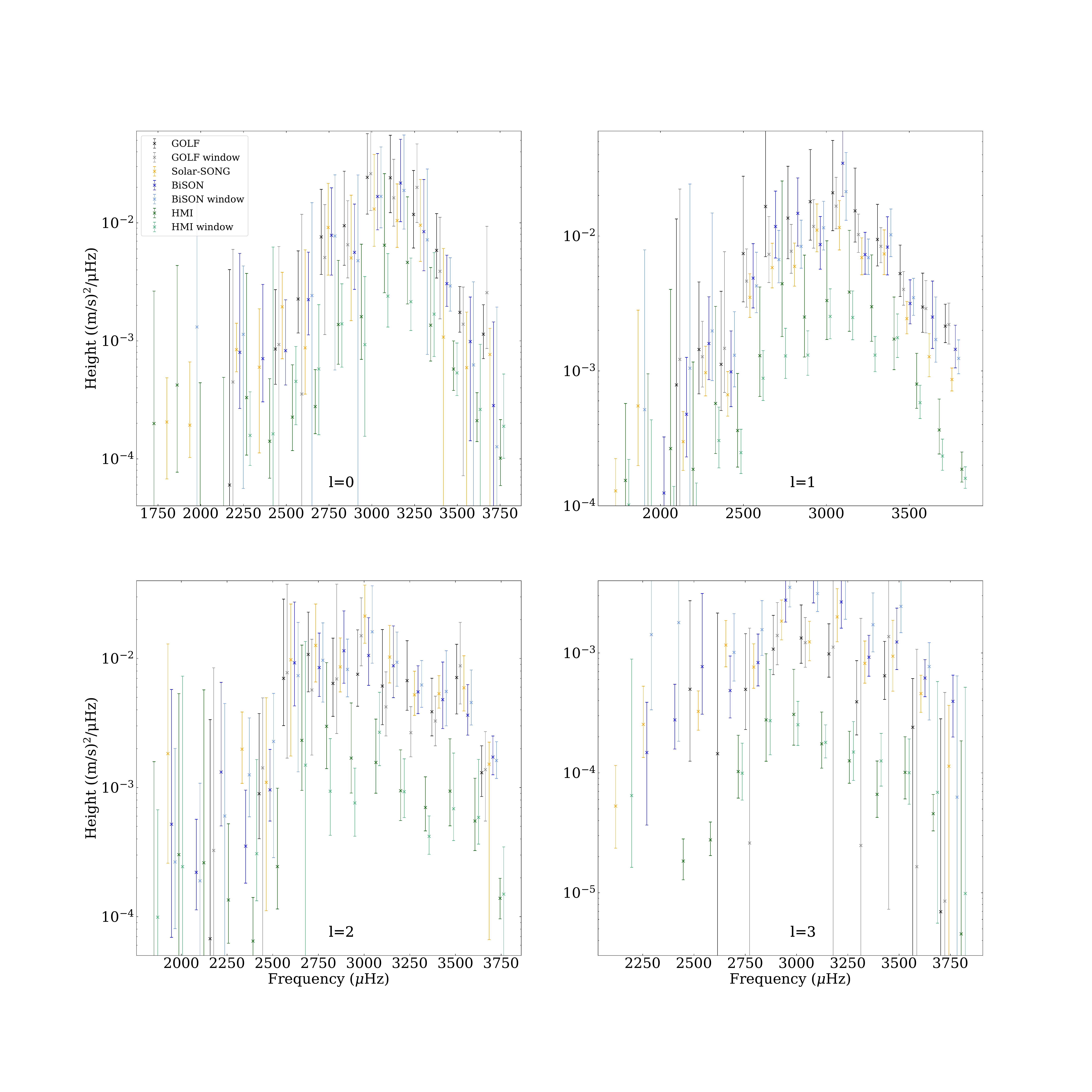}
    \caption{Heights $H$ of the fitted modes for GOLF (black), GOLF with the Solar-SONG window (grey), Solar-SONG (orange), BiSON (blue), BiSON with the Solar-SONG window (light blue), HMI (green), and HMI with the Solar-SONG window (light green) spectra. The horizontal position of the markers has been slightly shifted for visualisation convenience.}
    \label{fig:heights_all_instruments}
\end{figure*}

\begin{figure*}[ht]
    \centering
    \includegraphics[width=1.\textwidth]{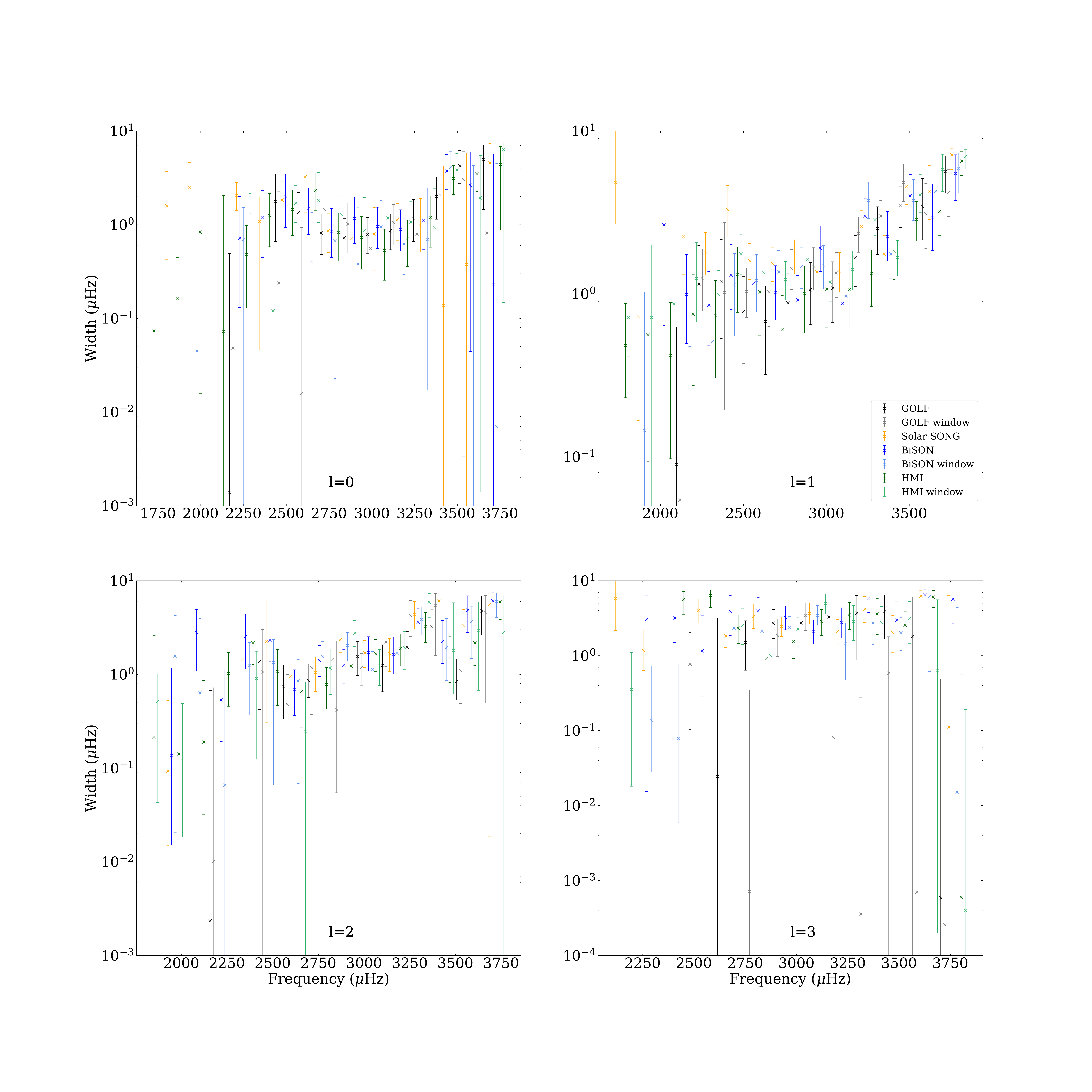}
    \caption{Same as in Fig.~\ref{fig:heights_all_instruments} but for mode widths $\Gamma$.}
    \label{fig:widths_all_instruments}
\end{figure*}

Here, the study of the HMI and BiSON contemporaneous 30-day series (spanning from from 3 June to 2 July) is performed and compared to the Solar-SONG data similar to what was done in Sect.~\ref{sec:peakbagging}. The BiSON subseries are extracted from the January 1976 to March 2020 optimised-for-fill time series\footnote{Available on the BiSON website at:\\ \texttt{http://bison.ph.bham.ac.uk/portal/timeseries}} while the considered HMI time series is obtained from the $\ell=0$ reduction of the HMI full-disk dopplergrams\footnote{Available at:\\ \texttt{http://jsoc.stanford.edu/HMI/Dopplergrams.html}} \citep{2015SoPh..290.3221L}. The duty cycle of the complete BiSON series is 63.5\% while it is 100\% for HMI. For each instrument we did two analyses, one with the original duty cycle and the other using the Solar-SONG observational window function (that is by simply multiplying the time series with the Solar-SONG window function). As the duty cycle of the considered BiSON time series is significantly below 90\%, we used the method described in Sect.~\ref{sec:observational_window} to fit the PSD of this time series, similarly to what was done for all the time series with Solar-SONG observational window.
Table~\ref{tab:power_ratio_all_instruments} compares the mean power value in the 1000-1500 $\mu$Hz region and the power ratio in the 2000-3500 $\mu$Hz and 1700-2200 $\mu$Hz regions as explained in Sect.~\ref{sec:golf_aging}.
Figure~\ref{fig:echelle_diagrams_hmi_bison} shows the four \'{e}chelle diagrams.
HMI and BiSON fitted heights and widths are represented together with the values fitted for GOLF and Solar-SONG in Fig.~\ref{fig:heights_all_instruments} and \ref{fig:widths_all_instruments}, respectively. 
All the fitted parameters, uncertainties and the corresponding $\ln K$ are summarized in Tables~\ref{table:pkb_bison}, \ref{table:pkb_bison_wdw}, \ref{table:pkb_hmi}, and \ref{table:pkb_hmi_wdw}, respectively. 

We note that the mode heights in HMI are significantly lower than for the other instruments, which is expected as we considered only the $\ell = 0$ time series. For both HMI and BiSON, the ratio between the mean power density in the 2000-3500 $\mu$Hz region and the mean power density in the 1000-1500 $\mu$Hz is close to the 9.8 value obtained with Solar-SONG and well above the 3.6 value of GOLF in 2018. The ratio between the mean power density in the 1700-2200 $\mu$Hz region and the mean power density in the 1000-1500 $\mu$Hz is also similar for the four instruments: 1.3, 0.9, 1.1 and 1.0 for Solar-SONG, GOLF, BiSON and HMI, respectively. The characterisation of modes below 1700 $\mu$Hz is only possible using HMI data. The $n=11$, $\ell=0$ mode was properly fitted and a frequency of $1686.73 \pm 0.14$ $\mu$Hz was obtained.  

\begin{table*}[h!]
    \centering
    \caption{Mean power in the 1000=1500 $\mu$Hz for each considered instrument and power ratios in the 2000-3500 $\mu$Hz and 1700-2200 $\mu$Hz regions. The values are given for the 30-day time series spanning from 3 June to 2 July.}
    \begingroup
    \renewcommand{\arraystretch}{2}
    \begin{tabular}{ccccc}
\hline \hline
& Solar-SONG & GOLF & BiSON & HMI  \\
\hline
$<\rm PSD_{[1000-1500 \mu Hz]}>$ ($\rm m^2.s^{-2}Hz$) & 29.1 & 104 & 40.9 & 5.0 \\
$\rm \frac {<PSD_{[2000-3500 \mu Hz]}>}{<PSD_{[1000-1500 \mu Hz]}>}$ & 9.8 & 3.6 & 10.6 & 10.9 \\
$\rm \frac {<PSD_{[1700-2200 \mu Hz]}>}{<PSD_{[1000-1500 \mu Hz]}>}$ & 1.3 & 0.9 & 1.1 & 1.0 \\
\hline
    \end{tabular}
    \label{tab:power_ratio_all_instruments}
    \endgroup
\end{table*}

\begin{table*}[!ht]
\centering
\caption{Parameters of the modes fitted in the BiSON spectrum.} 
\begingroup
\renewcommand{\arraystretch}{1.3}
\begin{adjustbox}{totalheight = .8\textheight}
\begin{tabular}{rrrrrrrrr}
\hline\hline
  $n$ &  $\ell$ &  $\nu$ & $H$ & $\Gamma$ & $\ln K$  \\
  &   &  ($\mu$Hz) &  (${\rm m^2 s^{-2} \mu Hz^{-1}}$) &  ($\mu$Hz) &   \\
\hline
 12 &  2 &  $1946.31_{-1.00}^{+0.16}$ &  $5.20 \times 10^{-4} ~_{-4.50 \times 10^{-4}}^{+5.23 \times 10^{-3}}$ &  $0.14_{-0.12}^{+1.04}$ &  2.29 \\
 13 &  1 &  $2021.10_{-1.08}^{+0.88}$ &  $1.24 \times 10^{-4} ~_{-6.28 \times 10^{-5}}^{+2.00 \times 10^{-4}}$ &  $2.66_{-2.02}^{+2.56}$ &   > 6 \\
 13 &  2 &  $2081.79_{-1.00}^{+0.80}$ &  $2.20 \times 10^{-4} ~_{-1.07 \times 10^{-4}}^{+3.48 \times 10^{-4}}$ &  $2.82_{-1.73}^{+2.12}$ &  3.73 \\
 14 &  1 &  $2157.22_{-0.36}^{+0.37}$ &  $4.77 \times 10^{-4} ~_{-2.47 \times 10^{-4}}^{+7.83 \times 10^{-4}}$ &  $0.99_{-0.50}^{+0.76}$ &   > 6 \\
 14 &  2 &  $2217.60_{-0.25}^{+0.21}$ &  $1.32 \times 10^{-3} ~_{-8.13 \times 10^{-4}}^{+5.21 \times 10^{-3}}$ &  $0.53_{-0.34}^{+0.55}$ &   > 6 \\
 15 &  0 &  $2228.39_{-0.29}^{+0.29}$ &  $7.99 \times 10^{-4} ~_{-5.32 \times 10^{-4}}^{+4.71 \times 10^{-3}}$ &  $0.72_{-0.59}^{+1.29}$ &   > 6 \\
 14 &  3 &  $2270.21_{-1.14}^{+1.81}$ &  $1.48 \times 10^{-4} ~_{-1.11 \times 10^{-4}}^{+2.40 \times 10^{-4}}$ &  $3.06_{-3.05}^{+3.24}$ &  2.34 \\
 15 &  1 &  $2292.02_{-0.19}^{+0.20}$ &  $1.60 \times 10^{-3} ~_{-7.35 \times 10^{-4}}^{+1.95 \times 10^{-3}}$ &  $0.85_{-0.37}^{+0.52}$ &   > 6 \\
 15 &  2 &  $2351.95_{-0.79}^{+0.60}$ &  $3.52 \times 10^{-4} ~_{-1.70 \times 10^{-4}}^{+6.03 \times 10^{-4}}$ &  $2.56_{-1.42}^{+1.86}$ &  4.42 \\
 16 &  0 &  $2363.02_{-0.41}^{+0.38}$ &  $7.07 \times 10^{-4} ~_{-4.03 \times 10^{-4}}^{+2.30 \times 10^{-3}}$ &  $1.20_{-0.75}^{+1.14}$ &   > 6 \\
 15 &  3 &  $2406.72_{-0.92}^{+0.99}$ &  $2.77 \times 10^{-4} ~_{-1.19 \times 10^{-4}}^{+2.72 \times 10^{-4}}$ &  $3.19_{-1.69}^{+2.20}$ &  3.62 \\
 16 &  1 &  $2425.37_{-0.38}^{+0.40}$ &  $9.84 \times 10^{-4} ~_{-4.40 \times 10^{-4}}^{+9.95 \times 10^{-4}}$ &  $1.30_{-0.50}^{+0.71}$ &   > 6 \\
 16 &  2 &  $2485.03_{-0.44}^{+0.47}$ &  $9.60 \times 10^{-4} ~_{-4.10 \times 10^{-4}}^{+1.02 \times 10^{-3}}$ &  $2.31_{-0.93}^{+1.32}$ &   > 6 \\
 17 &  0 &  $2495.90_{-0.42}^{+0.39}$ &  $8.26 \times 10^{-4} ~_{-4.03 \times 10^{-4}}^{+1.40 \times 10^{-3}}$ &  $1.98_{-1.05}^{+1.51}$ &   > 6 \\
 16 &  3 &  $2540.45_{-0.67}^{+0.38}$ &  $7.69 \times 10^{-4} ~_{-4.61 \times 10^{-4}}^{+2.36 \times 10^{-3}}$ &  $1.16_{-0.87}^{+2.31}$ &  3.72 \\
 17 &  1 &  $2558.69_{-0.22}^{+0.23}$ &  $4.87 \times 10^{-3} ~_{-1.94 \times 10^{-3}}^{+3.90 \times 10^{-3}}$ &  $1.16_{-0.37}^{+0.48}$ &   > 6 \\
 17 &  2 &  $2619.08_{-0.18}^{+0.22}$ &  $9.25 \times 10^{-3} ~_{-4.96 \times 10^{-3}}^{+1.81 \times 10^{-2}}$ &  $0.69_{-0.32}^{+0.44}$ &   > 6 \\
 18 &  0 &  $2629.37_{-0.36}^{+0.39}$ &  $2.23 \times 10^{-3} ~_{-1.11 \times 10^{-3}}^{+3.41 \times 10^{-3}}$ &  $1.48_{-0.69}^{+1.00}$ &   > 6 \\
 17 &  3 &  $2676.40_{-1.13}^{+1.28}$ &  $4.86 \times 10^{-4} ~_{-1.98 \times 10^{-4}}^{+4.57 \times 10^{-4}}$ &  $3.91_{-2.05}^{+2.49}$ &  4.47 \\
 18 &  1 &  $2693.64_{-0.20}^{+0.18}$ &  $1.18 \times 10^{-2} ~_{-4.88 \times 10^{-3}}^{+9.77 \times 10^{-3}}$ &  $1.02_{-0.33}^{+0.47}$ &   > 6 \\
 18 &  2 &  $2754.42_{-0.28}^{+0.29}$ &  $8.50 \times 10^{-3} ~_{-3.42 \times 10^{-3}}^{+7.23 \times 10^{-3}}$ &  $1.42_{-0.46}^{+0.61}$ &   > 6 \\
 19 &  0 &  $2764.30_{-0.25}^{+0.23}$ &  $7.85 \times 10^{-3} ~_{-4.25 \times 10^{-3}}^{+1.20 \times 10^{-2}}$ &  $0.84_{-0.39}^{+0.64}$ &   > 6 \\
 18 &  3 &  $2812.43_{-0.77}^{+0.64}$ &  $8.31 \times 10^{-4} ~_{-3.00 \times 10^{-4}}^{+6.07 \times 10^{-4}}$ &  $3.99_{-1.50}^{+1.96}$ &   > 6 \\
 19 &  1 &  $2827.83_{-0.16}^{+0.20}$ &  $1.47 \times 10^{-2} ~_{-6.30 \times 10^{-3}}^{+1.22 \times 10^{-2}}$ &  $0.92_{-0.28}^{+0.38}$ &   > 6 \\
 19 &  2 &  $2889.63_{-0.29}^{+0.25}$ &  $1.15 \times 10^{-2} ~_{-5.05 \times 10^{-3}}^{+1.19 \times 10^{-2}}$ &  $1.25_{-0.44}^{+0.59}$ &   > 6 \\
 20 &  0 &  $2898.82_{-0.29}^{+0.33}$ &  $5.62 \times 10^{-3} ~_{-2.89 \times 10^{-3}}^{+8.81 \times 10^{-3}}$ &  $1.16_{-0.53}^{+0.83}$ &   > 6 \\
 19 &  3 &  $2947.11_{-0.50}^{+0.54}$ &  $2.76 \times 10^{-3} ~_{-9.51 \times 10^{-4}}^{+1.71 \times 10^{-3}}$ &  $3.20_{-0.99}^{+1.41}$ &   > 6 \\
 20 &  1 &  $2963.22_{-0.30}^{+0.30}$ &  $8.64 \times 10^{-3} ~_{-2.97 \times 10^{-3}}^{+5.31 \times 10^{-3}}$ &  $1.91_{-0.54}^{+0.69}$ &   > 6 \\
 20 &  2 &  $3024.61_{-0.28}^{+0.30}$ &  $1.05 \times 10^{-2} ~_{-4.35 \times 10^{-3}}^{+1.01 \times 10^{-2}}$ &  $1.71_{-0.62}^{+0.82}$ &   > 6 \\
 21 &  0 &  $3033.56_{-0.22}^{+0.23}$ &  $1.67 \times 10^{-2} ~_{-7.99 \times 10^{-3}}^{+2.19 \times 10^{-2}}$ &  $0.96_{-0.40}^{+0.57}$ &   > 6 \\
 20 &  3 &  $3082.81_{-0.39}^{+0.40}$ &  $4.17 \times 10^{-3} ~_{-1.55 \times 10^{-3}}^{+2.94 \times 10^{-3}}$ &  $2.08_{-0.64}^{+0.90}$ &   > 6 \\
 21 &  1 &  $3098.64_{-0.14}^{+0.14}$ &  $3.46 \times 10^{-2} ~_{-1.50 \times 10^{-2}}^{+3.04 \times 10^{-2}}$ &  $0.87_{-0.29}^{+0.41}$ &   > 6 \\
 21 &  2 &  $3160.29_{-0.33}^{+0.33}$ &  $8.77 \times 10^{-3} ~_{-3.80 \times 10^{-3}}^{+9.11 \times 10^{-3}}$ &  $1.64_{-0.63}^{+0.88}$ &   > 6 \\
 22 &  0 &  $3168.22_{-0.19}^{+0.20}$ &  $2.17 \times 10^{-2} ~_{-1.15 \times 10^{-2}}^{+2.93 \times 10^{-2}}$ &  $0.88_{-0.36}^{+0.56}$ &   > 6 \\
 21 &  3 &  $3218.34_{-0.53}^{+0.49}$ &  $2.66 \times 10^{-3} ~_{-1.05 \times 10^{-3}}^{+2.29 \times 10^{-3}}$ &  $2.79_{-1.06}^{+1.57}$ &   > 6 \\
 22 &  1 &  $3233.12_{-0.39}^{+0.38}$ &  $7.28 \times 10^{-3} ~_{-2.08 \times 10^{-3}}^{+3.36 \times 10^{-3}}$ &  $3.00_{-0.70}^{+0.87}$ &   > 6 \\
 22 &  2 &  $3295.43_{-0.49}^{+0.54}$ &  $5.51 \times 10^{-3} ~_{-1.77 \times 10^{-3}}^{+3.26 \times 10^{-3}}$ &  $3.60_{-1.10}^{+1.42}$ &   > 6 \\
 23 &  0 &  $3304.07_{-0.32}^{+0.29}$ &  $8.42 \times 10^{-3} ~_{-4.50 \times 10^{-3}}^{+1.49 \times 10^{-2}}$ &  $1.11_{-0.56}^{+1.06}$ &   > 6 \\
 22 &  3 &  $3353.08_{-0.90}^{+1.32}$ &  $9.21 \times 10^{-4} ~_{-2.82 \times 10^{-4}}^{+4.82 \times 10^{-4}}$ &  $5.80_{-2.04}^{+1.50}$ &   > 6 \\
 23 &  1 &  $3368.29_{-0.35}^{+0.31}$ &  $8.28 \times 10^{-3} ~_{-3.13 \times 10^{-3}}^{+5.61 \times 10^{-3}}$ &  $2.26_{-0.66}^{+0.94}$ &   > 6 \\
 23 &  2 &  $3430.10_{-0.41}^{+0.45}$ &  $4.80 \times 10^{-3} ~_{-1.94 \times 10^{-3}}^{+4.59 \times 10^{-3}}$ &  $2.26_{-0.96}^{+1.49}$ &   > 6 \\
 24 &  0 &  $3438.66_{-0.64}^{+0.62}$ &  $3.06 \times 10^{-3} ~_{-1.10 \times 10^{-3}}^{+2.27 \times 10^{-3}}$ &  $3.74_{-1.38}^{+1.86}$ &   > 6 \\
 23 &  3 &  $3488.87_{-0.60}^{+0.62}$ &  $1.23 \times 10^{-3} ~_{-5.07 \times 10^{-4}}^{+1.13 \times 10^{-3}}$ &  $2.98_{-1.36}^{+2.25}$ &   > 6 \\
 24 &  1 &  $3504.56_{-0.54}^{+0.49}$ &  $3.16 \times 10^{-3} ~_{-9.28 \times 10^{-4}}^{+1.57 \times 10^{-3}}$ &  $4.00_{-1.07}^{+1.40}$ &   > 6 \\
 24 &  2 &  $3567.02_{-0.64}^{+0.81}$ &  $3.63 \times 10^{-3} ~_{-1.09 \times 10^{-3}}^{+2.63 \times 10^{-3}}$ &  $4.86_{-2.06}^{+2.04}$ &   > 6 \\
 25 &  0 &  $3575.69_{-1.81}^{+0.98}$ &  $9.86 \times 10^{-4} ~_{-8.43 \times 10^{-4}}^{+1.37 \times 10^{-3}}$ &  $2.65_{-2.60}^{+3.34}$ &   > 6 \\
 24 &  3 &  $3626.28_{-1.33}^{+1.30}$ &  $6.18 \times 10^{-4} ~_{-1.86 \times 10^{-4}}^{+2.62 \times 10^{-4}}$ &  $6.54_{-1.64}^{+1.04}$ &   > 6 \\
 25 &  1 &  $3639.81_{-0.51}^{+0.57}$ &  $2.51 \times 10^{-3} ~_{-1.04 \times 10^{-3}}^{+2.13 \times 10^{-3}}$ &  $2.93_{-1.08}^{+1.80}$ &   > 6 \\
 25 &  2 &  $3704.87_{-0.77}^{+0.47}$ &  $1.72 \times 10^{-3} ~_{-4.67 \times 10^{-4}}^{+7.87 \times 10^{-4}}$ &  $6.11_{-2.01}^{+1.34}$ &   > 6 \\
 26 &  0 &  $3711.30_{-1.61}^{+1.39}$ &  $2.83 \times 10^{-4} ~_{-2.82 \times 10^{-4}}^{+1.16 \times 10^{-3}}$ &  $0.23_{-0.23}^{+5.46}$ &   > 6 \\
 25 &  3 &  $3762.50_{-1.14}^{+1.66}$ &  $3.95 \times 10^{-4} ~_{-1.96 \times 10^{-4}}^{+2.59 \times 10^{-4}}$ &  $5.66_{-2.99}^{+1.66}$ &  3.33 \\
 26 &  1 &  $3777.49_{-0.90}^{+0.71}$ &  $1.44 \times 10^{-3} ~_{-3.91 \times 10^{-4}}^{+7.36 \times 10^{-4}}$ &  $5.48_{-1.74}^{+1.68}$ &   > 6 \\
\hline
\end{tabular}
\end{adjustbox}
\endgroup
\label{table:pkb_bison}
\end{table*}

\begin{table*}[!ht]
\centering
\caption{Same as Table~\ref{table:pkb_bison} but for the BiSON spectrum obtained with the series multiplied by the Solar-SONG-like window.} 
\begingroup
\renewcommand{\arraystretch}{1.3}
\begin{adjustbox}{totalheight = .8\textheight}
\begin{tabular}{rrrrrrrrr}
\hline\hline
  $n$ &  $\ell$ &  $\nu$ & $H$ & $\Gamma$ & $\ln K$  \\
  &   &  ($\mu$Hz) &  (${\rm m^2 s^{-2} \mu Hz^{-1}}$) &  ($\mu$Hz) &   \\
\hline
 12 &  1 &  $1884.54_{-0.92}^{+0.25}$ &  $5.16 \times 10^{-4} ~_{-4.75 \times 10^{-4}}^{+7.38 \times 10^{-3}}$ &  $0.14_{-0.14}^{+0.88}$ &   > 6 \\
 12 &  2 &  $1945.26_{-1.77}^{+1.13}$ &  $2.65 \times 10^{-4} ~_{-1.84 \times 10^{-4}}^{+1.74 \times 10^{-3}}$ &  $1.56_{-1.54}^{+2.70}$ &  2.21 \\
 13 &  0 &  $1957.92_{-1.27}^{+0.93}$ &  $1.31 \times 10^{-3} ~_{-1.30 \times 10^{-3}}^{+1.35 \times 10^{-2}}$ &  $0.04_{-0.04}^{+0.31}$ &   > 6 \\
 13 &  2 &  $2081.88_{-2.54}^{+1.73}$ &  $1.89 \times 10^{-4} ~_{-1.87 \times 10^{-4}}^{+8.91 \times 10^{-4}}$ &  $0.64_{-0.64}^{+3.33}$ &  1.32 \\
 14 &  1 &  $2157.75_{-0.71}^{+0.16}$ &  $1.05 \times 10^{-3} ~_{-9.82 \times 10^{-4}}^{+2.33 \times 10^{-2}}$ &  $0.05_{-0.05}^{+0.43}$ &   > 6 \\
 14 &  2 &  $2217.28_{-2.42}^{+2.73}$ &  $6.02 \times 10^{-4} ~_{-6.01 \times 10^{-4}}^{+3.88 \times 10^{-3}}$ &  $0.07_{-0.07}^{+1.08}$ &  0.97 \\
 15 &  0 &  $2228.95_{-0.62}^{+0.44}$ &  $1.13 \times 10^{-3} ~_{-1.08 \times 10^{-3}}^{+3.18 \times 10^{-3}}$ &  $0.69_{-0.69}^{+0.84}$ &   > 6 \\
 14 &  3 &  $2272.40_{-0.11}^{+0.30}$ &  $1.42 \times 10^{-3} ~_{-1.09 \times 10^{-3}}^{+5.62 \times 10^{-3}}$ &  $0.14_{-0.11}^{+0.59}$ &  2.71 \\
 15 &  1 &  $2291.92_{-0.10}^{+0.17}$ &  $1.98 \times 10^{-3} ~_{-1.13 \times 10^{-3}}^{+1.28 \times 10^{-2}}$ &  $0.51_{-0.38}^{+0.54}$ &   > 6 \\
 15 &  2 &  $2351.97_{-0.55}^{+0.35}$ &  $1.25 \times 10^{-3} ~_{-6.61 \times 10^{-4}}^{+2.21 \times 10^{-3}}$ &  $1.24_{-0.87}^{+0.96}$ &  3.07 \\
 15 &  3 &  $2405.49_{-0.11}^{+0.71}$ &  $1.79 \times 10^{-3} ~_{-1.61 \times 10^{-3}}^{+7.90 \times 10^{-3}}$ &  $0.08_{-0.07}^{+0.70}$ &  2.31 \\
 16 &  1 &  $2425.80_{-0.34}^{+0.30}$ &  $1.31 \times 10^{-3} ~_{-5.44 \times 10^{-4}}^{+1.44 \times 10^{-3}}$ &  $1.13_{-0.58}^{+0.63}$ &   > 6 \\
 16 &  2 &  $2484.50_{-0.81}^{+0.57}$ &  $2.27 \times 10^{-3} ~_{-1.99 \times 10^{-3}}^{+3.09 \times 10^{-3}}$ &  $1.34_{-1.28}^{+1.01}$ &  2.26 \\
 17 &  1 &  $2558.69_{-0.20}^{+0.21}$ &  $4.27 \times 10^{-3} ~_{-1.56 \times 10^{-3}}^{+3.31 \times 10^{-3}}$ &  $1.21_{-0.43}^{+0.54}$ &   > 6 \\
 17 &  2 &  $2618.91_{-0.37}^{+0.38}$ &  $7.36 \times 10^{-3} ~_{-6.04 \times 10^{-3}}^{+1.17 \times 10^{-2}}$ &  $0.85_{-0.78}^{+0.60}$ &  2.23 \\
 18 &  0 &  $2630.20_{-2.34}^{+1.52}$ &  $2.42 \times 10^{-3} ~_{-2.42 \times 10^{-3}}^{+1.24 \times 10^{-2}}$ &  $0.40_{-0.40}^{+0.95}$ &   > 6 \\
 17 &  3 &  $2675.32_{-0.44}^{+0.74}$ &  $1.01 \times 10^{-3} ~_{-4.28 \times 10^{-4}}^{+1.12 \times 10^{-3}}$ &  $2.33_{-1.50}^{+2.14}$ &   > 6 \\
 18 &  1 &  $2693.60_{-0.21}^{+0.20}$ &  $6.69 \times 10^{-3} ~_{-2.18 \times 10^{-3}}^{+4.31 \times 10^{-3}}$ &  $1.36_{-0.41}^{+0.48}$ &   > 6 \\
 18 &  2 &  $2754.27_{-0.45}^{+0.61}$ &  $9.68 \times 10^{-3} ~_{-5.09 \times 10^{-3}}^{+9.20 \times 10^{-3}}$ &  $1.55_{-0.54}^{+0.69}$ &   > 6 \\
 19 &  0 &  $2764.61_{-0.46}^{+0.61}$ &  $7.74 \times 10^{-3} ~_{-7.17 \times 10^{-3}}^{+1.78 \times 10^{-2}}$ &  $0.68_{-0.65}^{+1.04}$ &   > 6 \\
 18 &  3 &  $2811.81_{-0.41}^{+0.49}$ &  $1.56 \times 10^{-3} ~_{-6.10 \times 10^{-4}}^{+1.17 \times 10^{-3}}$ &  $2.12_{-0.92}^{+1.28}$ &   > 6 \\
 19 &  1 &  $2828.00_{-0.20}^{+0.20}$ &  $8.38 \times 10^{-3} ~_{-2.61 \times 10^{-3}}^{+4.73 \times 10^{-3}}$ &  $1.47_{-0.39}^{+0.46}$ &   > 6 \\
 19 &  2 &  $2889.05_{-0.50}^{+0.57}$ &  $8.22 \times 10^{-3} ~_{-3.18 \times 10^{-3}}^{+5.91 \times 10^{-3}}$ &  $2.04_{-0.65}^{+0.75}$ &   > 6 \\
 20 &  0 &  $2899.17_{-0.61}^{+0.69}$ &  $4.78 \times 10^{-3} ~_{-4.75 \times 10^{-3}}^{+2.07 \times 10^{-2}}$ &  $0.38_{-0.38}^{+1.16}$ &   > 6 \\
 19 &  3 &  $2947.12_{-0.34}^{+0.37}$ &  $3.53 \times 10^{-3} ~_{-1.11 \times 10^{-3}}^{+1.84 \times 10^{-3}}$ &  $2.38_{-0.70}^{+0.96}$ &   > 6 \\
 20 &  1 &  $2963.15_{-0.20}^{+0.20}$ &  $1.15 \times 10^{-2} ~_{-3.62 \times 10^{-3}}^{+6.59 \times 10^{-3}}$ &  $1.48_{-0.41}^{+0.49}$ &   > 6 \\
 20 &  2 &  $3024.53_{-0.32}^{+0.42}$ &  $1.61 \times 10^{-2} ~_{-6.90 \times 10^{-3}}^{+2.05 \times 10^{-2}}$ &  $1.13_{-0.62}^{+0.62}$ &   > 6 \\
 21 &  0 &  $3033.58_{-0.21}^{+0.49}$ &  $1.68 \times 10^{-2} ~_{-7.68 \times 10^{-3}}^{+2.72 \times 10^{-2}}$ &  $0.95_{-0.59}^{+0.86}$ &   > 6 \\
 20 &  3 &  $3082.54_{-0.51}^{+0.48}$ &  $3.13 \times 10^{-3} ~_{-9.10 \times 10^{-4}}^{+1.41 \times 10^{-3}}$ &  $3.45_{-0.90}^{+1.24}$ &   > 6 \\
 21 &  1 &  $3098.57_{-0.16}^{+0.13}$ &  $2.13 \times 10^{-2} ~_{-8.25 \times 10^{-3}}^{+2.03 \times 10^{-2}}$ &  $0.97_{-0.38}^{+0.47}$ &   > 6 \\
 21 &  2 &  $3160.08_{-0.32}^{+0.29}$ &  $9.34 \times 10^{-3} ~_{-3.25 \times 10^{-3}}^{+6.67 \times 10^{-3}}$ &  $1.68_{-0.54}^{+0.65}$ &   > 6 \\
 22 &  0 &  $3168.08_{-0.15}^{+0.21}$ &  $1.88 \times 10^{-2} ~_{-9.97 \times 10^{-3}}^{+3.67 \times 10^{-2}}$ &  $0.62_{-0.33}^{+0.54}$ &   > 6 \\
 21 &  3 &  $3218.58_{-0.57}^{+0.27}$ &  $4.10 \times 10^{-3} ~_{-2.19 \times 10^{-3}}^{+7.56 \times 10^{-3}}$ &  $1.44_{-0.97}^{+1.28}$ &   > 6 \\
 22 &  1 &  $3233.23_{-0.48}^{+0.44}$ &  $6.91 \times 10^{-3} ~_{-1.73 \times 10^{-3}}^{+2.59 \times 10^{-3}}$ &  $3.75_{-0.84}^{+1.13}$ &   > 6 \\
 22 &  2 &  $3295.00_{-0.88}^{+0.70}$ &  $6.23 \times 10^{-3} ~_{-2.05 \times 10^{-3}}^{+3.41 \times 10^{-3}}$ &  $3.86_{-1.24}^{+1.39}$ &   > 6 \\
 23 &  0 &  $3304.26_{-0.71}^{+0.36}$ &  $7.18 \times 10^{-3} ~_{-6.41 \times 10^{-3}}^{+2.14 \times 10^{-2}}$ &  $0.69_{-0.68}^{+1.76}$ &   > 6 \\
 22 &  3 &  $3352.85_{-0.46}^{+0.59}$ &  $1.72 \times 10^{-3} ~_{-6.97 \times 10^{-4}}^{+1.45 \times 10^{-3}}$ &  $2.73_{-1.31}^{+2.17}$ &   > 6 \\
 23 &  1 &  $3368.58_{-0.22}^{+0.21}$ &  $1.02 \times 10^{-2} ~_{-3.18 \times 10^{-3}}^{+5.66 \times 10^{-3}}$ &  $1.76_{-0.51}^{+0.64}$ &   > 6 \\
 23 &  2 &  $3429.83_{-0.43}^{+0.47}$ &  $5.55 \times 10^{-3} ~_{-2.54 \times 10^{-3}}^{+5.93 \times 10^{-3}}$ &  $1.92_{-1.06}^{+2.05}$ &  4.66 \\
 24 &  0 &  $3438.52_{-0.80}^{+1.17}$ &  $2.92 \times 10^{-3} ~_{-1.14 \times 10^{-3}}^{+2.14 \times 10^{-3}}$ &  $4.06_{-1.94}^{+2.01}$ &   > 6 \\
 23 &  3 &  $3489.18_{-0.41}^{+0.36}$ &  $2.45 \times 10^{-3} ~_{-9.70 \times 10^{-4}}^{+1.86 \times 10^{-3}}$ &  $2.03_{-0.86}^{+1.23}$ &   > 6 \\
 24 &  1 &  $3505.11_{-0.51}^{+0.46}$ &  $3.49 \times 10^{-3} ~_{-9.12 \times 10^{-4}}^{+1.37 \times 10^{-3}}$ &  $3.75_{-0.94}^{+1.28}$ &   > 6 \\
 24 &  2 &  $3566.40_{-0.49}^{+0.38}$ &  $4.56 \times 10^{-3} ~_{-1.51 \times 10^{-3}}^{+3.58 \times 10^{-3}}$ &  $3.64_{-2.16}^{+1.71}$ &   > 6 \\
 25 &  0 &  $3574.19_{-0.94}^{+1.79}$ &  $6.25 \times 10^{-4} ~_{-6.24 \times 10^{-4}}^{+2.54 \times 10^{-3}}$ &  $0.06_{-0.06}^{+4.24}$ &   > 6 \\
 24 &  3 &  $3627.15_{-2.05}^{+0.89}$ &  $7.70 \times 10^{-4} ~_{-4.94 \times 10^{-4}}^{+4.48 \times 10^{-4}}$ &  $6.11_{-5.50}^{+1.40}$ &  2.52 \\
 25 &  1 &  $3639.30_{-0.40}^{+0.75}$ &  $1.71 \times 10^{-3} ~_{-5.47 \times 10^{-4}}^{+1.83 \times 10^{-3}}$ &  $4.27_{-3.17}^{+2.44}$ &   > 6 \\
 25 &  2 &  $3704.43_{-0.91}^{+0.73}$ &  $1.62 \times 10^{-3} ~_{-4.42 \times 10^{-4}}^{+6.48 \times 10^{-4}}$ &  $6.04_{-1.99}^{+1.33}$ &  2.21 \\
 26 &  0 &  $3710.90_{-1.30}^{+1.54}$ &  $1.27 \times 10^{-4} ~_{-1.26 \times 10^{-4}}^{+1.81 \times 10^{-3}}$ &  $0.01_{-0.01}^{+4.48}$ &  2.23 \\
 25 &  3 &  $3762.71_{-1.28}^{+1.54}$ &  $6.27 \times 10^{-5} ~_{-6.24 \times 10^{-5}}^{+5.81 \times 10^{-4}}$ &  $0.02_{-0.02}^{+4.39}$ &  0.53 \\
 26 &  1 &  $3777.16_{-0.97}^{+0.84}$ &  $1.24 \times 10^{-3} ~_{-2.84 \times 10^{-4}}^{+4.60 \times 10^{-4}}$ &  $5.90_{-1.75}^{+1.42}$ &   > 6 \\
\hline
\end{tabular}
\end{adjustbox}
\endgroup
\label{table:pkb_bison_wdw}
\end{table*}

\begin{table*}[!ht]
\centering
\caption{Parameters of the modes fitted in the HMI spectrum.} 
\begingroup
\renewcommand{\arraystretch}{1.3}
\begin{adjustbox}{totalheight = .8\textheight}
\begin{tabular}{rrrrrrrrr}
\hline\hline
  $n$ &  $\ell$ &  $\nu$ & $H$ & $\Gamma$ & $\ln K$  \\
  &   &  ($\mu$Hz) &  (${\rm m^2 s^{-2} \mu Hz^{-1}}$) &  ($\mu$Hz) &   \\
\hline
 11 &  0 &  $1686.73_{-0.14}^{+0.14}$ &  $1.99 \times 10^{-4} ~_{-1.75 \times 10^{-4}}^{+2.44 \times 10^{-3}}$ &  $0.07_{-0.06}^{+0.25}$ &   > 6 \\
 11 &  1 &  $1749.09_{-0.17}^{+0.19}$ &  $1.54 \times 10^{-4} ~_{-9.55 \times 10^{-5}}^{+4.19 \times 10^{-4}}$ &  $0.48_{-0.25}^{+0.39}$ &   > 6 \\
 11 &  2 &  $1810.20_{-0.32}^{+0.23}$ &  $4.34 \times 10^{-5} ~_{-3.63 \times 10^{-5}}^{+1.54 \times 10^{-3}}$ &  $0.21_{-0.19}^{+2.40}$ &  3.39 \\
 12 &  0 &  $1822.26_{-0.17}^{+0.10}$ &  $4.22 \times 10^{-4} ~_{-3.45 \times 10^{-4}}^{+3.93 \times 10^{-3}}$ &  $0.16_{-0.12}^{+0.28}$ &   > 6 \\
 12 &  1 &  $1884.73_{-0.28}^{+0.23}$ &  $4.84 \times 10^{-5} ~_{-3.35 \times 10^{-5}}^{+9.03 \times 10^{-4}}$ &  $0.56_{-0.47}^{+0.78}$ &   > 6 \\
 12 &  2 &  $1946.32_{-0.14}^{+0.16}$ &  $3.02 \times 10^{-4} ~_{-2.54 \times 10^{-4}}^{+5.04 \times 10^{-3}}$ &  $0.14_{-0.11}^{+0.39}$ &   > 6 \\
 13 &  0 &  $1957.03_{-0.54}^{+0.79}$ &  $1.87 \times 10^{-5} ~_{-1.34 \times 10^{-5}}^{+4.23 \times 10^{-4}}$ &  $0.84_{-0.82}^{+1.88}$ &   > 6 \\
 13 &  1 &  $2020.94_{-0.24}^{+0.15}$ &  $2.66 \times 10^{-4} ~_{-1.86 \times 10^{-4}}^{+3.76 \times 10^{-3}}$ &  $0.42_{-0.32}^{+0.47}$ &   > 6 \\
 13 &  2 &  $2082.97_{-0.20}^{+0.14}$ &  $2.61 \times 10^{-4} ~_{-2.14 \times 10^{-4}}^{+5.45 \times 10^{-3}}$ &  $0.19_{-0.16}^{+0.67}$ &   > 6 \\
 14 &  0 &  $2093.65_{-2.01}^{+1.60}$ &  $2.15 \times 10^{-5} ~_{-2.05 \times 10^{-5}}^{+4.69 \times 10^{-4}}$ &  $0.07_{-0.07}^{+1.98}$ &   > 6 \\
 14 &  1 &  $2156.89_{-0.23}^{+0.25}$ &  $1.86 \times 10^{-4} ~_{-1.10 \times 10^{-4}}^{+9.78 \times 10^{-4}}$ &  $0.75_{-0.48}^{+0.56}$ &   > 6 \\
 14 &  2 &  $2217.63_{-0.28}^{+0.26}$ &  $1.35 \times 10^{-4} ~_{-7.24 \times 10^{-5}}^{+3.89 \times 10^{-4}}$ &  $1.02_{-0.56}^{+0.69}$ &   > 6 \\
 15 &  0 &  $2228.15_{-0.21}^{+0.16}$ &  $3.30 \times 10^{-4} ~_{-2.23 \times 10^{-4}}^{+3.40 \times 10^{-3}}$ &  $0.48_{-0.35}^{+0.50}$ &   > 6 \\
 15 &  1 &  $2292.01_{-0.20}^{+0.20}$ &  $5.74 \times 10^{-4} ~_{-3.30 \times 10^{-4}}^{+2.44 \times 10^{-3}}$ &  $0.73_{-0.43}^{+0.47}$ &   > 6 \\
 15 &  2 &  $2352.09_{-0.49}^{+0.48}$ &  $6.47 \times 10^{-5} ~_{-2.81 \times 10^{-5}}^{+7.61 \times 10^{-5}}$ &  $2.18_{-0.90}^{+1.20}$ &   > 6 \\
 16 &  0 &  $2363.03_{-0.31}^{+0.31}$ &  $1.41 \times 10^{-4} ~_{-7.25 \times 10^{-5}}^{+3.37 \times 10^{-4}}$ &  $1.25_{-0.66}^{+0.91}$ &   > 6 \\
 15 &  3 &  $2407.89_{-1.06}^{+1.18}$ &  $1.84 \times 10^{-5} ~_{-5.56 \times 10^{-6}}^{+9.75 \times 10^{-6}}$ &  $5.59_{-2.01}^{+1.61}$ &   > 6 \\
 16 &  1 &  $2425.06_{-0.26}^{+0.26}$ &  $3.61 \times 10^{-4} ~_{-1.68 \times 10^{-4}}^{+5.96 \times 10^{-4}}$ &  $1.32_{-0.55}^{+0.62}$ &   > 6 \\
 16 &  2 &  $2485.73_{-0.29}^{+0.27}$ &  $2.44 \times 10^{-4} ~_{-1.30 \times 10^{-4}}^{+7.44 \times 10^{-4}}$ &  $1.08_{-0.62}^{+0.77}$ &   > 6 \\
 17 &  0 &  $2495.95_{-0.32}^{+0.31}$ &  $2.26 \times 10^{-4} ~_{-1.08 \times 10^{-4}}^{+4.01 \times 10^{-4}}$ &  $1.46_{-0.69}^{+0.90}$ &   > 6 \\
 16 &  3 &  $2540.31_{-0.87}^{+0.86}$ &  $2.76 \times 10^{-5} ~_{-7.18 \times 10^{-6}}^{+1.13 \times 10^{-5}}$ &  $6.33_{-1.96}^{+1.20}$ &   > 6 \\
 17 &  1 &  $2558.69_{-0.22}^{+0.21}$ &  $1.30 \times 10^{-3} ~_{-6.50 \times 10^{-4}}^{+2.89 \times 10^{-3}}$ &  $1.03_{-0.47}^{+0.49}$ &   > 6 \\
 17 &  2 &  $2619.30_{-0.20}^{+0.17}$ &  $2.32 \times 10^{-3} ~_{-1.37 \times 10^{-3}}^{+1.04 \times 10^{-2}}$ &  $0.66_{-0.39}^{+0.45}$ &   > 6 \\
 18 &  0 &  $2629.42_{-0.37}^{+0.38}$ &  $2.77 \times 10^{-4} ~_{-1.14 \times 10^{-4}}^{+2.93 \times 10^{-4}}$ &  $2.31_{-0.90}^{+1.25}$ &   > 6 \\
 17 &  3 &  $2675.94_{-0.39}^{+0.40}$ &  $1.03 \times 10^{-4} ~_{-4.09 \times 10^{-5}}^{+1.03 \times 10^{-4}}$ &  $2.34_{-0.88}^{+1.16}$ &   > 6 \\
 18 &  1 &  $2693.32_{-0.18}^{+0.16}$ &  $4.43 \times 10^{-3} ~_{-2.63 \times 10^{-3}}^{+2.11 \times 10^{-2}}$ &  $0.60_{-0.36}^{+0.37}$ &   > 6 \\
 18 &  2 &  $2754.49_{-0.18}^{+0.19}$ &  $2.98 \times 10^{-3} ~_{-1.58 \times 10^{-3}}^{+6.30 \times 10^{-3}}$ &  $0.78_{-0.35}^{+0.41}$ &   > 6 \\
 19 &  0 &  $2764.34_{-0.21}^{+0.20}$ &  $1.38 \times 10^{-3} ~_{-7.42 \times 10^{-4}}^{+3.43 \times 10^{-3}}$ &  $0.83_{-0.42}^{+0.51}$ &   > 6 \\
 18 &  3 &  $2811.70_{-0.23}^{+0.27}$ &  $2.77 \times 10^{-4} ~_{-1.52 \times 10^{-4}}^{+7.07 \times 10^{-4}}$ &  $0.92_{-0.50}^{+0.75}$ &   > 6 \\
 19 &  1 &  $2827.82_{-0.21}^{+0.21}$ &  $2.51 \times 10^{-3} ~_{-1.24 \times 10^{-3}}^{+4.71 \times 10^{-3}}$ &  $1.01_{-0.43}^{+0.47}$ &   > 6 \\
 19 &  2 &  $2889.53_{-0.23}^{+0.24}$ &  $1.69 \times 10^{-3} ~_{-7.84 \times 10^{-4}}^{+2.83 \times 10^{-3}}$ &  $1.23_{-0.51}^{+0.56}$ &   > 6 \\
 20 &  0 &  $2898.80_{-0.19}^{+0.20}$ &  $1.60 \times 10^{-3} ~_{-9.06 \times 10^{-4}}^{+4.99 \times 10^{-3}}$ &  $0.73_{-0.40}^{+0.49}$ &   > 6 \\
 19 &  3 &  $2946.52_{-0.30}^{+0.33}$ &  $3.07 \times 10^{-4} ~_{-1.37 \times 10^{-4}}^{+4.22 \times 10^{-4}}$ &  $1.55_{-0.63}^{+0.79}$ &   > 6 \\
 20 &  1 &  $2962.89_{-0.21}^{+0.21}$ &  $3.32 \times 10^{-3} ~_{-1.61 \times 10^{-3}}^{+5.90 \times 10^{-3}}$ &  $1.07_{-0.45}^{+0.48}$ &   > 6 \\
 20 &  2 &  $3024.62_{-0.27}^{+0.28}$ &  $1.56 \times 10^{-3} ~_{-6.60 \times 10^{-4}}^{+1.83 \times 10^{-3}}$ &  $1.66_{-0.60}^{+0.69}$ &   > 6 \\
 21 &  0 &  $3033.74_{-0.16}^{+0.15}$ &  $6.47 \times 10^{-3} ~_{-3.91 \times 10^{-3}}^{+1.97 \times 10^{-2}}$ &  $0.53_{-0.28}^{+0.36}$ &   > 6 \\
 20 &  3 &  $3082.68_{-0.44}^{+0.46}$ &  $1.75 \times 10^{-4} ~_{-6.51 \times 10^{-5}}^{+1.47 \times 10^{-4}}$ &  $2.84_{-0.98}^{+1.30}$ &   > 6 \\
 21 &  1 &  $3098.27_{-0.21}^{+0.21}$ &  $3.84 \times 10^{-3} ~_{-1.87 \times 10^{-3}}^{+7.18 \times 10^{-3}}$ &  $1.06_{-0.45}^{+0.48}$ &   > 6 \\
 21 &  2 &  $3159.85_{-0.31}^{+0.32}$ &  $9.44 \times 10^{-4} ~_{-3.89 \times 10^{-4}}^{+1.02 \times 10^{-3}}$ &  $1.90_{-0.67}^{+0.79}$ &   > 6 \\
 22 &  0 &  $3168.35_{-0.19}^{+0.17}$ &  $4.62 \times 10^{-3} ~_{-2.57 \times 10^{-3}}^{+1.20 \times 10^{-2}}$ &  $0.70_{-0.35}^{+0.41}$ &   > 6 \\
 21 &  3 &  $3217.52_{-0.50}^{+0.55}$ &  $1.26 \times 10^{-4} ~_{-4.41 \times 10^{-5}}^{+9.63 \times 10^{-5}}$ &  $3.49_{-1.25}^{+1.62}$ &   > 6 \\
 22 &  1 &  $3232.84_{-0.24}^{+0.23}$ &  $2.99 \times 10^{-3} ~_{-1.33 \times 10^{-3}}^{+4.25 \times 10^{-3}}$ &  $1.34_{-0.50}^{+0.53}$ &   > 6 \\
 22 &  2 &  $3295.71_{-0.45}^{+0.50}$ &  $7.01 \times 10^{-4} ~_{-2.39 \times 10^{-4}}^{+5.10 \times 10^{-4}}$ &  $3.22_{-1.04}^{+1.36}$ &   > 6 \\
 23 &  0 &  $3303.97_{-0.29}^{+0.26}$ &  $1.36 \times 10^{-3} ~_{-6.83 \times 10^{-4}}^{+2.82 \times 10^{-3}}$ &  $1.20_{-0.63}^{+0.82}$ &   > 6 \\
 22 &  3 &  $3352.82_{-0.60}^{+0.86}$ &  $6.62 \times 10^{-5} ~_{-2.37 \times 10^{-5}}^{+5.94 \times 10^{-5}}$ &  $3.61_{-1.68}^{+2.21}$ &   > 6 \\
 23 &  1 &  $3368.01_{-0.28}^{+0.27}$ &  $1.72 \times 10^{-3} ~_{-7.00 \times 10^{-4}}^{+1.81 \times 10^{-3}}$ &  $1.82_{-0.60}^{+0.65}$ &   > 6 \\
 23 &  2 &  $3430.18_{-0.32}^{+0.43}$ &  $9.38 \times 10^{-4} ~_{-4.33 \times 10^{-4}}^{+1.45 \times 10^{-3}}$ &  $1.52_{-0.70}^{+1.04}$ &   > 6 \\
 24 &  0 &  $3437.74_{-0.42}^{+0.53}$ &  $5.77 \times 10^{-4} ~_{-1.97 \times 10^{-4}}^{+4.22 \times 10^{-4}}$ &  $3.12_{-1.02}^{+1.18}$ &   > 6 \\
 23 &  3 &  $3489.13_{-0.48}^{+0.53}$ &  $1.01 \times 10^{-4} ~_{-4.03 \times 10^{-5}}^{+9.90 \times 10^{-5}}$ &  $2.55_{-0.98}^{+1.30}$ &   > 6 \\
 24 &  1 &  $3504.49_{-0.35}^{+0.34}$ &  $7.99 \times 10^{-4} ~_{-2.71 \times 10^{-4}}^{+5.51 \times 10^{-4}}$ &  $2.86_{-0.75}^{+0.83}$ &   > 6 \\
 24 &  2 &  $3566.57_{-0.34}^{+0.38}$ &  $5.51 \times 10^{-4} ~_{-2.26 \times 10^{-4}}^{+6.29 \times 10^{-4}}$ &  $2.18_{-0.93}^{+1.18}$ &   > 6 \\
 25 &  0 &  $3574.68_{-0.68}^{+0.57}$ &  $2.11 \times 10^{-4} ~_{-7.06 \times 10^{-5}}^{+1.53 \times 10^{-4}}$ &  $3.51_{-1.25}^{+1.90}$ &   > 6 \\
 24 &  3 &  $3626.53_{-1.00}^{+0.96}$ &  $4.57 \times 10^{-5} ~_{-1.29 \times 10^{-5}}^{+2.03 \times 10^{-5}}$ &  $6.06_{-1.68}^{+1.31}$ &   > 6 \\
 25 &  1 &  $3640.28_{-0.42}^{+0.40}$ &  $3.65 \times 10^{-4} ~_{-1.23 \times 10^{-4}}^{+2.54 \times 10^{-4}}$ &  $3.20_{-0.92}^{+1.09}$ &   > 6 \\
 25 &  2 &  $3704.18_{-1.25}^{+0.95}$ &  $1.39 \times 10^{-4} ~_{-4.25 \times 10^{-5}}^{+5.94 \times 10^{-5}}$ &  $5.95_{-2.10}^{+1.43}$ &   > 6 \\
 26 &  0 &  $3711.80_{-1.62}^{+0.97}$ &  $1.01 \times 10^{-4} ~_{-4.21 \times 10^{-5}}^{+1.14 \times 10^{-4}}$ &  $4.41_{-3.53}^{+2.44}$ &   > 6 \\
 25 &  3 &  $3762.86_{-1.31}^{+1.47}$ &  $4.54 \times 10^{-6} ~_{-4.46 \times 10^{-6}}^{+1.80 \times 10^{-4}}$ &  $0.00_{-0.00}^{+0.57}$ &  0.03 \\
 26 &  1 &  $3776.36_{-0.61}^{+0.60}$ &  $1.87 \times 10^{-4} ~_{-3.67 \times 10^{-5}}^{+6.38 \times 10^{-5}}$ &  $6.54_{-1.20}^{+0.96}$ &   > 6 \\
\hline
\end{tabular}
\end{adjustbox}
\endgroup
\label{table:pkb_hmi}
\end{table*}

\begin{table*}[!ht]
\centering
\caption{Same as Table~\ref{table:pkb_hmi} but for the HMI spectrum obtained with the series multiplied by the Solar-SONG-like window.} 
\begingroup
\renewcommand{\arraystretch}{1.3}
\begin{adjustbox}{totalheight = .8\textheight}
\begin{tabular}{rrrrrrrrr}
\hline\hline
  $n$ &  $\ell$ &  $\nu$ & $H$ & $\Gamma$ & $\ln K$  \\
  &   &  ($\mu$Hz) &  (${\rm m^2 s^{-2} \mu Hz^{-1}}$) &  ($\mu$Hz) &   \\
\hline
 11 &  1 &  $1748.95_{-0.17}^{+0.14}$ &  $1.01 \times 10^{-4} ~_{-4.39 \times 10^{-5}}^{+1.19 \times 10^{-4}}$ &  $0.72_{-0.31}^{+0.42}$ &   > 6 \\
 11 &  2 &  $1810.41_{-0.32}^{+0.28}$ &  $9.90 \times 10^{-5} ~_{-6.19 \times 10^{-5}}^{+5.74 \times 10^{-4}}$ &  $0.52_{-0.47}^{+0.49}$ &  2.50 \\
 12 &  1 &  $1884.78_{-0.39}^{+0.39}$ &  $2.76 \times 10^{-5} ~_{-1.77 \times 10^{-5}}^{+4.07 \times 10^{-4}}$ &  $0.72_{-0.70}^{+1.28}$ &   > 6 \\
 12 &  2 &  $1946.44_{-0.22}^{+0.11}$ &  $2.44 \times 10^{-4} ~_{-1.98 \times 10^{-4}}^{+7.05 \times 10^{-3}}$ &  $0.13_{-0.11}^{+0.36}$ &  2.54 \\
 13 &  1 &  $2020.82_{-0.22}^{+0.24}$ &  $6.07 \times 10^{-5} ~_{-2.68 \times 10^{-5}}^{+7.86 \times 10^{-5}}$ &  $0.87_{-0.40}^{+0.53}$ &   > 6 \\
 13 &  3 &  $2136.06_{-0.19}^{+0.65}$ &  $6.47 \times 10^{-5} ~_{-4.84 \times 10^{-5}}^{+8.24 \times 10^{-4}}$ &  $0.36_{-0.34}^{+0.75}$ &  1.97 \\
 14 &  1 &  $2156.68_{-0.24}^{+0.24}$ &  $7.38 \times 10^{-5} ~_{-2.99 \times 10^{-5}}^{+7.38 \times 10^{-5}}$ &  $1.24_{-0.57}^{+0.82}$ &   > 6 \\
 15 &  0 &  $2228.33_{-0.29}^{+0.26}$ &  $1.58 \times 10^{-4} ~_{-6.99 \times 10^{-5}}^{+2.12 \times 10^{-4}}$ &  $1.32_{-0.77}^{+0.84}$ &   > 6 \\
 15 &  1 &  $2292.08_{-0.15}^{+0.16}$ &  $3.04 \times 10^{-4} ~_{-1.13 \times 10^{-4}}^{+2.34 \times 10^{-4}}$ &  $0.99_{-0.32}^{+0.40}$ &   > 6 \\
 15 &  2 &  $2352.28_{-0.54}^{+0.28}$ &  $3.07 \times 10^{-4} ~_{-1.75 \times 10^{-4}}^{+1.34 \times 10^{-3}}$ &  $0.90_{-0.78}^{+0.86}$ &  3.69 \\
 16 &  0 &  $2362.69_{-1.60}^{+2.51}$ &  $1.63 \times 10^{-4} ~_{-1.57 \times 10^{-4}}^{+6.08 \times 10^{-3}}$ &  $0.12_{-0.12}^{+1.95}$ &   > 6 \\
 16 &  1 &  $2425.22_{-0.23}^{+0.23}$ &  $2.48 \times 10^{-4} ~_{-7.43 \times 10^{-5}}^{+1.21 \times 10^{-4}}$ &  $1.77_{-0.45}^{+0.54}$ &   > 6 \\
 17 &  0 &  $2496.18_{-0.52}^{+0.29}$ &  $4.53 \times 10^{-4} ~_{-2.59 \times 10^{-4}}^{+4.43 \times 10^{-4}}$ &  $1.70_{-0.64}^{+0.93}$ &   > 6 \\
 17 &  1 &  $2558.82_{-0.17}^{+0.17}$ &  $8.82 \times 10^{-4} ~_{-2.79 \times 10^{-4}}^{+5.32 \times 10^{-4}}$ &  $1.35_{-0.35}^{+0.40}$ &   > 6 \\
 17 &  2 &  $2619.16_{-3.09}^{+1.12}$ &  $1.49 \times 10^{-3} ~_{-1.48 \times 10^{-3}}^{+1.20 \times 10^{-2}}$ &  $0.25_{-0.25}^{+0.58}$ &  1.15 \\
 18 &  0 &  $2630.12_{-2.09}^{+0.51}$ &  $5.78 \times 10^{-4} ~_{-4.18 \times 10^{-4}}^{+1.45 \times 10^{-3}}$ &  $1.81_{-0.75}^{+1.80}$ &   > 6 \\
 17 &  3 &  $2675.32_{-0.46}^{+0.46}$ &  $9.94 \times 10^{-5} ~_{-4.00 \times 10^{-5}}^{+7.74 \times 10^{-5}}$ &  $2.51_{-1.11}^{+1.71}$ &   > 6 \\
 18 &  1 &  $2693.47_{-0.16}^{+0.15}$ &  $1.29 \times 10^{-3} ~_{-4.12 \times 10^{-4}}^{+7.78 \times 10^{-4}}$ &  $1.22_{-0.31}^{+0.35}$ &   > 6 \\
 18 &  2 &  $2754.85_{-0.62}^{+0.44}$ &  $9.37 \times 10^{-4} ~_{-5.10 \times 10^{-4}}^{+1.46 \times 10^{-3}}$ &  $1.17_{-0.56}^{+0.70}$ &   > 6 \\
 19 &  0 &  $2765.13_{-0.32}^{+0.31}$ &  $1.39 \times 10^{-3} ~_{-7.95 \times 10^{-4}}^{+1.65 \times 10^{-3}}$ &  $1.28_{-0.49}^{+0.70}$ &   > 6 \\
 18 &  3 &  $2811.62_{-0.23}^{+0.21}$ &  $2.72 \times 10^{-4} ~_{-1.31 \times 10^{-4}}^{+4.55 \times 10^{-4}}$ &  $1.01_{-0.62}^{+1.13}$ &   > 6 \\
 19 &  1 &  $2828.13_{-0.18}^{+0.18}$ &  $1.31 \times 10^{-3} ~_{-3.81 \times 10^{-4}}^{+6.74 \times 10^{-4}}$ &  $1.63_{-0.38}^{+0.43}$ &   > 6 \\
 19 &  2 &  $2889.17_{-0.67}^{+0.73}$ &  $7.60 \times 10^{-4} ~_{-3.41 \times 10^{-4}}^{+6.54 \times 10^{-4}}$ &  $2.76_{-0.77}^{+1.00}$ &   > 6 \\
 20 &  0 &  $2899.29_{-0.48}^{+0.53}$ &  $9.29 \times 10^{-4} ~_{-7.74 \times 10^{-4}}^{+2.58 \times 10^{-3}}$ &  $0.87_{-0.85}^{+1.08}$ &   > 6 \\
 19 &  3 &  $2946.88_{-0.32}^{+0.33}$ &  $2.52 \times 10^{-4} ~_{-8.25 \times 10^{-5}}^{+1.43 \times 10^{-4}}$ &  $2.26_{-0.70}^{+0.95}$ &   > 6 \\
 20 &  1 &  $2963.09_{-0.14}^{+0.14}$ &  $2.53 \times 10^{-3} ~_{-8.04 \times 10^{-4}}^{+1.52 \times 10^{-3}}$ &  $1.18_{-0.29}^{+0.32}$ &   > 6 \\
 20 &  2 &  $3024.11_{-0.26}^{+0.27}$ &  $2.68 \times 10^{-3} ~_{-1.20 \times 10^{-3}}^{+2.79 \times 10^{-3}}$ &  $1.26_{-0.49}^{+0.59}$ &   > 6 \\
 21 &  0 &  $3033.76_{-0.30}^{+0.42}$ &  $2.39 \times 10^{-3} ~_{-1.08 \times 10^{-3}}^{+3.09 \times 10^{-3}}$ &  $1.18_{-0.59}^{+0.69}$ &   > 6 \\
 20 &  3 &  $3081.56_{-0.64}^{+0.80}$ &  $1.80 \times 10^{-4} ~_{-4.63 \times 10^{-5}}^{+7.18 \times 10^{-5}}$ &  $5.00_{-1.38}^{+1.68}$ &  5.97 \\
 21 &  1 &  $3098.60_{-0.17}^{+0.16}$ &  $2.49 \times 10^{-3} ~_{-7.86 \times 10^{-4}}^{+1.43 \times 10^{-3}}$ &  $1.41_{-0.35}^{+0.42}$ &   > 6 \\
 21 &  2 &  $3159.62_{-0.38}^{+0.34}$ &  $9.29 \times 10^{-4} ~_{-3.43 \times 10^{-4}}^{+7.40 \times 10^{-4}}$ &  $1.95_{-0.82}^{+0.94}$ &   > 6 \\
 22 &  0 &  $3168.28_{-0.24}^{+0.28}$ &  $2.15 \times 10^{-3} ~_{-9.22 \times 10^{-4}}^{+2.88 \times 10^{-3}}$ &  $1.07_{-0.53}^{+0.69}$ &   > 6 \\
 21 &  3 &  $3217.63_{-0.63}^{+0.73}$ &  $1.49 \times 10^{-4} ~_{-6.27 \times 10^{-5}}^{+1.18 \times 10^{-4}}$ &  $2.85_{-1.25}^{+1.80}$ &   > 6 \\
 22 &  1 &  $3233.10_{-0.28}^{+0.28}$ &  $1.31 \times 10^{-3} ~_{-3.22 \times 10^{-4}}^{+4.90 \times 10^{-4}}$ &  $2.86_{-0.58}^{+0.71}$ &   > 6 \\
 22 &  2 &  $3295.92_{-1.24}^{+0.84}$ &  $4.19 \times 10^{-4} ~_{-1.15 \times 10^{-4}}^{+1.84 \times 10^{-4}}$ &  $5.90_{-1.82}^{+1.39}$ &  4.04 \\
 23 &  0 &  $3304.24_{-0.32}^{+0.27}$ &  $1.68 \times 10^{-3} ~_{-9.47 \times 10^{-4}}^{+3.91 \times 10^{-3}}$ &  $0.94_{-0.58}^{+1.52}$ &   > 6 \\
 22 &  3 &  $3352.46_{-0.46}^{+0.58}$ &  $1.26 \times 10^{-4} ~_{-4.82 \times 10^{-5}}^{+8.75 \times 10^{-5}}$ &  $2.78_{-1.11}^{+1.77}$ &   > 6 \\
 23 &  1 &  $3368.59_{-0.18}^{+0.18}$ &  $1.76 \times 10^{-3} ~_{-5.05 \times 10^{-4}}^{+8.81 \times 10^{-4}}$ &  $1.67_{-0.38}^{+0.45}$ &   > 6 \\
 23 &  2 &  $3429.73_{-0.41}^{+0.43}$ &  $6.85 \times 10^{-4} ~_{-2.97 \times 10^{-4}}^{+1.16 \times 10^{-3}}$ &  $1.79_{-1.18}^{+4.03}$ &  4.08 \\
 24 &  0 &  $3437.90_{-0.55}^{+0.72}$ &  $5.35 \times 10^{-4} ~_{-1.91 \times 10^{-4}}^{+4.23 \times 10^{-4}}$ &  $3.84_{-2.37}^{+1.92}$ &   > 6 \\
 23 &  3 &  $3489.16_{-0.79}^{+0.77}$ &  $1.00 \times 10^{-4} ~_{-4.55 \times 10^{-5}}^{+9.16 \times 10^{-5}}$ &  $3.12_{-1.68}^{+2.13}$ &  4.16 \\
 24 &  1 &  $3504.99_{-0.45}^{+0.43}$ &  $5.81 \times 10^{-4} ~_{-1.38 \times 10^{-4}}^{+2.01 \times 10^{-4}}$ &  $4.06_{-0.89}^{+1.34}$ &   > 6 \\
 24 &  2 &  $3566.27_{-0.47}^{+0.33}$ &  $5.87 \times 10^{-4} ~_{-2.23 \times 10^{-4}}^{+1.07 \times 10^{-3}}$ &  $2.97_{-2.29}^{+2.97}$ &   > 6 \\
 25 &  0 &  $3573.71_{-0.51}^{+1.65}$ &  $2.62 \times 10^{-4} ~_{-2.45 \times 10^{-4}}^{+6.72 \times 10^{-4}}$ &  $1.93_{-1.93}^{+3.56}$ &   > 6 \\
 24 &  3 &  $3627.25_{-1.67}^{+0.53}$ &  $6.89 \times 10^{-5} ~_{-6.33 \times 10^{-5}}^{+5.09 \times 10^{-4}}$ &  $0.63_{-0.63}^{+4.96}$ &  1.77 \\
 25 &  1 &  $3639.98_{-0.62}^{+0.64}$ &  $2.34 \times 10^{-4} ~_{-5.06 \times 10^{-5}}^{+7.84 \times 10^{-5}}$ &  $5.78_{-1.53}^{+1.42}$ &   > 6 \\
 25 &  2 &  $3703.97_{-1.66}^{+1.12}$ &  $1.49 \times 10^{-4} ~_{-1.46 \times 10^{-4}}^{+1.98 \times 10^{-4}}$ &  $2.83_{-2.83}^{+4.22}$ &  1.27 \\
 26 &  0 &  $3710.92_{-1.35}^{+1.88}$ &  $1.89 \times 10^{-4} ~_{-8.76 \times 10^{-5}}^{+3.35 \times 10^{-4}}$ &  $6.34_{-6.19}^{+1.29}$ &   > 6 \\
 25 &  3 &  $3762.92_{-1.45}^{+1.41}$ &  $9.86 \times 10^{-6} ~_{-9.71 \times 10^{-6}}^{+5.08 \times 10^{-4}}$ &  $0.00_{-0.00}^{+0.19}$ &  0.16 \\
 26 &  1 &  $3777.40_{-0.83}^{+0.76}$ &  $1.60 \times 10^{-4} ~_{-2.51 \times 10^{-5}}^{+3.48 \times 10^{-5}}$ &  $6.94_{-1.11}^{+0.74}$ &   > 6 \\
\hline
\end{tabular}
\end{adjustbox}
\endgroup
\label{table:pkb_hmi_wdw}
\end{table*}

\section{Solar-SONG data reduction module \label{appendic:apollinaire_songlib}}

The reduction process described in Sect.~\ref{section:reduction} can be performed with \texttt{songlib} submodule of \texttt{apollinaire}. The \texttt{standard\_correction} function has been designed to process the \texttt{iSONG} cube outputs. The default settings of the function arguments are the ones that have been used to obtain the data used in this paper.  

\end {appendix}

\end{document}